\documentclass[useAMS,usenatbib]{mn2e}

\usepackage{graphicx}
\usepackage{amssymb}
\usepackage{subfigure}
\usepackage{captcont}
\usepackage{multirow}
\usepackage{longtable}
\usepackage{comment}
\usepackage{url}
\usepackage{lscape}
\usepackage{booktabs,multirow}
\usepackage{float}
\newcommand{\kepler}{\emph{Kepler}}

\title[WASP high-frequency A-type pulsators]{High frequency A-type pulsators discovered using SuperWASP\thanks{Based on service observations made with the WHT operated on the island of La Palma by the Isaac Newton Group in the Spanish Observatorio del Roque de los Muchachos of the Instituto de Astrof\'{i}sica de Canarias.}\thanks{Based on observations made with the Southern African Large Telescope (SALT)}}
\author[D. L. Holdsworth et al.]{Daniel L. Holdsworth$^{1}$\thanks{E-mail:d.l.holdsworth@keele.ac.uk},
B. Smalley$^{1}$,
M. Gillon$^{2}$,
K. I. Clubb$^{3}$,
J. Southworth$^{1}$,
\newauthor P. F. L. Maxted$^{1}$,
D. R. Anderson$^{1}$,
S. C. C. Barros$^{4}$,
A. Collier Cameron$^{5}$,
\newauthor L. Delrez$^{2}$,
F. Faedi$^{6}$,
C. A. Haswell$^{7}$,
C. Hellier$^{1}$,
K. Horne$^{5}$,
E. Jehin$^{2}$,
\newauthor A. J. Norton$^{7}$,
D. Pollacco$^{6}$,
I. Skillen$^{8}$,
A. M. S. Smith$^{9}$,
R. G. West$^{6}$,
\newauthor and P. J. Wheatley$^{6}$\\
$^{1}$ Astrophysics Group, Keele University, Staffordshire, ST5 5BG, United Kingdom\\
$^{2}$ Institut d'Astrophysique et de G\'eophysique, Universit\'e de Li\`ege, All\'ee du 6 Ao\^ut, 17, Bat. B5C, Li\`ege 1, Belgium\\
$^{3}$ Department of Astronomy, University of California, Berkeley, CA 94720-3411, USA\\
$^{4}$ Aix-Marseille Université, CNRS, LAM (Laboratoire d'Astrophysique de Marseille) UMR 7326, 13388 Marseille, France\\
$^{5}$ SUPA, School of Physics \& Astronomy, University of St. Andrews, North Haugh, Fife, KY16 9SS, UK\\
$^{6}$ Department of Physics, University of Warwick, Coventry, CV4 7AL, UK\\
$^{7}$ Department of Physical Sciences, The Open University, Walton Hall, Milton Keynes, MK7 6AA, UK\\
$^{8}$ Isaac Newton Group of Telescopes, Apartado de Correos 321, 38700 Santa Cruz de la Palma, Tenerife, Spain\\
$^{9}$ N. Copernicus Astronomical Centre, Polish Academy of Sciences, Bartycka 18, 00-716 Warsaw, Poland\\
}
\begin{document}

\date{\today}

\pagerange{\pageref{firstpage}--19} \pubyear{2014}

\maketitle

\label{firstpage}

\begin{abstract}
We present the results of a survey using the WASP archive to search for high frequency pulsations in F-, A- and B-type stars. Over $1.5$~million targets have been searched for pulsations with amplitudes greater than $0.5$ millimagnitude. We identify over $350$~stars which pulsate with periods less than $30$ min. Spectroscopic follow-up of selected targets has enabled us to confirm $10$~new rapidly oscillating Ap stars, $13$~pulsating Am stars and the fastest known $\delta$~Scuti star. We also observe stars which show pulsations in both the high-frequency domain and in the low-frequency $\delta$ Scuti range. This work shows the power of the WASP photometric survey to find variable stars with amplitudes well below the nominal photometric precision per observation.

\end{abstract}

\begin{keywords}
asteroseismology -- stars: chemically peculiar -- stars: oscillations -- stars: variables: $\delta$~Scuti -- techniques: photometric -- surveys.
\end{keywords}

\section{Introduction}
\label{sec:intro}

With the advent of large ground-based photometric surveys (e.g. WASP, \citealt{pollacco06}; HATnet, \citealt{bakos04}; ASAS, \citealt{pojmanski97}; OGLE, \citealt{udalski92}) there is a wealth of photometric data on millions of stars. Despite not being the prime science goal, these surveys can be exploited to probe stellar variability across the entire sky \citep[e.g.][]{norton11,smalley13}. The ability of these surveys to achieve millimagnitude (mmag) precision provides a vast database in which to search for low-amplitude stellar variability.

The WASP project is a wide-field survey for transiting exoplanets. The project is a two-site campaign: the first instrument is located at the Observatorio del Roque de los Muchachos on La Palma and achieved first light in November $2003$, the second is located at the Sutherland Station of the SAAO and achieved first light in December $2005$. Each instrument consists of eight $200$ mm, f$/1.8$~Canon telephoto lenses backed by Andor CCDs of $2048\times2048$~pixels observing $\sim61$~deg$^{2}$~each through broadband filters covering a wavelength range of $400-700$~nm \citep{pollacco06}. This set-up enables simultaneous observations of up to $8$~fields with a pixel size of $13.7$\arcsec. The instruments capture two consecutive $30$ s integrations at a given pointing, then move to the next observable field. Typically, fields are revisited every $10$ min. 

The images collected are passed through the reduction pipeline, where the data are corrected for primary and secondary extinction, the instrumental colour response and the system zero-point. The atmospheric extinction correction uses a network of stars with a known ($B-V$) colour to determine the extinction terms, which are then applied to all extracted stars using an assumed colour of G-type stars. This process results in a `WASP $V$' magnitude which is comparable to the Tycho-$2$ $V_t$~passband. The data are also corrected for systematic errors using the SysRem algorithm of \cite{tamuz05}. Aperture photometry is performed at stellar positions provided by the USNO-B$1.0$~input catalogue \citep{monet03}. Stars brighter than $\sim15^{th}$~magnitude are extracted and provided with a unique WASP ID of the format $1$SWASPJ$hhmmss.ss\pm ddmmss.s$. Data are stored in {\sc{fits}} format with labels of the observed field, camera and date of observation. Such a configuration and extended time-base allows the extraction of multiple lightcurves for each object based on either date, field or camera. 

To date there are over $428$~billion data points in the archive covering over $31$~million unique objects. With such a large database of objects it is possible to search for a wide variety of stellar variability.

In this paper we present the results from a survey of the WASP archive in the search of rapidly varying stars. We focus on stars hotter than mid F-type in the region of the Hertzsprung-Russell (HR) diagram where the classical instability strip intersects the main-sequence. In this region we find the classical $\delta$~Scuti ($\delta$~Sct) pulsators, the non-magnetic metallic-lined (Am) stars, the strongly-magnetic, chemically peculiar, rapidly oscillating Ap (roAp) stars, the metal deficient $\lambda$~Bo\"{o}tes stars, the $\gamma$~Doradus ($\gamma$~Dor) pulsators and the SX Phoenicis variables.

Most of the stars in this region of the HR diagram are $\delta$~Sct stars which show normal chemical abundances in their atmospheres. However, at about A$8$, $50$~per cent of the A-type stars are in fact Am stars \citep{smith73}. These stars show a discrepancy of at least five spectral subclasses between their hydrogen line spectral class, their Ca K line strength, and their metallic line strengths \citep{conti70}. If the differences are less than five subclasses they are designated `marginal' Am stars (denoted Am:). It was previously thought that the Am stars cannot pulsate with amplitudes greater than $2.0$ mmag, if at all \citep{breger70}, however a study by \citet{smalley11} has shown that $\sim14$~per cent of known Am stars pulsate at the mmag level. Their study concluded that the pulsations in Am stars must be laminar so as not to produce sub-cm s$^{-1}$~turbulence which would homogenize the star. The driving mechanism in the Am stars is the opacity ($\kappa$) mechanism acting on helium in the He {\sc{ii}} ionisation zone, resulting in both radial and non-radial pressure modes (p-modes) \citep[see][]{aerts10}.

Rarer than the Am stars are the chemically peculiar Ap stars. Constituting only about $10$~per cent of the A-type stars \citep{wolff68}, the Ap stars are strongly-magnetic and have over-abundances of rare-earth elements. Around $50$~Ap stars are known to be roAp stars \citep{kurtz82}. These stars show high-overtone p-mode pulsations in the range of $5-24$~min. The driving mechanism is the same as that for the Am stars, the $\kappa$~mechanism, but it is acting in the hydrogen ionization zone instead. The pulsational axis in the roAp stars is thought to be aligned with the magnetic field axis rather than the rotational axis, leading to the oblique pulsator model of \citet{kurtz82}. Due to the strong global magnetic fields, the Ap stars are subject to stratification in their atmospheres, often leading to surface brightness anomalies in the form of chemical spots \citep{stibbs50}. These spots can be stable for many decades, allowing for an accurate determination of their rotation period from the lightcurve.
The roAp stars provide the best test-bed, beyond the Sun, to study the interactions between pulsations and strong global magnetic fields, as well as testing gravitational settling and radiative levitation theories \citep[e.g.][]{gautschy98,balmforth01,kurtz11}.

Hybrid pulsators, which show both p-mode and gravity (g-) mode pulsations, are also found in this region of the HR diagram. At the base of the instability strip there is an area where the $\gamma$~Dor stars, g-mode pulsators, overlap with the p-mode $\delta$~Sct stars. These objects are of great scientific interest as it becomes possible to probe both the core and atmosphere simultaneously. Observations made with the {\textit{Kepler}}\, spacecraft of these hybrid stars \citep{grigahcene10,catanzaro11} have shown that these objects are more common than previously thought, with the possibility that nearly all stars in this region of the HR diagram show both $\gamma$~Dor and $\delta$~Sct pulsations \citep{grigahcene10}.

Finally, we also have the pre-main-sequence A-type stars in this region of the HR diagram. These stars are among the fastest known $\delta$~Sct pulsators, with the fastest being HD~$34282$~with a period of $18$ min \citep{amado04}. These targets often show multi-mode periodograms, with `noisy' lightcurves due to dusty circumstellar environments. As a result of this environment many of these objects are heavily reddened.

Previous surveys for pulsations in the A-type stars have targeted objects that were already known to be spectroscopically interesting A stars \citep[e.g.][]{kochukhov13,paunzen12,smalley11}. This approach limits the results to specific types of pulsators. However, the approach we have adopted in our study requires no previous knowledge of the targets, except for a rough photometric spectral type. This has allowed us to search for all types of pulsations in F-, A- and B-type stars and will permit the possible discovery of new types, thus the aim of our study is to identify stars which show variability with a period of less than $30$ min. This blind survey enables us to approach the search for different types of pulsators in a novel way, opening our results to pulsating Am stars, roAp stars, fast $\delta$~Sct stars and pre-main-sequence stars.

\section{Archive Survey}
\label{initial}

\subsection{Determining WASP detection limits}
\label{limits}

Before a full archive survey is conducted, we need to understand the capabilities of the WASP data in detecting pulsations. In order to do this, we employ the micromagnitude ($\mu$mag) precision of \kepler~data.

The \kepler~mission, launched in March $2009$, observed over $150~000$ stars in two cadence modes - the long cadence (LC) mode with an effective exposure time of $30$ min was used for the majority of stars, with $512$~stars observed in the short cadence (SC) mode with an effective exposure time of $1$ min \citep{gilliland10}. \kepler\, consisted of an array of $42$~CCDs covering $115$~deg$^2$ of sky in the direction of the constellations Cygnus and Lyra. Observations were made through fixed CCD filters covering $423-897$~nm~\citep{koch10} which is slightly redder than WASP. \kepler~achieved a photometric precision of up to $84$~parts per million \citep{borucki10,koch10} resulting in a large quantity of high cadence data at $\mu$mag precision.

\citet{debosscher11} have conducted a variability study on the first Quarter of \kepler~data.  We used their results to select $59~737$~\kepler~targets which showed a principal frequency with an amplitude of $\ge0.01$ mmag. Corresponding WASP data were extracted from the archive as long as their ($J-H$) colour was less than $0.4$~so to target G stars and hotter, ensuring we account for reddened objects to maximise our sample size. This resulted in a final sample size of $13~060$~stars.

Periodograms for these targets were calculated using the {\sc{fasper}} method of \citet{press89,press92}. A small selection of the resulting periodograms were inspected manually where it was found that periodograms suffered greatly from low-frequency noise and a high `Fourier grass' level -- the approximate background level of the periodogram which resembles mown grass -- at higher frequencies. Examination of the lightcurves revealed many points which significantly deviated from the mean (Fig. \ref{res_mn} top left). To remove these outliers we have adopted a resistant mean algorithm \citep{huber81} which removes data points from the lightcurve which deviate from the median by more than $4\sigma$. A recalculation of the periodograms resulted in a much smoother periodogram with clearly defined peaks (Fig. \ref{res_mn} top right).

\setcounter{figure}{0}
\begin{figure*}
  \centering
  \begin{minipage}{170mm}
    \centering
  \includegraphics[angle=180,width=80mm,trim=20mm 10mm 0mm 10mm,clip]{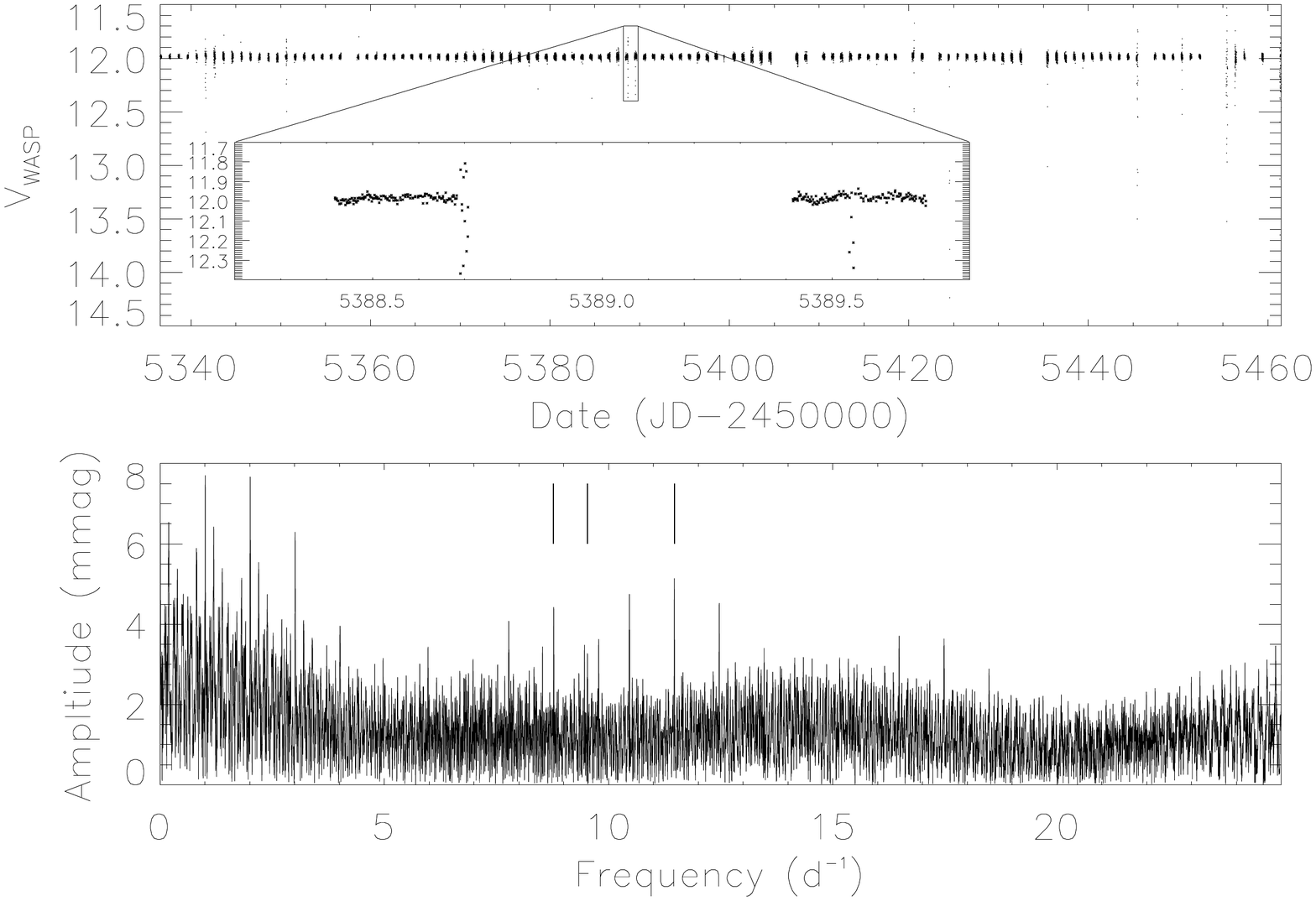}
  \includegraphics[angle=180,width=80mm,trim=20mm 10mm 0mm 10mm,clip]{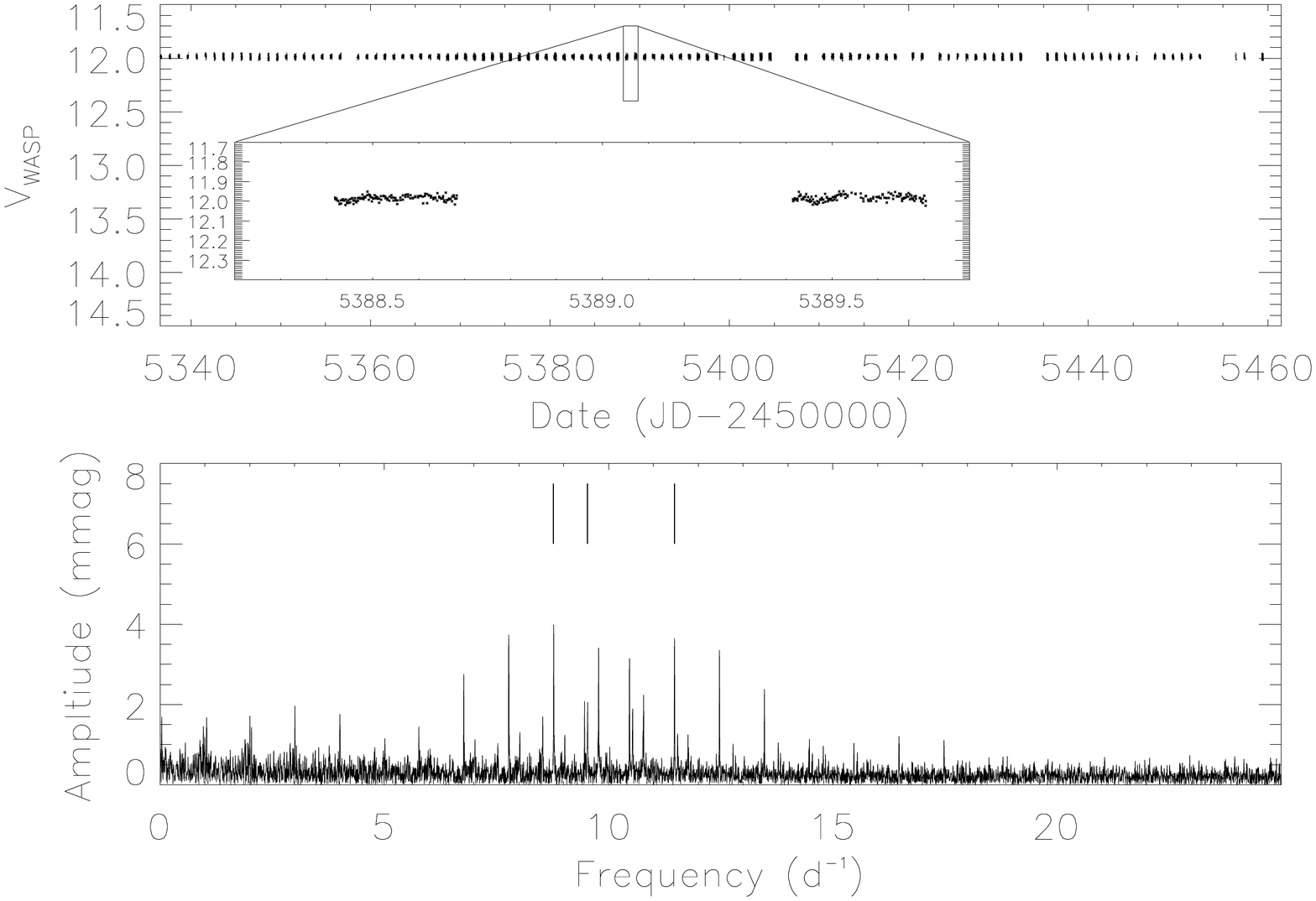}
  \includegraphics[angle=180,width=80mm, trim= 23mm 21mm 38mm 110mm,clip]{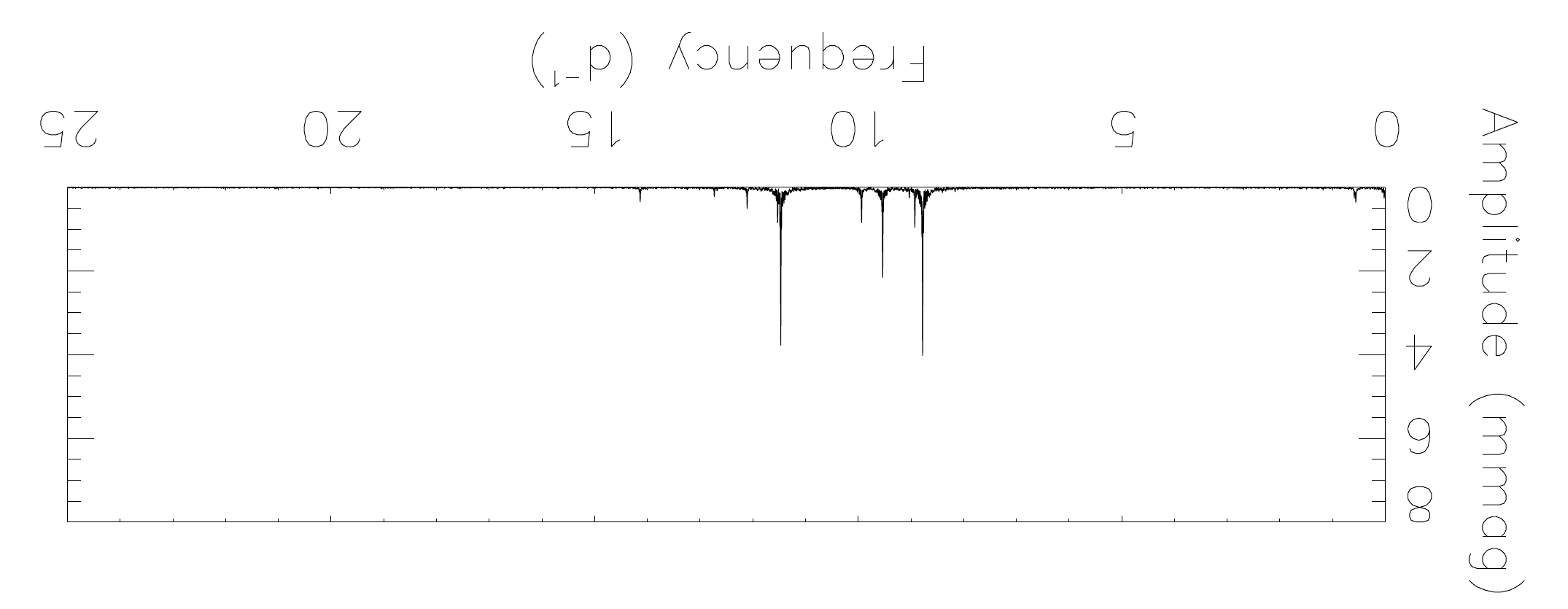}
  \caption{A comparison of the raw (top left), $14~659$~data points, and cut (top right), $13~970$~data points, lightcurves and corresponding periodograms for $1$SWASPJ$192444.63+490052.2$. The much cleaner lightcurve shown in the top right plot results in a clean periodogram, with a Fourier grass level of about $0.5$ mmag, which clearly shows the principal frequency at $8.77$ d$^{-1}$, a secondary peak at $11.47$ d$^{-1}$~and a third peak at $9.53$ d$^{-1}$. The top left periodogram does not identify the principal frequency correctly due to high-amplitude, low-frequency noise; the Fourier grass level is about $2.5$ mmag with the amplitudes of the peaks are also much higher here too. It is clear that the use of resistant mean clipping greatly enhances detection probability. The three pulsation frequencies are marked with vertical lines. The bottom plot shows the corresponding \kepler~data of KIC$~11295729$, the same star as $1$SWASPJ$192444.63+490052.2$.}
  \label{res_mn}
  \end{minipage}
\end{figure*}

All peaks from each WASP periodogram are compared with the corresponding \kepler~target data to determine if the pulsation has been detected. We require a maximum deviation of $0.01$ d$^{-1}$~between the two data sets to conclude that they agree (the Rayleigh criterion for a typical season of WASP data i.e. $\Delta\nu={1}/{\Delta T}$, where $\Delta T$ is the season duration). We find agreement between $11~188$~target stars. Of the non-matches, we determined that if the data set has fewer than $1~000$~data points the frequency is not recovered, and if we have a high level of target blending, due to the large WASP pixel sizes, the amplitudes are diluted to below our detection threshold. Of the targets that do match, we are able to conclude that the minimum detection limit is $0.5$ mmag.

\subsection{Target selection}
\label{sec:selection}

Using $2$MASS colours for stars in the WASP database we select over $1.5$~million F-, A- and B-type stars, requiring ($J-H$) $< 0.25$ (corresponding to F$7$). We also stipulate that the targets have a USNO magnitude of $14\leq r \leq 7$, which are the approximate detection limits of the WASP instruments. Finally we reject stars with fewer than $1~000$~data points as we know from Section \ref{limits} that signals cannot be extracted with so few data.

With our subset of stars we extract each season of data and calculate a periodogram, to a maximum frequency of $300$ d$^{-1}$, as described above. We choose this upper limit to balance computational time with expected detection rate. In all, we calculate over $9$~million individual periodograms, indicating an average of $6$ seasons per star. To increase the speed of later peak identification we do not store the entire periodogram, only the significant peaks. A peak is deemed significant if it has a S/N $>2.5$. To determine the noise level, the periodogram is split into sections of $10$ d$^{-1}$~with the mean of each section calculated. The noise level is then taken to be the lowest value calculated. That is to say, the noise is estimated to be half of the `Fourier grass' level of the most stable section. This value is then used for the entire periodogram. Although not the conventional method to calculate Fourier noise, this method is computationally fast. We have accounted for any over estimates of the noise by lowering the S/N criteria from the widely accepted value of $4$~\citep{breger93,koen10}. For later analysis of individual targets, for which we use P{\sc{eriod}}$04$ \citep{lenz05}, we use a threshold S/N of $4$.

To identify pulsation candidates in the data with confidence, we require certain criteria to be met. A single object must show a peak at the same frequency in more than one season, within a tolerance of $0.01$ d$^{-1}$. We implement this criterion to reduce the possibility of spurious noise being identified as a true signal. We also consider the window function to eliminate any sampling aliases. Targets which satisfy these criteria have their full periodograms calculated and stored for human inspection.

During this selection process we do not consider the effects of blending or overcrowding of the WASP aperture. Due to the large pixel size of the WASP detectors this can be a common occurrence, with an estimated $\sim12$\% of targets suffering $\ge50$\% dilution. However, at the detection stage, we are solely interested in the detection of variability, with later diagnostics to resolve blends or confusion being conducted to ensure the variability is attributed to the correct target (see Section \ref{sec:1940} for an example).

\subsection{Renson \& Manfroid catalogue search}

To understand the capabilities of the WASP data with regards to finding roAp stars, we conducted an initial test using the catalogue of \citet{renson09}. All stars identified as Ap were cross-checked with the WASP database, and extracted with the same criteria as in Section \ref{sec:selection}. This amounted to $543$~Ap stars. Periodograms were calculated using the aforementioned method, and examined if the above criteria were fulfilled. The automatic search resulted in the extraction of just $1$~known roAp star (HD~$12932$), out of a possible $15$~known roAp stars in the subset we searched. 

To decipher why only one object was recovered from the WASP data we first consider the spectral response of the WASP instruments. Designed for exoplanet detection, the broad spectral response dilutes the pulsations which are strongest in {\textit{B}}-band photometry \citep{kurtz90}. Comparing  WASP observations of HD~$12932$~with those in the literature, we see that WASP data suffers an amplitude decrease of $\sim75$~per cent. Given the WASP detection threshold is between $0.5-1$~mmag, we expect the lower limit in {\textit{B}}-band photometry to be between $2-4$~mmag, thus accounting for the non-detection of $10$~targets.

We further investigate why we detect no pulsations for the remaining $4$ targets by calculating for each season of data the weighted reduced-$\chi^2$ \citep{bevington69} which aims to characterize the lightcurve by accounting for the number of data points and the scatter in the lightcurve i.e.
\[
\chi^2/n=\displaystyle{\frac{\Sigma(({\rm{mag}}-\rm{median}({\rm{mag}}))/\sigma)^2}{(n-1)}}.
\]
With this taken into account, it becomes more clear why we do not detect the roAp stars with amplitudes above our expected amplitude threshold. For the four non-detected targets above the threshold, the $\chi^2/n$ value suggests the data are not of high enough quality to consistently detect pulsations. 

This suggests we will only detect a small fraction of the roAp stars that exist in the WASP database. The main reason for the lack of detections of the known targets is the colour response of the WASP instruments.  For the target that was automatically extracted, the amplitude dilution of $75$~per cent indicates that we will only observe the highest amplitude pulsators. 

Of the $543$~Ap stars we studied, we detected no new roAp stars amongst the sample. We are, of course, limited by our threshold of $0.5$ mmag which provides an upper limit on any pulsations in the WASP {\textit{V}}-band photometry.

\setcounter{table}{0}
\begin{table*}
  \tiny
  \centering
  \begin{minipage}{\textwidth}
    \centering
    \centering
    \caption{Renson and WASP coincident roAp stars. Columns $9-16$~are a measure of the quality of the WASP data for each season.}
  \label{tab:renson}
    \begin{tabular}{ccccccccccccccccc}

    \hline
    \hline

      \multicolumn{1}{c}{\multirow{4}{*}{HD}} & 
      \multicolumn{3}{c}{Literature Data}&
      \multicolumn{1}{c}{}&
      \multicolumn{12}{c}{WASP Data}\\

      \cline{2-4}
      \cline{6-17}\\

      \multicolumn{1}{c}{} & 
      \multicolumn{1}{c}{Mag}&
      \multicolumn{1}{c}{$\nu$}&
      \multicolumn{1}{c}{$\Delta~B$}&
      \multicolumn{1}{c}{}&
      \multicolumn{1}{c}{Mag}&
      \multicolumn{1}{c}{$\nu$}&
      \multicolumn{1}{c}{$\Delta~V$}&
      \multicolumn{1}{c}{$N^{\circ}$~of}&
      \multicolumn{8}{c}{\multirow{2}{*}{${\chi^2}/{n}$}}\\

      \multicolumn{1}{c}{} & 
      \multicolumn{1}{c}{(\textit{B})}&
      \multicolumn{1}{c}{(d$^{-1}$)}&
      \multicolumn{1}{c}{(mmag)}&
      \multicolumn{1}{c}{}&
      \multicolumn{1}{c}{(\textit{V}$_{\rm WASP}$)}&
      \multicolumn{1}{c}{(d$^{-1}$)}&
      \multicolumn{1}{c}{(mmag)}&
      \multicolumn{1}{c}{Seasons}&
      \multicolumn{8}{c}{} \\

      \hline
    6532    & 8.60  & 202.82 & 5   && 8.38  &  --    & --   & 2     &  42.43 & 2.33  &       &       &       &      &       & \\
    9289    & 9.63  & 137.14 & 3.5 && 9.42  &  --    & --   & 4     &  7.46  & 14.93 & 3.73  & 2.66  &       &      &       & \\
    12098   & 8.46  & 189.22 & 3   && 8.23  &  --    &  --  & 1     &  1.82  &       &       &       &       &      &       & \\
    12932   & 10.56 & 124.14 & 4   && 10.28 & 124.10 & 1.03 & 5     &  1.18  & 1.07  & 0.99  & 1.17  & 0.97  &      &       & \\
    84041   & 9.74  & 96.00  & 6   && 9.25  &  --    &  --  & 5     &  44.65 & 9.52  & 3.95  & 7.76  & 6.29  &      &       & \\
    99563   & 8.90  & 134.58 & 10  && 8.50  &  --    &  --  & 4     &  12.28 & 204.96& 11.33 & 2.65  &       &      &       & \\
    101065  & 8.73  & 119.01 & 13  && 8.27  &  --    &  --  & 2     &  2.44  & 5.65  &       &       &       &      &       & \\
    119027  & 10.41 & 165.52 & 2   && 10.19 &  --    &  --  & 4     &  1.58  & 1.27  & 8.69  & 1.79  &       &      &       & \\
    122970  & 8.70  & 129.73 & 2   && 8.33  &  --    &  --  & 4     &  2.89  & 4.60  & 6.95  & 3.90  &       &      &       & \\
    185256  & 10.37 & 141.18 & 3   && 10.10 &  --    &  --  & 7     &  3.77  & 2.18  & 0.934 & 1.47  & 2.91  & 1.43 & 37.82 & \\
    193756  & 9.56  & 110.77 & 1.5 && 9.27  &  --    &  --  & 5     &  3.29  & 0.95  & 1.82  & 4.25  & 3.10  &      &       & \\
    196470  & 10.14 & 133.33 & 0.7 && 9.84  &  --    &  --  & 6     &  3.02  & 4.10  & 3.53  & 4.91  & 2.58  & 4.48 &       & \\
    203932  & 9.10  & 244.07 & 2   && 8.92  &  --    &  --  & 4     &  23.25 & 3.80  & 0.66  & 17.62 &       &      &       & \\
    213637  & 10.05 & 125.22 & 1.5 && 9.73  &  --    &  --  & 8     &  1.52  & 1.63  & 1.53  & 2.06  & 1.52  & 1.42 & 0.89  & 1.36 \\
    218495  & 9.62  & 194.59 & 1   && 9.43  &  --    &  --  & 4     &  7.95  & 6.41  & 5.62  & 7.81  &       &      &       & \\
      \hline

    \end{tabular}
 \end{minipage}
  \normalsize
\end{table*}

\section{Candidate Targets}
\label{targets}

Of the $1.5$~million F-, A- and B-type stars extracted from the WASP archive, we find $375$~stars which show variations on the order of $30$ min or less which are present in two or more seasons of observations. Of these $375$~targets, we obtained spectral follow-up for $37$~stars. The targets were selected for follow-up based on their frequency and amplitude of pulsation. Focusing initially on the objects whose periodograms look like that of the rare roAp stars, then moving to lower frequencies and periodograms which show a more complex pulsation spectrum. In Table \ref{tab:photom} we present photometric information on the $37$~stars for which we obtained spectra. 

\setcounter{table}{1}
\begin{table*}
  \centering
  \begin{minipage}{170mm}
    \centering
    \caption{Photometric data for the $37$~high-frequency WASP pulsators for which spectra were obtained. Column $8$~is the number of data points in the WASP archive for that object after applying our cleaning routine, column $9$~is the principal pulsation frequency above $50$ d$^{-1}$, and column $10$~is the pulsational amplitude in the WASP data.}
    \label{tab:photom}
    \begin{tabular}{cccccccccc}
      \hline
      \hline
      
      \multicolumn{1}{c}{\multirow{1}{*}{Abbreviated}} & 
      \multicolumn{1}{c}{\multirow{1}{*}{Other}} & 
      \multicolumn{1}{c}{\multirow{1}{*}{R.A.}} & 
      \multicolumn{1}{c}{\multirow{1}{*}{Dec}} & 
      \multicolumn{1}{c}{\multirow{2}{*}{\textit{V}}} &
      \multicolumn{1}{c}{\multirow{2}{*}{($J-H$)}}&
      \multicolumn{1}{c}{\multirow{1}{*}{$N^{\circ}$ of}} &
      \multicolumn{1}{c}{\multirow{2}{*}{npts}} & 
      \multicolumn{1}{c}{\multirow{1}{*}{$\nu_{osc}$}} & 
      \multicolumn{1}{c}{\multirow{1}{*}{Amp}} \\
      
      \multicolumn{1}{c}{\multirow{1}{*}{ID}} & 
      \multicolumn{1}{c}{\multirow{1}{*}{ID}} & 
      \multicolumn{1}{c}{\multirow{1}{*}{J2000}} & 
      \multicolumn{1}{c}{\multirow{1}{*}{J2000}} & 
      \multicolumn{1}{c}{} &
      \multicolumn{1}{c}{} &
      \multicolumn{1}{c}{\multirow{1}{*}{Seasons}} &
      \multicolumn{1}{c}{} & 
      \multicolumn{1}{c}{\multirow{1}{*}{(d$^{-1}$)}} & 
      \multicolumn{1}{c}{\multirow{1}{*}{(mmag)}} \\
      
      \hline
      
      J0004  & HD 225186              & 00:04:15.12 &  $-$17:25:29.6 &  9.05   &     $0.152$   &    4   &  13814 & 60.08          &   3.40  \\	
      J0008  & TYC 4-562-1            & 00:08:30.50 &  $+$04:28:18.2 &  10.16  &     $0.087$   &    3   &  23076 & 150.26	  &   0.76  \\    
   J0026$^a$ & TYC 2269-996-1         & 00:26:04.18 &  $+$34:47:32.9 &  10.04  &     $0.232$   &    3   &  13037 & 79.13          &   2.05  \\    
      J0051  & TYC 5270-1900-1        & 00:51:07.36 &  $-$11:08:31.9 &  11.52  &     $0.059$   &    3   &  21913 & 58.04	  &   4.77  \\    
   J0206$^b$ & HD 12932               & 02:06:15.80 &  $-$19:07:26.2 &  10.17  &     $0.105$   &    4   &  15269 & 124.10	  &   1.38  \\    
      J0353  & HD 24355	              & 03:53:23.09 &  $+$25:38:33.3 &  9.65   &     $0.054$   &    3   &  7951  & 224.31	  &   1.65  \\    
      J0410  & HD 26400	              & 04:10:45.78 &  $+$07:17:17.2 &  9.54   &     $0.017$   &    3   &  28629 & 68.22	  &   2.70  \\    
      J0429  & HD 28548	              & 04:29:27.24 &  $-$15:01:51.1 &  9.22   &     $0.038$   &    2   &  22195 & 65.65	  &   4.41  \\    
      J0629  & HD 258048	      & 06:29:56.85 &  $+$32:24:46.9 &  10.52  &     $0.239$   &    3   &  15343 & 169.54	  &   1.49  \\    
      J0651  & TYC 8912-1407-1	      & 06:51:42.17 &  $-$63:25:49.6 &  11.51  &     $0.062$   &    3   &  36597 & 132.38	  &   0.79  \\    
      J0855  & TYC 2488-1241-1 	      & 08:55:22.22 &  $+$32:42:36.3 &  10.80  &     $0.014$   &    3   &  13203 & 197.27	  &   1.40  \\    
      J1110  & HD 97127	              & 11:10:53.91 &  $+$17:03:47.5 &  9.43   &     $0.172$   &    3   &  11184 & 106.61         &   0.66  \\    
      J1215  & HD 106563	      & 12:15:28.17 &  $-$11:24:41.3 &  10.55  &     $-0.021$  &    3   &  16354 &  65.46	  &   1.48  \\    
      J1250  & TYC 297-328-1          & 12:50:56.15 &  $+$05:32:12.9 &  11.25  &     $0.168$   &    4   &  22902 & 68.99	  &   4.30  \\    
      J1403  & HD 122570	      & 14:03:41.51 &  $-$40:51:08.9 &  10.41  &     $0.079$   &    4   &  19613 & 99.12	  &   1.27  \\    
      J1430  & TYC 2553-480-1	      & 14:30:49.64 &  $+$31:47:55.1 &  11.56  &     $0.094$   &    3   &  21181 & 235.54	  &   1.06  \\    
      J1625  & HD 147911	      & 16:25:24.10 &  $-$21:41:18.6 &  9.17   &     $0.186$   &    5   &  21514 & 68.52	  &   6.37  \\    
      J1640  & 2MASS J16400299-0737293 & 16:40:02.99 &  $-$07:37:29.7 &  12.67  &     $0.150$   &    2   &  14511 & 151.93	  &   3.52  \\    
      J1648  & TYC 2062-1188-1	      & 16:48:36.99 &  $+$25:15:48.6 &  9.98   &     $0.123$   &    5   &  43024 & 105.12	  &   0.60  \\    
      J1757  & TYC 2612-1843-1	      & 17:57:26.48 &  $+$32:25:23.7 &  11.58  &     $0.173$   &    5   &  35794 & 63.71	  &   2.46  \\    
      J1758  & 2MASS J17584421+3458339 & 17:58:44.20 &  $+$34:58:33.9 &  12.93  &     $0.11 $   &    5   &  74131 & 71.28	  &   2.49  \\    
      J1844  & TYC 3130-2480-1 	      & 18:44:12.27 &  $+$43:17:51.9 &  11.25  &     $0.072$   &    5   &  39389 & 181.73	  &   1.45  \\    
      J1917  & TYC 7926-99-1	      & 19:17:33.42 &  $-$42:42:07.3 &  11.18  &     $0.144$   &    3   &  12120 & 164.47	  &   1.85  \\    
      J1940  & 2MASS J19400781-4420093 & 19:40:07.81 &  $-$44:20:09.2 &  13.02  &     $0.066$   &    5   &  25071 & 176.39	  &   4.16  \\    
      J1951  & 2MASS J19512756-6446247 & 19:51:27.55 &  $-$64:46:24.5 &  13.26  &     $-0.010$  &    4   &  28567 & 58.43	  &   4.29  \\    
      J2022  & TYC 9311-73-1	      & 20:22:36.50 &  $-$70:11:00.2 &  12.77  &     $0.022$   &    4   &  27249 & 62.54	  &   3.30  \\    
      J2026  & TYC 5762-828-1	      & 20:26:42.62 &  $-$11:52:45.1 &  11.80  &     $0.232$   &    4   &  15957 & 212.66         &   1.62  \\
      J2029  & HD 195061	      & 20:29:33.21 &  $-$18:13:12.2 &  9.81   &     $0.072$   &    6   &  24743 & 28.86          &   2.50  \\    
      J2054  & TYC 525-2319-1	      & 20:54:19.84 &  $+$07:13:11.0 &  9.94   &     $0.020$   &    5   &  31892 & 104.86	  &   1.10  \\    
      J2155  & 2MASS J21553126+0849170 & 21:55:31.26 &  $+$08:49:17.0 &  12.83  &     $0.214$   &    4   &  29563 & 61.34	  &   6.38  \\    
      J2241  & TYC 3218-888-2         & 22:41:54.21 &  $+$40:30:39.1 &  11.45  &     $0.063$   &    3   &  10389 & 75.54	  &   3.32  \\    
      J2254  & TYC 569-353-1	      & 22:54:34.21 &  $+$00:52:46.3 &  12.72  &     $0.163$   &    3   &  33997 & 52.59 	  &   5.06  \\    
      J2255  & TYC 6390-339-1	      & 22:55:20.44 &  $-$18:36:35.3 &  11.95  &     $-0.029$  &    4   &  17232 & 66.96	  &   2.68  \\    
      J2305  & TYC 9131-119-1	      & 23:05:45.31 &  $-$67:19:03.0 &  11.43  &     $-0.019$  &    2   &  16265 & 92.75	  &   1.61  \\    
      J2313  & TYC 577-322-1          & 23:13:26.32 &  $+$02:27:49.5 &  10.72  &     $0.109$   &    3   &  24001 & 60.42	  &   5.97  \\    
      J2345  & 2MASS J23455445-3932085 & 23:45:54.44 &  $-$39:32:08.2 &  13.60  &     $0.006$   &    2   &  11594 & 60.47	  &   20.33 \\	
      
      \hline
      \multicolumn{9}{l}{$^{a}$This target is classed as one star in the WASP archive but is in fact a visual binary.}\\
      \multicolumn{9}{l}{$^{b}$This is a known roAp star included for comparison.}\\

    \end{tabular}
  \end{minipage}
\end{table*}

\subsection{Spectroscopy}

To obtain spectra for our candidate stars, we make use of the Intermediate dispersion Spectrograph and Imaging System (ISIS) mounted on the $4.2$-m William Herschel Telescope (WHT) in service mode for our northern targets, and the Robert Stobie Spectrograph (RSS) mounted on the $10$-m Southern African Large Telescope (SALT) to observe our southern targets. We require only low resolution classification spectra for our targets, achieving a spectral resolution of $\sim5000$~for SALT/RSS observations and $\sim2000$~for WHT/ISIS observations. In addition, two objects were observed with the Hamilton Echelle Spectrograph \citep[HamSpec;][]{vogt87} on the Shane $3$-m telescope at Lick Observatory, achieving a resolution of $R\sim37~000$.

All spectra have been extracted from their {\sc{fits}} images and have been flat-field corrected, de-biased, cleaned of cosmic-rays and have had wavelength calibrations applied. Tools from the {\sc{starlink}} project\footnote{\url{http://starlink.jach.hawaii.edu/starlink/}} were used to perform these tasks, with the exception of the HamSpec spectra, which were reduced using routines written in IDL. Finally, the spectra were intensity rectified using the {\sc{uclsyn}} spectral synthesis package \citep{smalley01}. 

To determine the stellar temperatures from the spectra we used {\sc{uclsyn}}, setting $\log g$~to a constant $4.0$~and synthesised the Balmer lines. Due to the low resolution classification spectra, we are only able to attain the temperatures to within $\pm200$~K. We estimate the S/N for each spectrum using the {\sc{der\_snr}} code of \citet{stoehr08}. The spectral types of the stars were determined via comparison with MK Standard stars using the method of \citet{gray09}. We present the spectroscopic information and results in Table \ref{tab:spec}.

\setcounter{table}{2}
\begin{table*}
  \centering
  \begin{minipage}{155mm}
    \centering
    \caption{Spectroscopic information on the 37 targets for which we obtained follow-up spectra.}
    \label{tab:spec}
    \begin{tabular}{cccccccc}
      \hline
      \hline
      
      \multicolumn{1}{c}{\multirow{2}{*}{ID}} & 
      \multicolumn{1}{c}{\multirow{1}{*}{Telescope/}} & 
      \multicolumn{1}{c}{\multirow{1}{*}{Exposure}} &
      \multicolumn{1}{c}{\multirow{2}{*}{Observation Date}}&
      \multicolumn{1}{c}{\multirow{2}{*}{S/N}}&
      \multicolumn{1}{c}{\multirow{1}{*}{Balmer $T_{\rm eff}$}}&
      \multicolumn{1}{c}{\multirow{1}{*}{SED $T_{\rm eff}$}}&
      \multicolumn{1}{c}{\multirow{1}{*}{Spectral}}\\
      
      \multicolumn{1}{c}{} & 
      \multicolumn{1}{c}{\multirow{1}{*}{Instrument}} & 
      \multicolumn{1}{c}{\multirow{1}{*}{(s)}} &
      \multicolumn{1}{c}{} &
      \multicolumn{1}{c}{} & 
      \multicolumn{1}{c}{\multirow{1}{*}{(K)}}&
      \multicolumn{1}{c}{\multirow{1}{*}{(K)}}&
      \multicolumn{1}{c}{\multirow{1}{*}{Type}}\\
      \hline

      J0004                                       &  SALT/RSS      &  $15$	  &  2012-05-31	   &  180   & 7900 & 7518 $\pm$ 377 &  A5    \\
      J0008                                       &  SALT/RSS      &  $99$	  &  2013-06-15	   &   60   & 7300 & 7484 $\pm$ 336 &  A9p SrEu(Cr)  \\
      J0026S$^{a}$                                &  WHT/ISIS      &  $30$	  &  2011-11-22	   &  100   & 6650 & \multicolumn{1}{c}{\multirow{2}{*}{6100$\pm$318}} &  F4    \\
      J0026P                                      &  WHT/ISIS      &  $30$	  &  2011-11-22	   &  115   & 8100 &               &  A2m   \\
      J0051                                       &  SALT/RSS      &  $200$	  &  2013-06-06	   &   50   & 7850 & 8144 $\pm$ 479  &  A6    \\
      J0206                                       &  WHT/ISIS      &  $30$	  &  2011-11-22	   &   70   & 7600 & 7310 $\pm$ 318  &  A4p   \\
      J0353                                       &  WHT/ISIS      &  $15$	  &  2012-10-24	   &   50   & 8250 & 7417 $\pm$ 331  &  A5p SrEu  \\
      J0410                                       &  WHT/ISIS      &  $20$	  &  2011-11-22	   &  125   & 8150 & 7712 $\pm$ 346  &  A3m:    \\
      J0429                                       &  WHT/ISIS      &  $15$	  &  2013-10-24	   &  120   & 8200 & 8770 $\pm$ 334  &  A2m    \\
      J0629                                       &  WHT/ISIS      &  $40$	  &  2012-10-24	   &   40   & 6600 & 6211 $\pm$ 281  &  F4p EuCr(Sr)  \\
      J0651                                       &  SALT/RSS      &  $1300$	  &  2012-09-11	   &   70   & 7400 & 7843 $\pm$ 491  &  F0p SrEu(Cr)  \\
      J0855                                       &  WHT/ISIS      &  $50$	  &  2011-11-22	   &   80   & 7800 & 8287 $\pm$ 475  &  A6p SrEu  \\
      J1110                                       &  WHT/ISIS      &  $15$	  &  2012-12-25	   &   55   & 6300 & 6707 $\pm$ 294  &  F3p SrEu(Cr) \\
      J1215                                       &  WHT/ISIS      &  $40$	  &  2012-12-25	   &   70   & 8100 & 8437 $\pm$ 412  &  A3m:    \\
      J1250                                       &  WHT/ISIS      &  $100$	  &  2012-12-25	   &  100   & 8300 & 7099 $\pm$ 370  &  A4m   \\
      J1403                                       &  SALT/RSS      &  $500$	  &  2012-05-08	   &  140   & 7550 & 7500 $\pm$ 350  &  A9    \\
      J1430                                       &  WHT/ISIS      &  $100$	  &  2012-12-25	   &   85   & 7100 & 7479 $\pm$ 366  &  A9p SrEu  \\
      J1625                                       &  SALT/RSS      &  $230$	  &  2012-05-08	   &  235   & 8200 & 6147 $\pm$ 272  &  A1m    \\
      \multicolumn{1}{c}{\multirow{1}{*}{J1640}}                                                                      
      & $\left\{\begin{tabular}{@{}l} WHT/ISIS \\ SALT/RSS \end{tabular}\right.$                                      
      & $ {\begin{tabular}{@{\ }@{}l} {\ }1200 \\ {\ }{\ }600 \end{tabular}}$                                         
      & $ {\begin{tabular}{@{\ }@{}l} {\ }2013-02-01 \\ {\ }2013-03-22 \end{tabular}}$                                
      & $ {\begin{tabular}{@{\ }@{}l} {\ }95 \\ {\ }75 \end{tabular}}$                                                
      & $ {\begin{tabular}{@{\ }@{}l} {\ }7400 \\ {\ }7400 \end{tabular}}$                                            
      & \multicolumn{1}{c}{\multirow{1}{*}{6282 $\pm$ 314}}                                             
      & $ {\begin{tabular}{@{\ }@{}l} {\ }A8p SrEu \\ {\ }A8p SrEu~ \end{tabular}}$  \\                               
      J1648                                       &  WHT/ISIS      &  $200$	  &  2013-02-01	   &   80   & 7100 &  6939 $\pm$ 307 &   F0    \\
      J1757                                       &  WHT/ISIS      &  $600$	  &  2013-02-01	   &  105   & 7900 &  7425 $\pm$ 377 &   A7m:    \\
      J1758                                       &  WHT/ISIS      &  $1200$	  &  2013-02-01	   &  110   & 7700 &  7260 $\pm$ 480 &   A7m:    \\
      J1844                                       &  Shane/HamSpec & $1800$	  &  2012-07-24	   &   40   & 7000 &  8043 $\pm$ 435 &   A7p EuCr  \\
      J1917                                       &  SALT/RSS      &  $249$	  &  2013-04-22	   &  100   & 7800 &  6989 $\pm$ 293 &   A7m  \\
      J1940                                       &  SALT/RSS      &  $759$	  &  2012-11-03	   &   20   & 6900 &  7623 $\pm$ 411 &   F2(p Cr)    \\
      J1951                                       &  SALT/RSS      &  $900$	  &  2013-04-25	   &   75   & 8100 &  8076 $\pm$ 438 &   A5    \\
      J2022                                       &  SALT/RSS      &  $500$	  &  2013-04-25	   &   85   & 8200 &  7973 $\pm$ 507 &   A4m    \\
      J2026                                       &  WHT/ISIS      &  $150$	  &  2013-04-25	   &   60   & 6000 &  6528 $\pm$ 433 &     F8    \\
      J2029                                       &  SALT/RSS      &  $25$	  &  2013-04-27	   &   40   & 8000 &  7754 $\pm$ 339 &    A4m    \\
      J2054                                       &  Shane/HamSpec & $1800$	  &  2012-07-24	   &   60   & 7000 &  8372 $\pm$ 438 &    A3m:    \\
      J2155                                       &  SALT/RSS      &  $800$	  &  2013-06-07	   &   50   & 8100 &  6681 $\pm$ 555 &    A3    \\
      J2241                                       &  WHT/ISIS      &  $100$	  &  2012-10-25	   &   70   & 8100 &  7771 $\pm$ 391 &    A3m   \\
      J2254                                       &  SALT/RSS      &  $600$	  &  2013-06-17	   &   55   & 8150 &  7059 $\pm$ 450 &    A4    \\
      J2255                                       &  SALT/RSS      &  $250$	  &  2013-05-13	   &   50   & 8200 &  8564 $\pm$ 627 &    A3    \\
      J2305                                       &  SALT/RSS      &  $139$	  &  2012-11-04	   &   35   & 8050 &  8146 $\pm$ 374 &    A5m:    \\
      J2313                                       &  SALT/RSS      &  $90$	  &  2013-06-19	   &   65   & 8000 &  7794 $\pm$ 376 &    A5    \\
      J2345                                       &  SALT/RSS      &	$1250$	  &  2013-06-07	   &   75   & 9700 &  9772 $\pm$ 964 &    A3    \\
      
      \hline
      \multicolumn{8}{l}{$^a$We identify this target as a spectroscopic binary, with a radial velocity shift of $172\pm21$~km~s$^{-1}$. }\\

    \end{tabular}
  \end{minipage}
\end{table*}

\subsection{Stellar temperatures from spectral energy distributions}

Effective temperatures can be determined from the stellar spectral energy distribution (SED). For our target stars these were constructed from literature photometry, using 2MASS \citep{sktrutskie06}, DENIS \citep{forque00}, Tycho $B$ and $V$ magnitudes \citep{hog97}, USNO-B1 $R$ magnitudes \citep{monet03}, TASS $V$ and $I$ magnitudes \citep{droege06} CMC14 $r'$ magnitudes \citep{evans02} as available. 

The stellar $T_{\rm eff}$ values were determined by fitting solar-composition \cite{kurucz93} model fluxes to the de-reddening SEDs. The model fluxes were convolved with photometric filter response functions. A weighted Levenberg-Marquardt non-linear least-squares fitting procedure was used to find the solution that minimized the difference between the observed and model fluxes. Since $\log g$ is poorly constrained by our SEDs, we fixed $\log g =4.0$ for all the fits. Stellar energy distributions can be significantly affected by interstellar reddening. However, in the absence of measured reddening values, we have assumed $E(B-V) = 0.02 \pm 0.02$ in our fitting. The uncertainties in $T_{\rm eff}$ includes the formal least-squares error and adopted uncertainties in $E(B-V)$ of $\pm$0.02 and $\log g$ of $\pm$0.5 added in quadrature. We present the SED derived $T_{\rm eff}$ for our spectroscopically observed targets in Table~\ref{tab:spec}.

\section{Results}
\label{sec:results}

\subsection{New roAp stars}

We present here the $10$~new roAp stars discovered in the SuperWASP archive. Some objects show a low frequency signature in their periodogram which is attributed to rotational modulation. We present both the low frequency periodograms and the phase folded lightcurves alongside a discussion of each object, with the details of the modulations shown in Table \ref{tab:rot_mod}. The periodograms indicate the frequency ($\nu$) on which the data are folded, as well as labels of other prominent peaks. In this frequency range the periodograms show the reflection of the $-1$~d$^{-1}$ aliases, labelled as $-\nu$, which must not be confused with the true peak. Each periodogram is calculated with a single season of WASP data for clarity (the peaks are also present in all other data sets for each target). The solid line on each phase plot represents the harmonic fit. The false-alarm probability (FAP) is calculated using the method of \citet{maxted11}.

\setcounter{table}{3}
\begin{table}
  \centering
  \begin{minipage}{65mm}
    \caption{Rotationally modulated lightcurve data for the newly detected roAp stars.}
    \label{tab:rot_mod}
    \centering
    \begin{tabular}{ccccc}
      \hline
      \hline
      
      \multicolumn{1}{c}{\multirow{2}{*}{ID}} & 
      \multicolumn{1}{c}{\multirow{1}{*}{$\nu_{rot}$}} & 
      \multicolumn{1}{c}{\multirow{1}{*}{Period}} & 
      \multicolumn{1}{c}{\multirow{1}{*}{Amp}} &
      \multicolumn{1}{c}{\multirow{2}{*}{FAP}} \\
      
      \multicolumn{1}{c}{} & 
      \multicolumn{1}{c}{\multirow{1}{*}{(d$^{-1}$)}} & 
      \multicolumn{1}{c}{\multirow{1}{*}{(d)}} & 
      \multicolumn{1}{c}{\multirow{1}{*}{(mmag)}} &
      \multicolumn{1}{c}{} \\ 
      \hline
      
      J$0353$ & 0.0717 & 13.95 & 6.37 & $<$0.001 \\
      J$0855$ & 0.3234 & 3.09  & 2.55 & 0.096 \\
      J$1640$ & 0.2722 & 3.67  & 4.17 & 0.003 \\
      J$1844$ & 0.0495 & 20.20 & 7.65 & 0.000 \\
      J$1940$ & 0.1044 & 9.58  & 5.87 & 0.017 \\

      \hline

    \end{tabular}
    \end{minipage}
\end{table}

\subsubsection{J$0008$}

J$0008$~shows roAp  pulsations at $150.26$ d$^{-1}$ with an amplitude of $0.76$ mmag (Fig. \ref{fig:J0008} top). WASP has observed the target for three consecutive seasons with slight discrepancies in the pulsation frequency which is attributed to the $1$-d aliases. The spectrum obtained for this star (Fig. \ref{fig:J0008} bottom) has been classified as A$9$p, with strong enhancements of Sr~{\sc{ii}} and Eu~{\sc{ii}}. The spectrum confirms this to be a new roAp star.

The pulsations in J$0008$ are similar to those in HD~$119027$, which pulsates at $165.52$ d$^{-1}$ with an amplitude in the blue of $2.0$ mmag \citep{martinez94}. HD~$119027$ is a hotter star, classified as A$3$p SrEu(Cr), and is also known to show amplitude modulation as a result of closely spaced frequencies.

\setcounter{figure}{1}
\begin{figure}
  \centering
  \includegraphics[angle=180,width=80mm, trim= 23mm 21mm 37mm 110mm,clip]{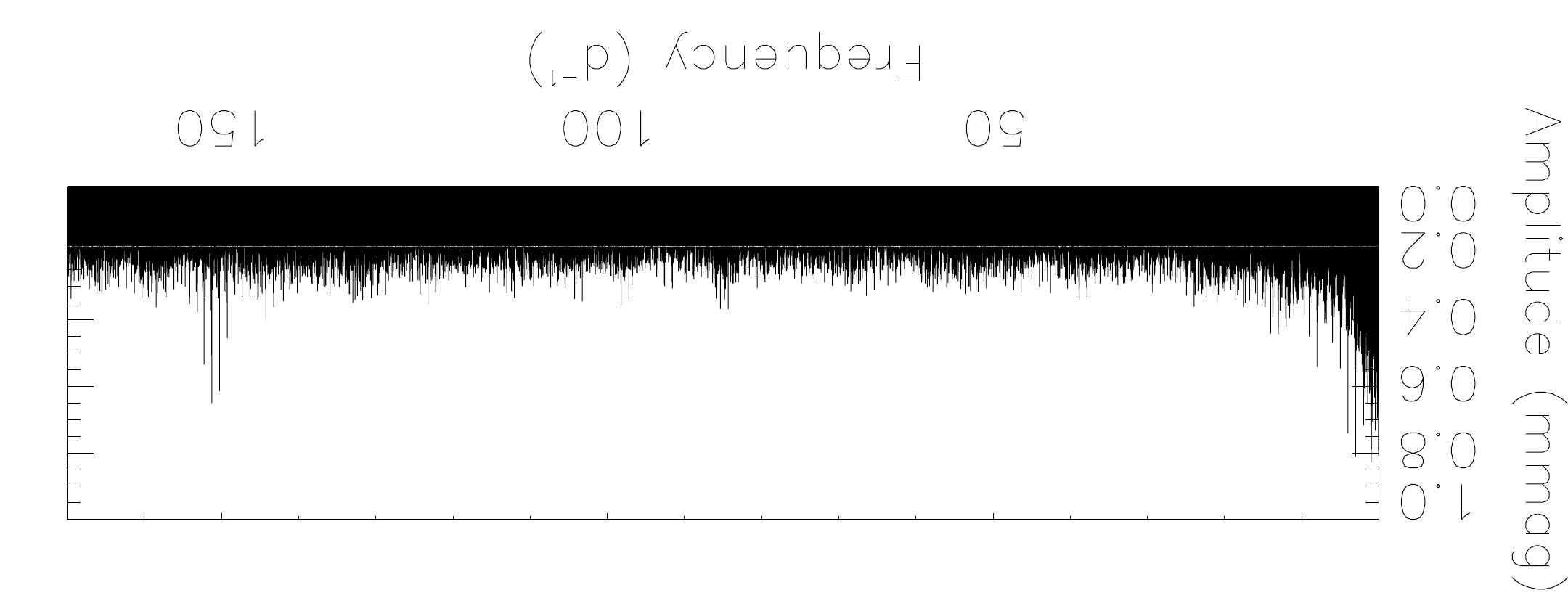}
  \includegraphics[angle=180,width=80mm, trim= 23mm 30mm 37mm 110mm,clip]{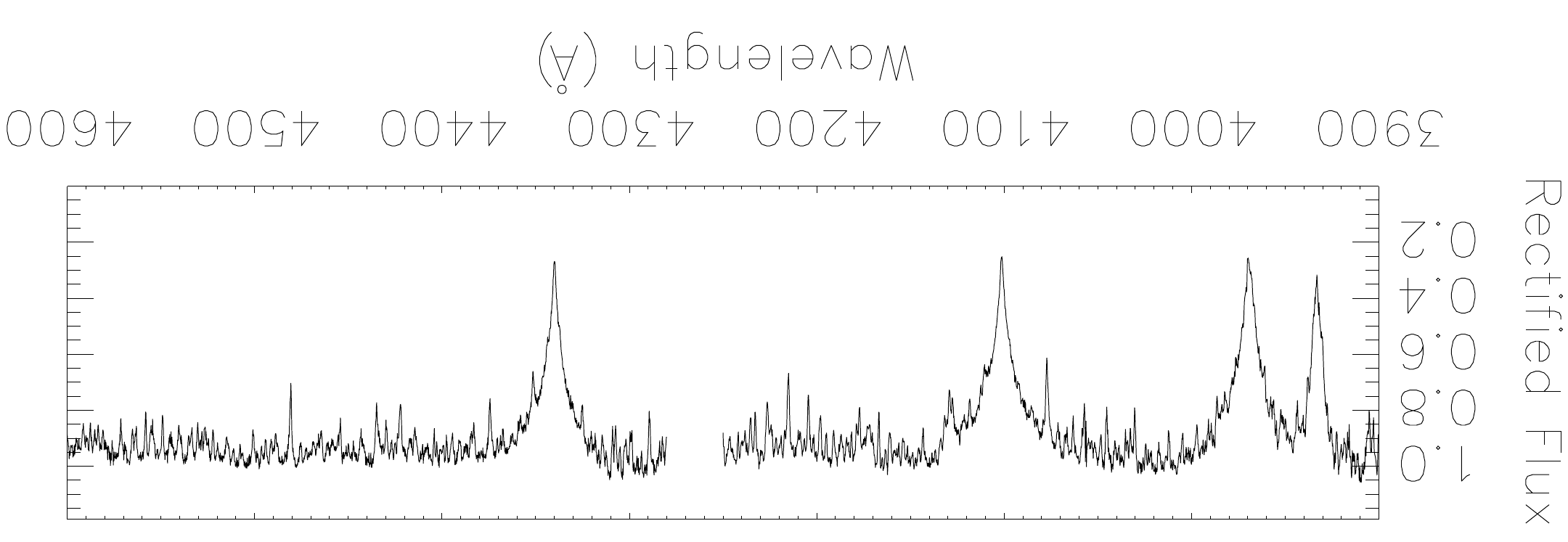}
  \caption{Periodogram and SALT/RSS spectrum of J$0008$. The low frequency peaks in the periodogram are due to noise.}
  \label{fig:J0008}
\end{figure}

\subsubsection{J$0353$}

The star J$0353$~displays pulsations at about $224$ d$^{-1}$~with an amplitude of $1.65$ mmag as well as a low-frequency variation corresponding to $13.95$ d (Fig. \ref{fig:J0353}~\& Fig. \ref{fig:J0353-lc}). The spectrum of this star is classified as A$5$p with an enhancement of Sr~{\sc{ii}} and Eu~{\sc{ii}}, confirming it as a new roAp star.

Similar to J$0353$, HR $1217$ shows pulsations at about the same frequency \citep[$232.26$ d$^{-1}$;][]{kurtz81} and is classified as an A$9$p SrEu(Cr) star. The rotation period of HR~$1217$ has been discussed at length in the literature, recently \cite{rusomarov13} present a period of $12.45812$ d derived from $81$ longitudinal magnetic field data points spanning over 4 decades. \cite{balona02} present spectra of HR~$1217$ which show a core-wing anomaly in the H$_\alpha$ line, a feature which we also note in our ISIS red-arm spectrum. 

\setcounter{figure}{2}
\begin{figure}
  \centering
  \includegraphics[angle=180,width=80mm, trim= 23mm 21mm 37mm 110mm,clip]{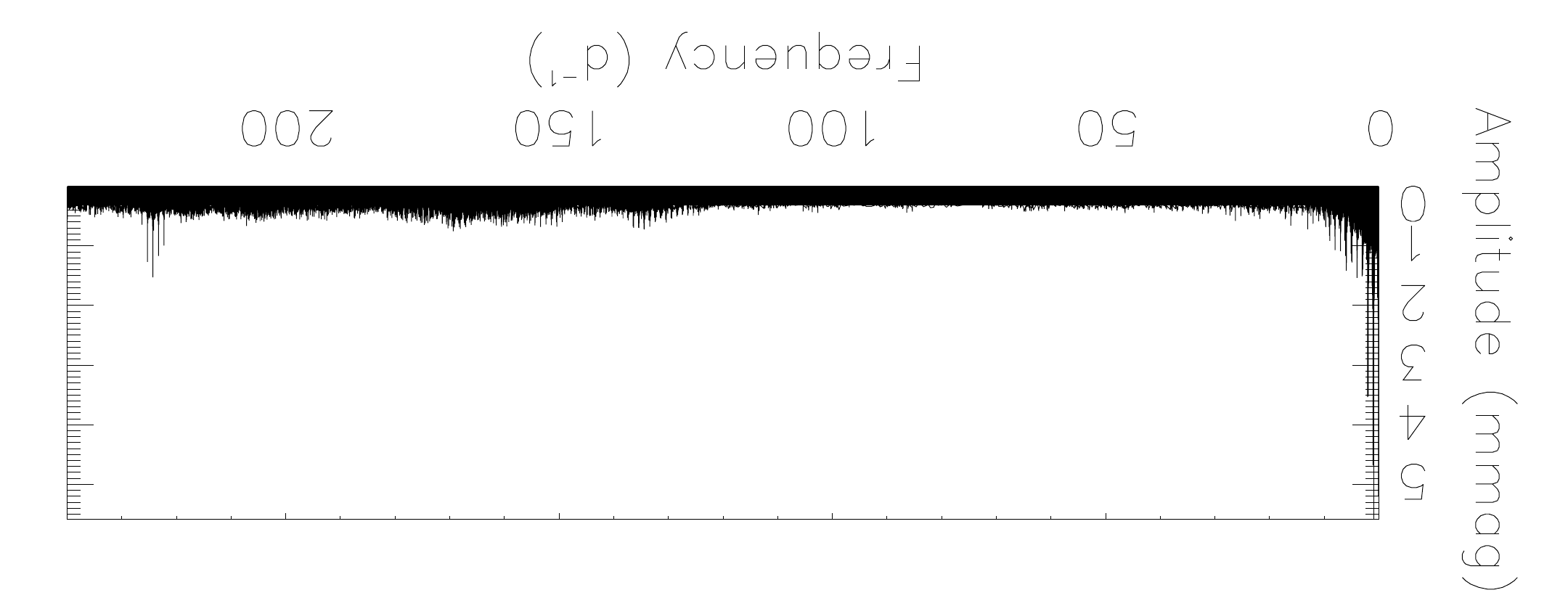}
  \includegraphics[angle=180,width=80mm, trim= 23mm 30mm 37mm 110mm,clip]{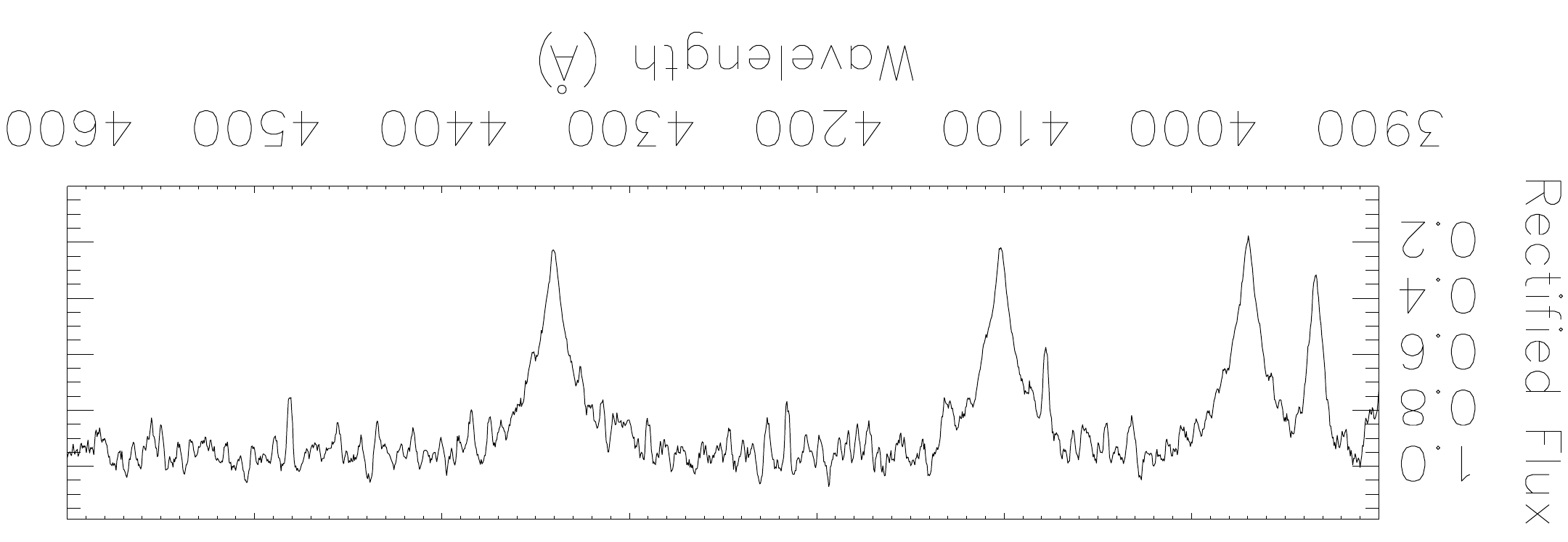}
  \caption{Periodogram and WHT/ISIS spectrum of J$0353$.}
  \label{fig:J0353}
\end{figure}

\setcounter{figure}{3}
\begin{figure}
  \centering
  \includegraphics[angle=180,width=80mm, trim= 23mm 21mm 37mm 110mm,clip]{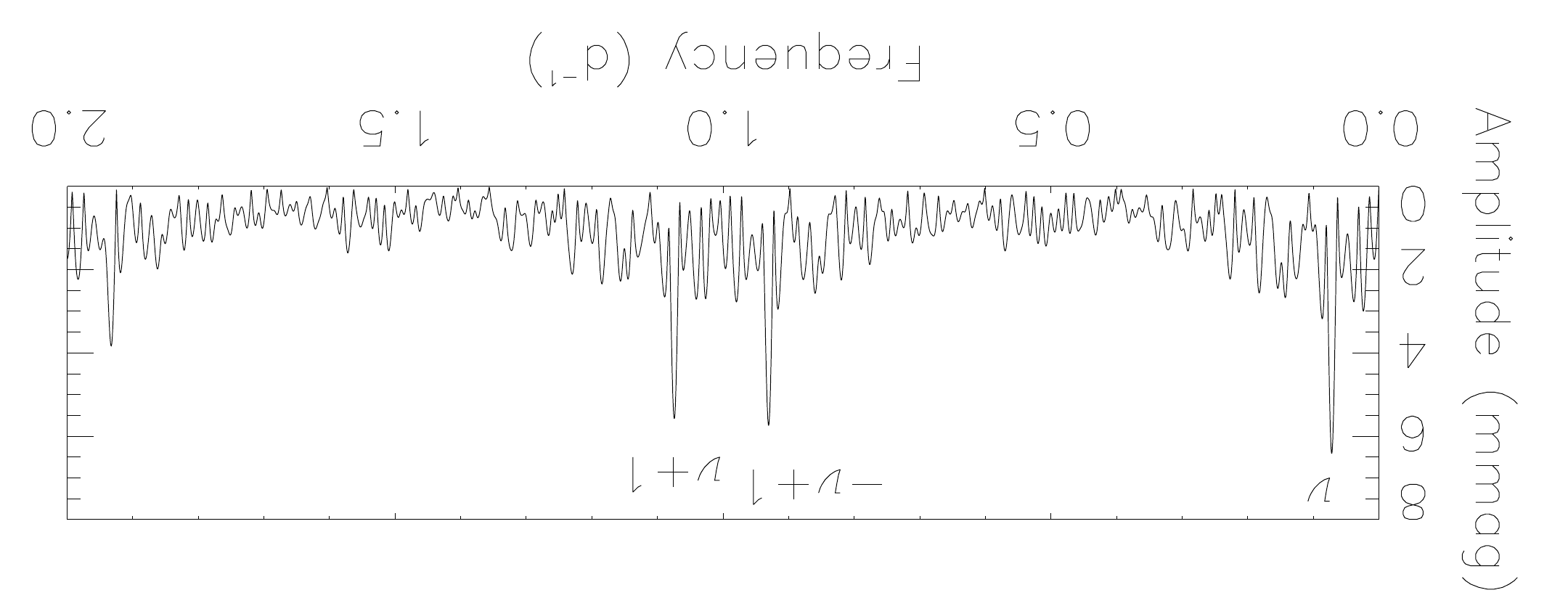}
  \includegraphics[angle=180,width=80mm, trim= 23mm 30mm 27mm 110mm,clip]{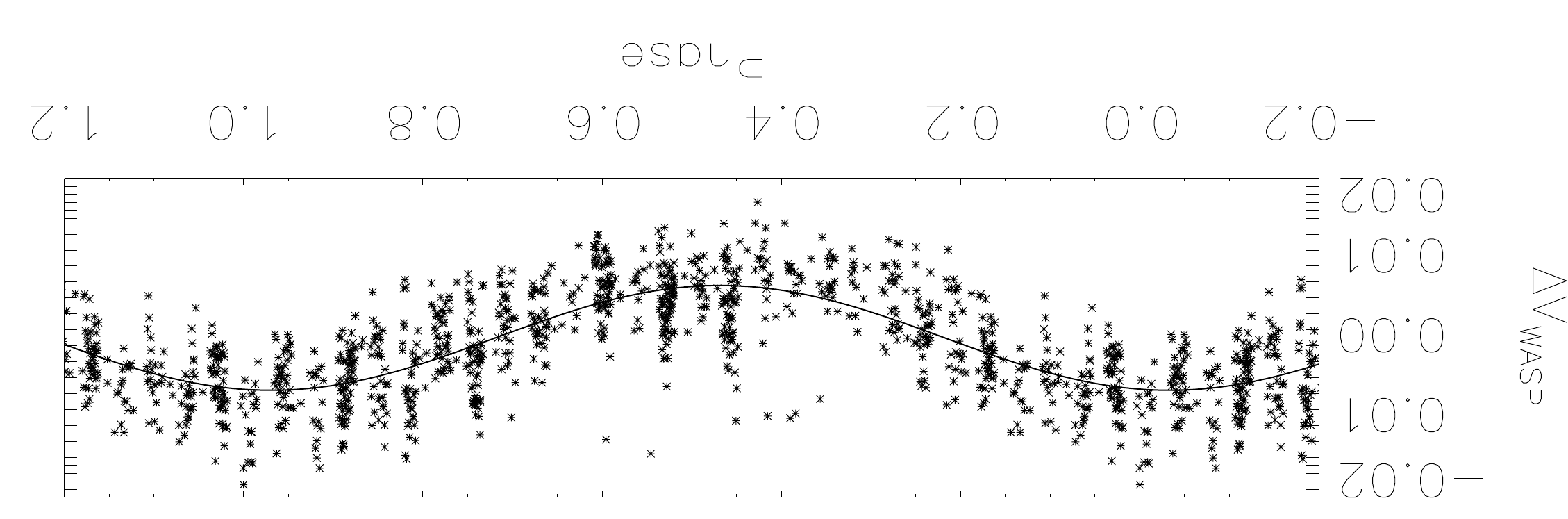}
  \caption{Low frequency periodogram and phase folded WASP lightcurve for J$0353$. The data are folded on a period of $13.95$ d, and shown in phase bins of $0.001$. The labelled peaks are the true frequency ($\nu$) and the positive and negative aliases.}
  \label{fig:J0353-lc}
\end{figure}

\subsubsection{J$0629$}

J$0629$~has pulsations at $169.54$ d$^{-1}$ with an amplitude of $1.49$ mmag (Fig. \ref{fig:J0629} top). The spectrum (Fig. \ref{fig:J0629} bottom) shows strong over-abundances of Sr~{\sc{ii}}, Cr~{\sc{ii}} and Eu~{\sc{ii}}. The photometric observations give no indication of rotational modulation. We classify the star as F$4$p.

Through both Balmer line analysis and SED fitting, we conclude that J$0629$ is a very cool Ap star with a $T_{\rm eff}$ similar to the roAp star HD~$213637$ \citep[$6400$~K;][]{kochukhov03}, and thus placing J$0629$ amongst the coolest roAp stars.

\setcounter{figure}{4}
\begin{figure}
  \centering
  \includegraphics[angle=180,width=80mm, trim= 23mm 21mm 37mm 110mm,clip]{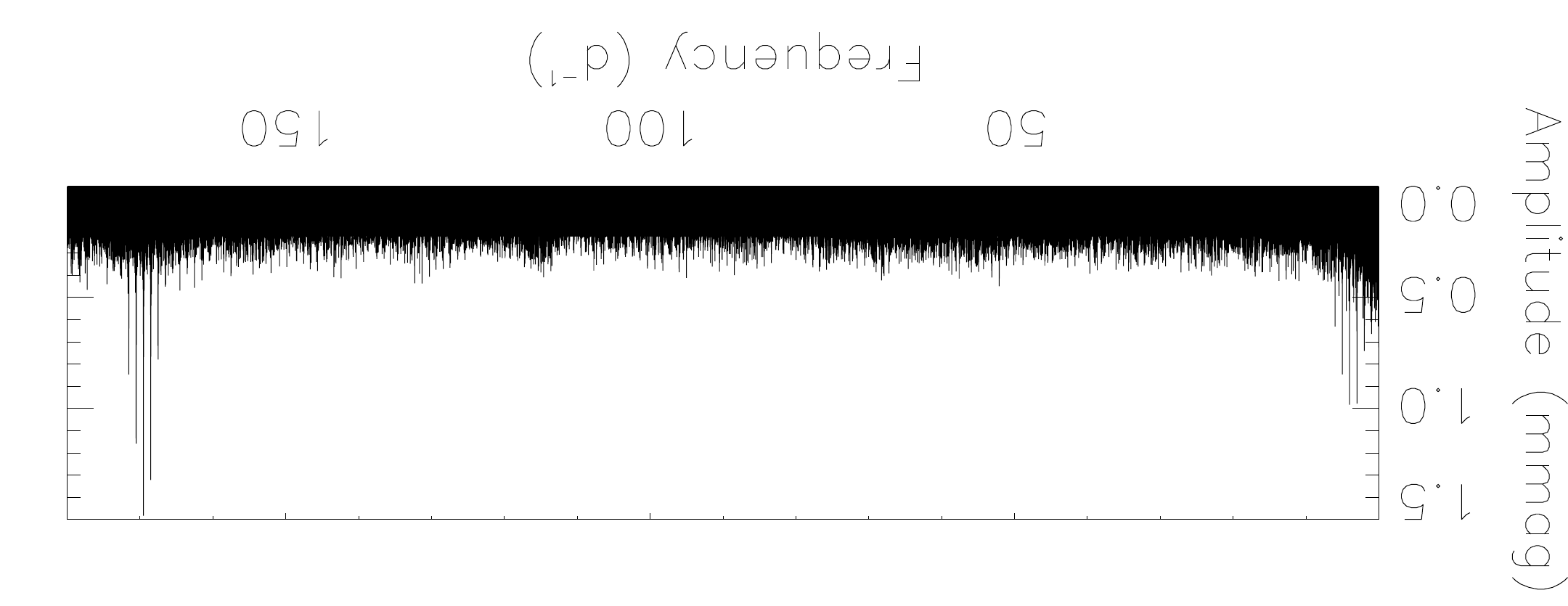}
  \includegraphics[angle=180,width=80mm, trim= 23mm 30mm 37mm 110mm,clip]{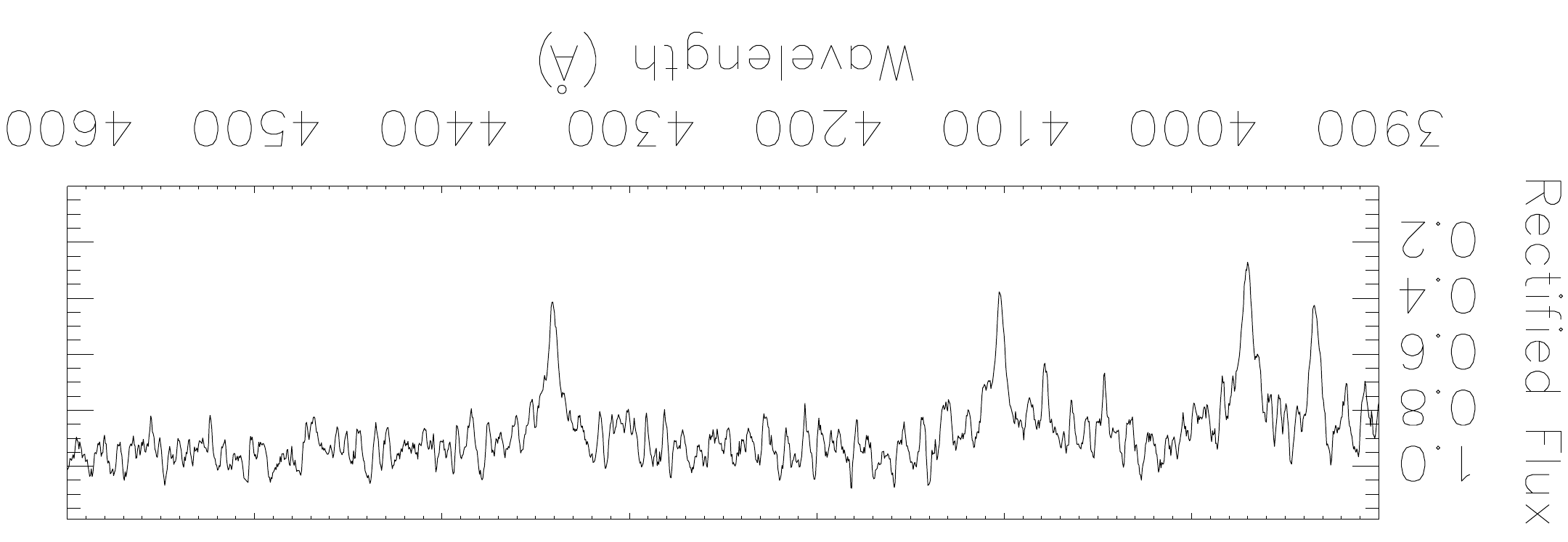}
  \caption{Periodogram and WHT/ISIS spectrum of J$0629$. The low frequency peaks in the periodogram are due to $1$-d aliasing.}
  \label{fig:J0629}
\end{figure}

\subsubsection{J$0651$}

We classify J$0651$~as an F$0$p star whose spectrum shows a strong over-abundance of Sr~{\sc{ii}} at $\lambda\lambda~4077$~\& $4216$~(Fig. \ref{fig:J0651}). We also see enhanced features of Eu~{\sc{ii}} at $\lambda\lambda~4128$~\& $4205$ and Cr~{\sc{ii}} at $\lambda~4111$. Our photometry shows pulsations at $132.38$ d$^{-1}$ (Fig. \ref{fig:J0651}), with no clear indications of rotational modulation in the lightcurve.

In the literature we find a very similar roAp star to J$0651$. Pulsating at a frequency of $137.17$ d$^{-1}$ with an amplitude in the blue of $3.5$ mmag, HD~$9289$ is an Ap SrEu star \citep{kurtz93}. \cite{elkin08} present photometric and spectroscopic $T_{\rm eff}$ values for HD~$9289$, deriving $7700$ and $8000$~K respectively, which are similar to those we obtain for J$0651$.

\setcounter{figure}{5}
\begin{figure}
  \centering
  \includegraphics[angle=180,width=80mm, trim= 23mm 21mm 37mm 110mm,clip]{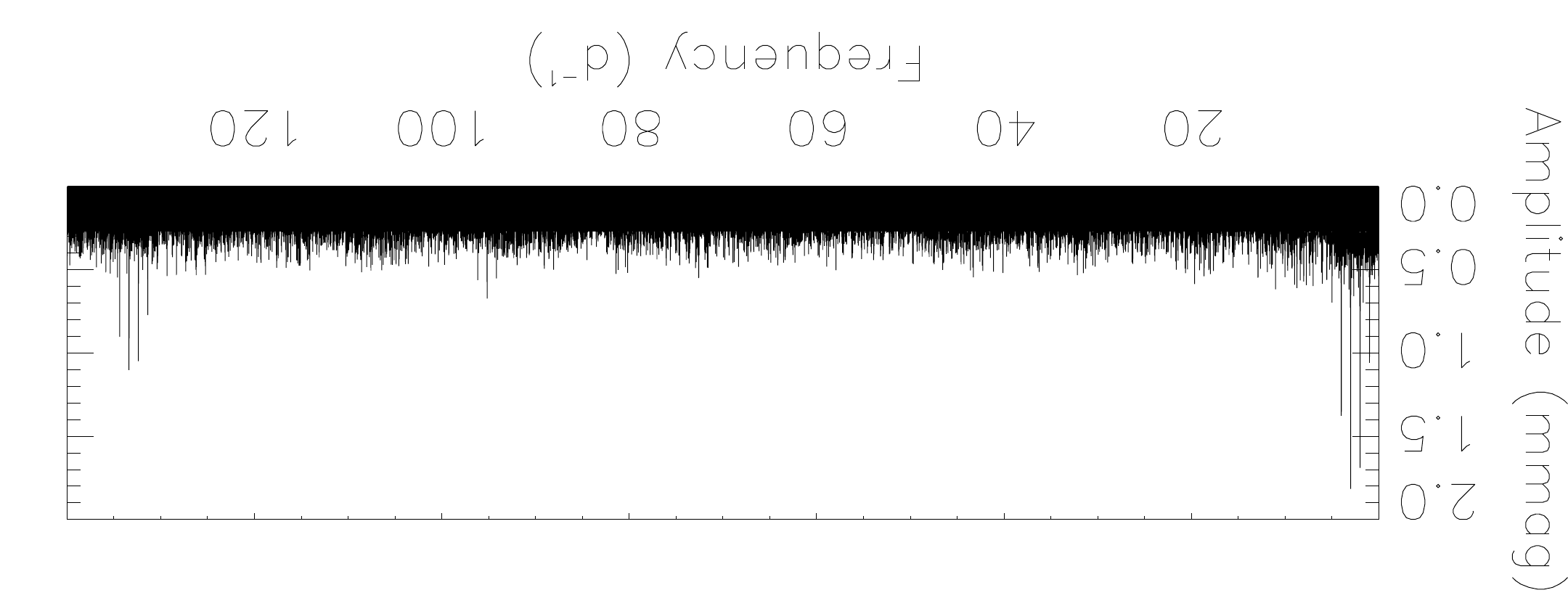}
  \includegraphics[angle=180,width=80mm, trim= 23mm 30mm 37mm 110mm,clip]{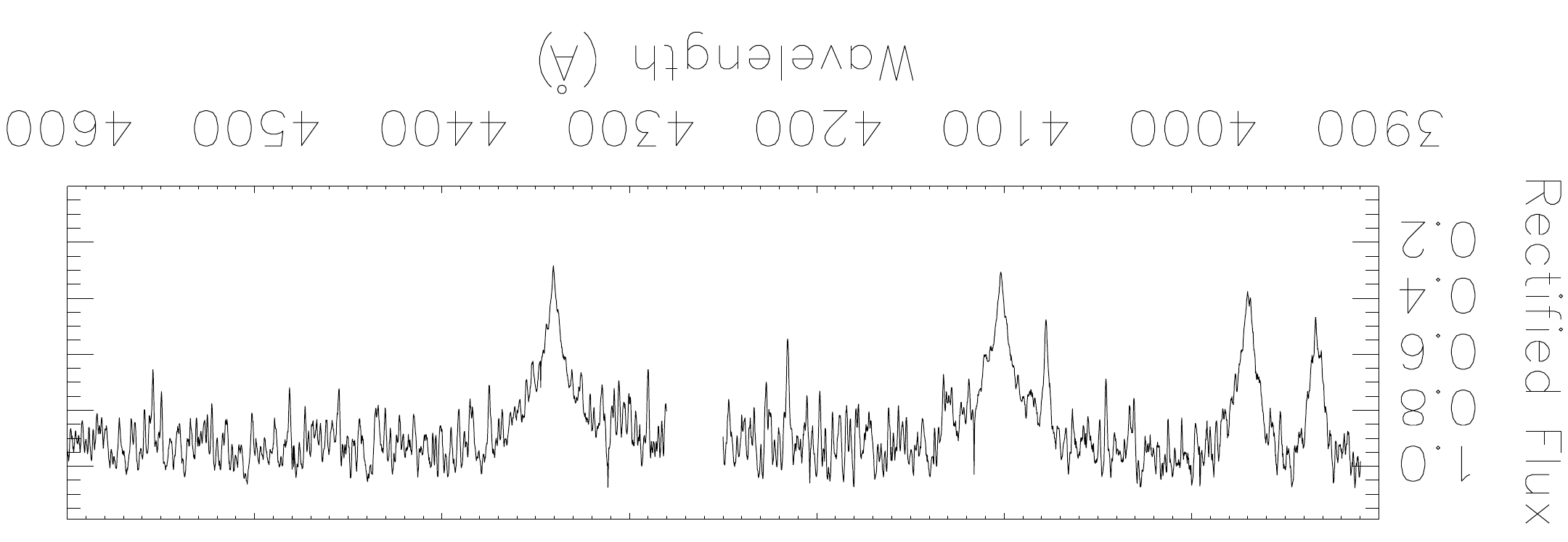}
  \caption{Periodogram and SALT/RSS spectrum of J$0651$. The low frequency peaks in the periodogram are due to $1$-d aliasing.}
  \label{fig:J0651}
\end{figure}

\subsubsection{J$0855$}

J$0855$~shows rapid pulsations at $197.27$ d$^{-1}$ with an amplitude of $1.4$ mmag (Fig. \ref{fig:J0855}). Balmer line fitting gives a $T_{\rm eff}$~of $7800$~K, and a spectral type of A$6$p when compared to MK standards. The spectrum also shows an over-abundance of Eu~{\sc{ii}} at $\lambda\lambda~4205$ \&~$4128$, with weak Ca K and Ca {\sc{i}} at $\lambda4266$~(Fig. \ref{fig:J0855}). As well as the high-frequency pulsation, the periodogram shows a low-frequency signature with a period of $3$ d (Fig. \ref{fig:J0855-lc}).

Our temperatures derived for J$0855$ vary greatly between methods, however the SED method is very poorly constrained for this target as indicated by the error bar. Assuming a $T_{\rm eff}$ of that derived through Balmer line fitting, J$0855$ is almost identical to the known roAp star HD~$190290$ \citep{martinez90}. The pulsations of the two are at the same frequency, with J$0855$ showing a larger undiluted amplitude. The ISIS red arm spectrum also shows a core-wing anomaly in the H$_\alpha$ line.

\setcounter{figure}{6}
\begin{figure}
  \centering
  \includegraphics[angle=180,width=80mm, trim= 23mm 21mm 37mm 110mm,clip]{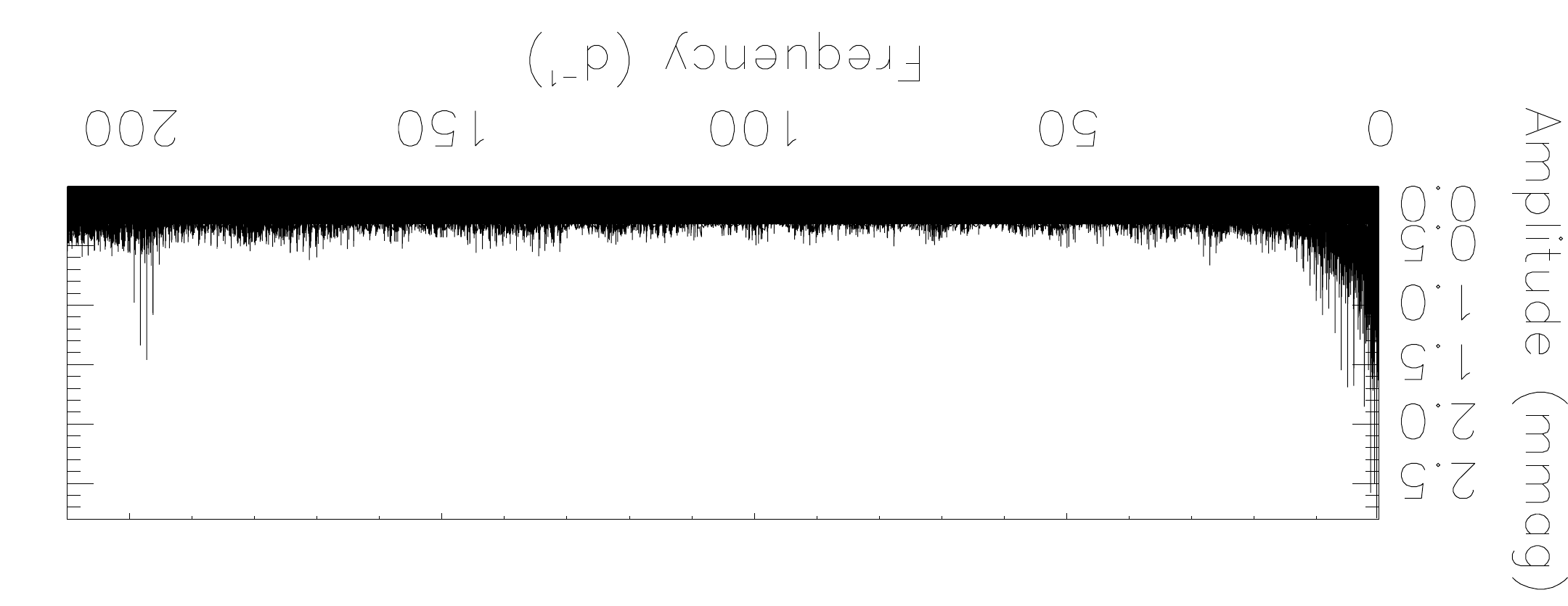}
  \includegraphics[angle=180,width=80mm, trim= 23mm 30mm 37mm 110mm,clip]{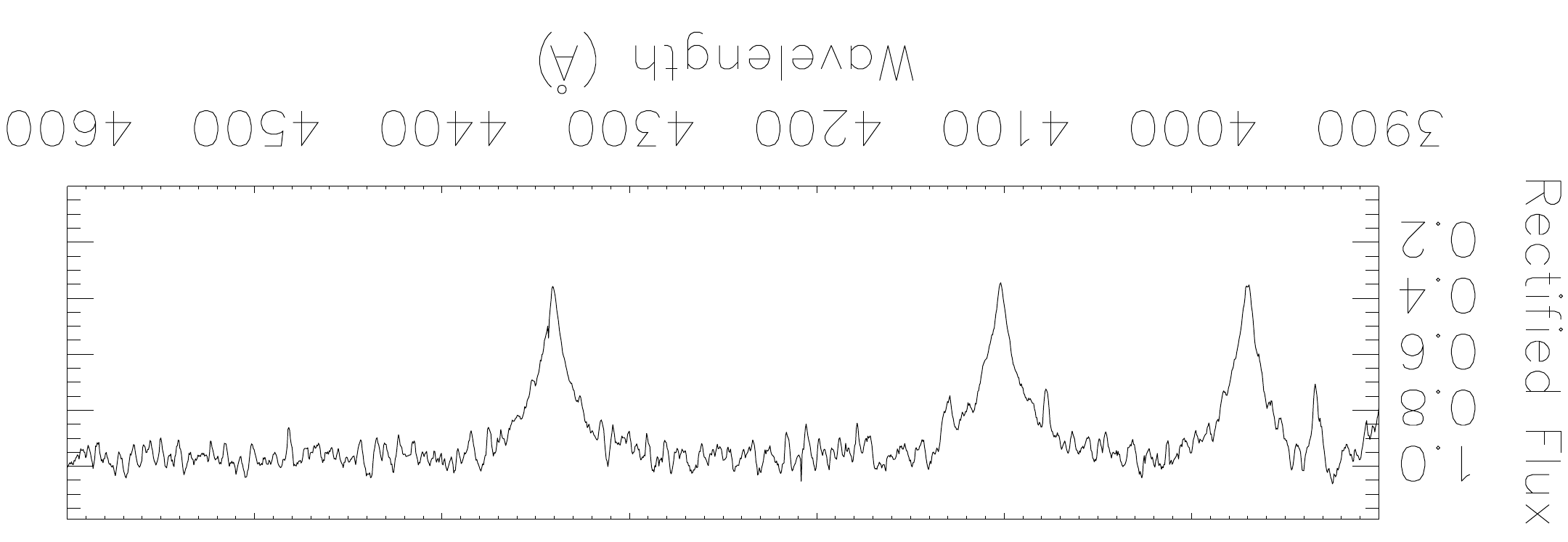}
  \caption{Periodogram and WHT/ISIS spectrum of J$0855$.}
  \label{fig:J0855}
\end{figure}

\setcounter{figure}{7}
\begin{figure}
  \centering
  \includegraphics[angle=180,width=80mm, trim= 23mm 21mm 37mm 110mm,clip]{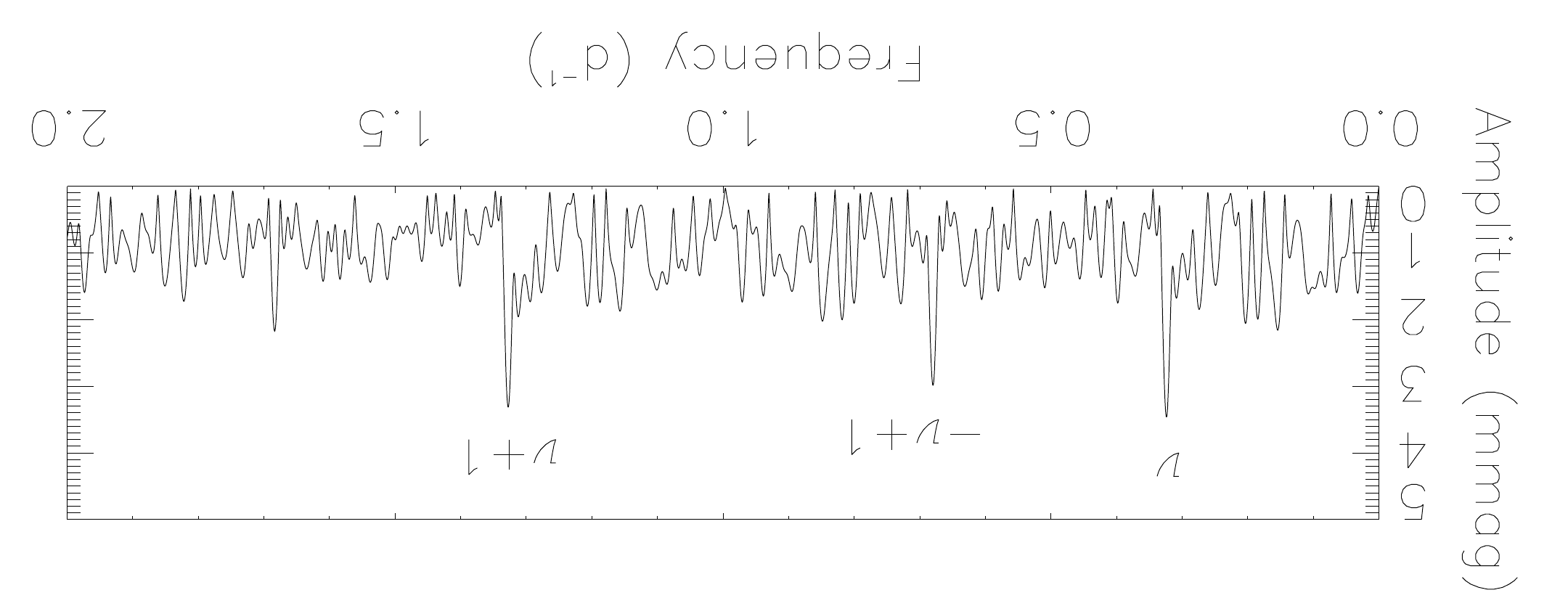}
  \includegraphics[angle=180,width=80mm, trim= 23mm 30mm 22mm 110mm,clip]{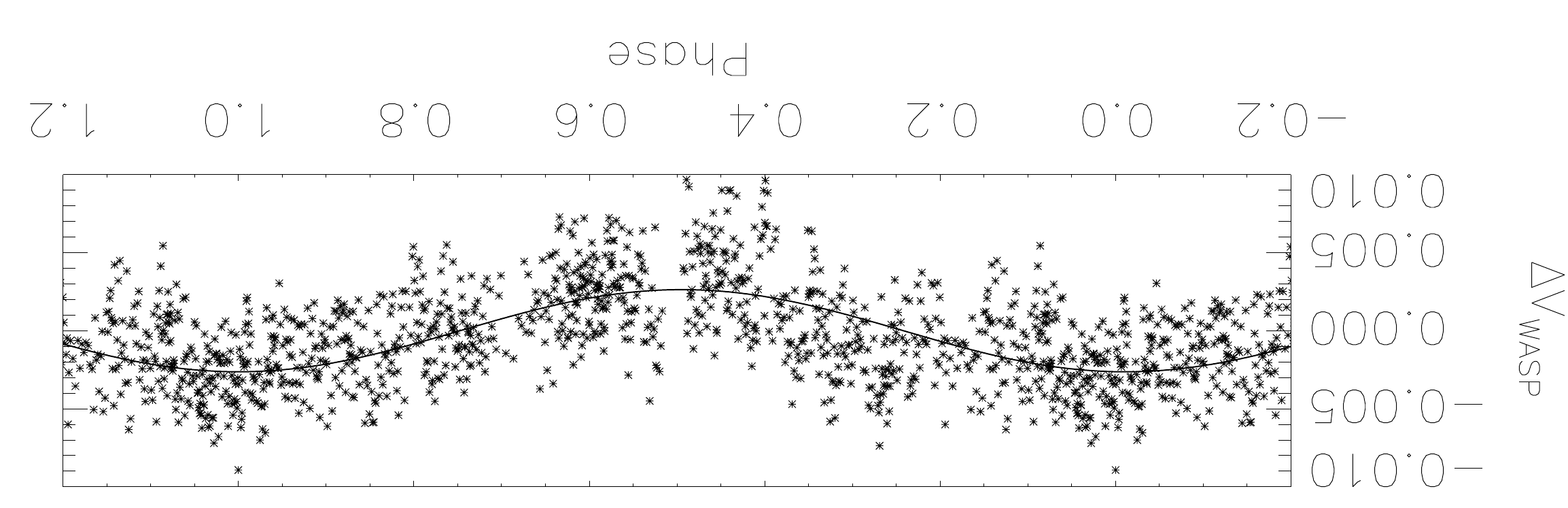}
  \caption{Low frequency periodogram and phased folded lightcurve of J$0855$~folded on a period of $3.09$ d and shown in phase bins of $0.001$. Thee labelled peaks are the true frequency ($\nu$) and the positive and negative aliases.}
  \label{fig:J0855-lc}
\end{figure}

\subsubsection{J$1110$}

J$1110$~exhibits low amplitude pulsations at $106$ d$^{-1}$~(Fig. \ref{fig:J1110}). We classify the spectrum as a cool F$3$p star with a $T_{\rm eff}$~measured from the Balmer lines of $6500$~K. The spectrum shows a slight over-abundance of Eu~{\sc{ii}} at $\lambda\lambda~4128$~\& $4205$, and a marginal over-abundance of Sr~{\sc{ii}} (Fig. \ref{fig:J1110}). We also note the weak Ca {\sc{i}} K and $\lambda4266$~lines which may be due to stratification in the atmosphere.

J$1110$ has the lowest pulsation frequency of our roAp stars and is also the coolest, as derived from the Balmer lines. J$1110$ is similar in amplitude and frequency to HD~$193756$. However, HD~$193756$ is classified as A$9$ with a temperature of $7500$~K derived from its H$_\alpha$ profile (Elkin et al. 2008).

\setcounter{figure}{8}
\begin{figure}
  \centering
  \includegraphics[angle=180,width=80mm, trim= 23mm 21mm 37mm 110mm,clip]{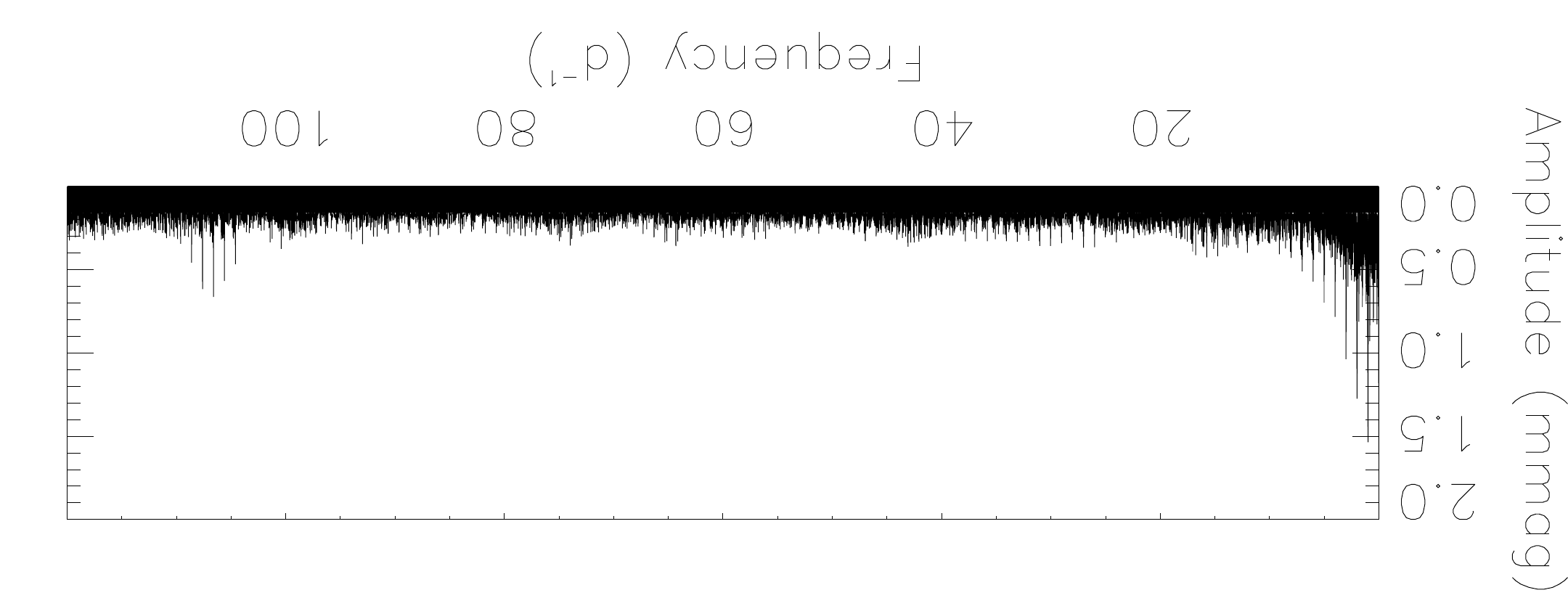}
  \includegraphics[angle=180,width=80mm, trim= 23mm 30mm 37mm 110mm,clip]{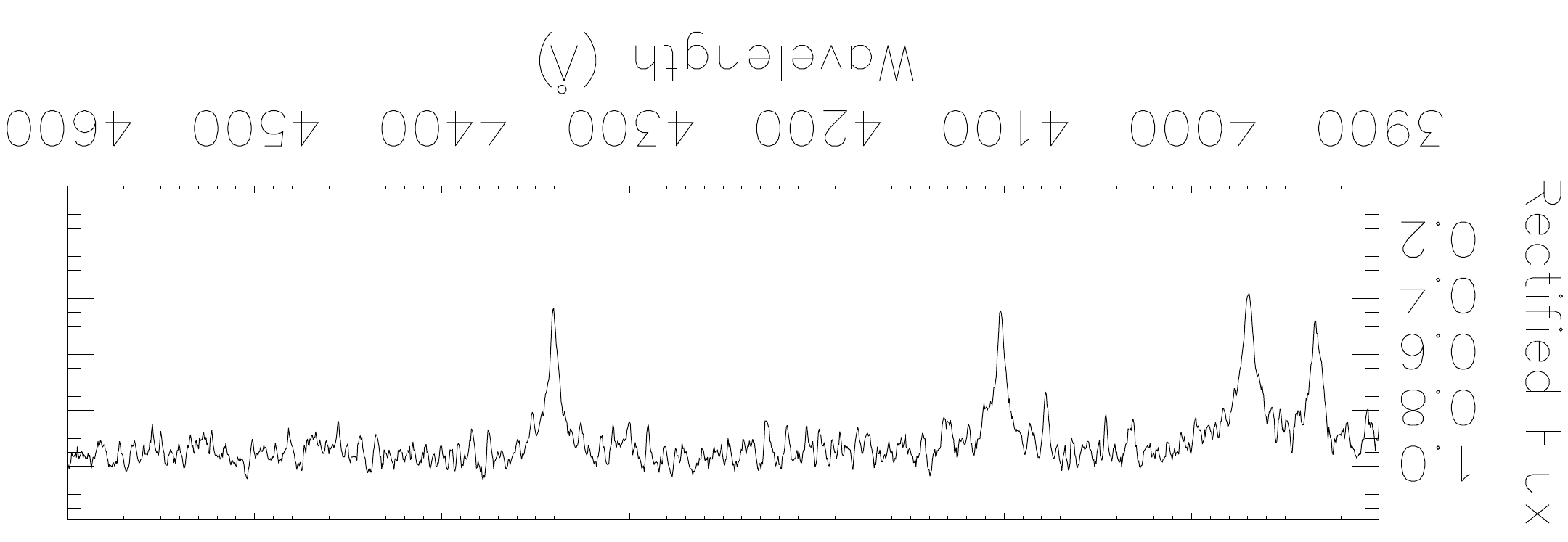}
  \caption{Periodogram and WHT/ISIS spectrum star J$1110$. The low frequency peaks in the periodogram are due to $1$-d aliasing.}
  \label{fig:J1110}
\end{figure}

\subsubsection{J$1430$}

Pulsating at a frequency of $235.5$ d$^{-1}$~with an amplitude of $1.06$ mmag (Fig. \ref{fig:J1430}), we classify J$1430$~as an A$9$p star with an effective temperature of $7100$~K derived from the Balmer lines. The spectrum shows an over-abundance of Eu~{\sc{ii}} (Fig. \ref{fig:J1430}). J$1430$ is the fastest roAp star we have found in the WASP archive, and is third fastest of all the roAp stars. 

HD~$86181$ is a similar object to J$1430$ in both pulsations and temperature. HD~$86181$ has a $T_{\rm eff}$ of $7900$~K \citep{balmforth01}, but is classified as only having an over-abundance of Sr. The H$_\alpha$ line profile of J$1430$ indicates the presence of a core-wing anomaly.

\setcounter{figure}{9}
\begin{figure}
  \centering
  \includegraphics[angle=180,width=80mm, trim= 23mm 21mm 37mm 110mm,clip]{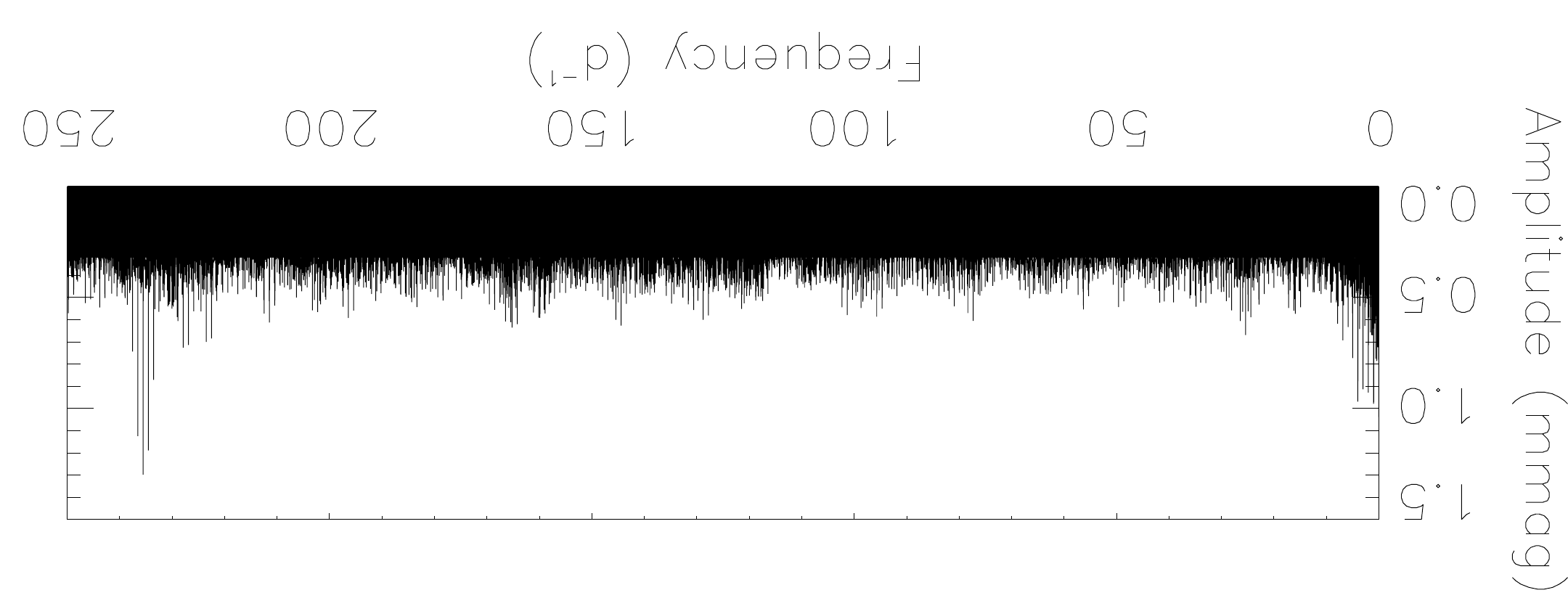}
  \includegraphics[angle=180,width=80mm, trim= 23mm 30mm 37mm 110mm,clip]{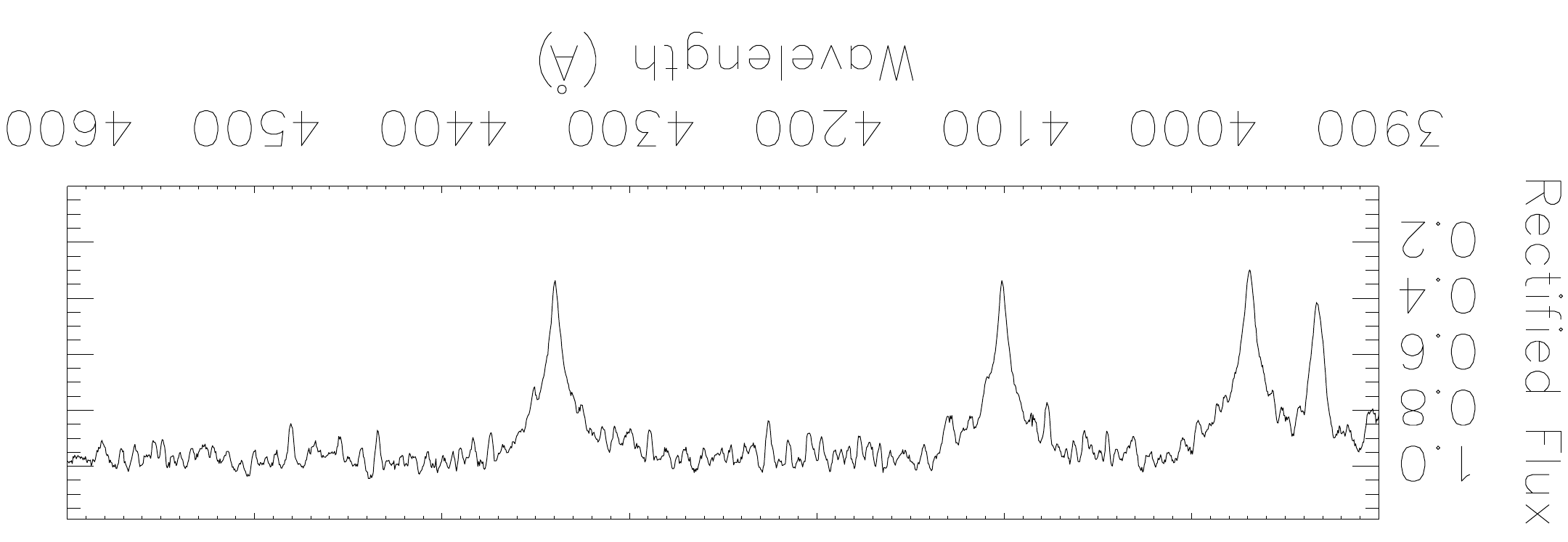}
  \caption{Periodogram and WHT/ISIS spectrum of J$1430$. The low frequency peaks in the periodogram are due to noise.}
  \label{fig:J1430}
\end{figure}

\subsubsection{J$1640$}

WASP photometry shows J$1640$ to pulsate at $151.93$ d$^{-1}$ with an amplitude of about $3.5$ mmag (Fig. \ref{fig:J1640}). The classification spectrum shows over-abundances of both Sr~{\sc{ii}} and Eu~{\sc{ii}} allowing us to classify this star as A$8$p (Fig. \ref{fig:J1640}). We also detect in the photometry a signature with a period of $3.67$ d, most likely due to the rotation of the star (Fig. \ref{fig:J1640-lc}).

We obtained two spectra of this target using two different instruments. The separate analysis of both spectra resulted in the same classification and $T_{\rm eff}$. There are no known roAp stars which exhibit a similar pulsation spectrum as J$1640$, in the sense that J$1640$ is exhibits pulsations of $3.52$ mmag in our diluted photometry, making it one of the highest amplitude pulsators.

\setcounter{figure}{10}
\begin{figure}
  \centering
  \includegraphics[angle=180,width=80mm, trim= 23mm 21mm 37mm 110mm,clip]{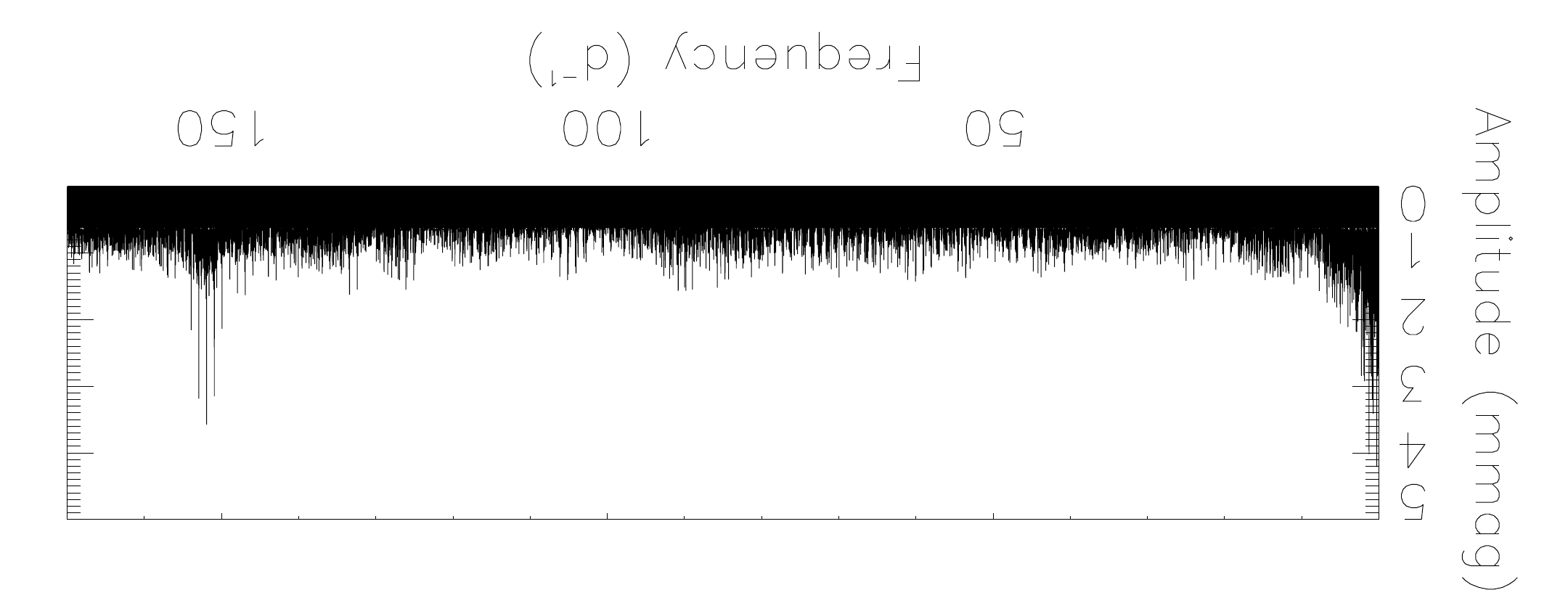}
  \includegraphics[angle=180,width=80mm, trim= 23mm 30mm 37mm 110mm,clip]{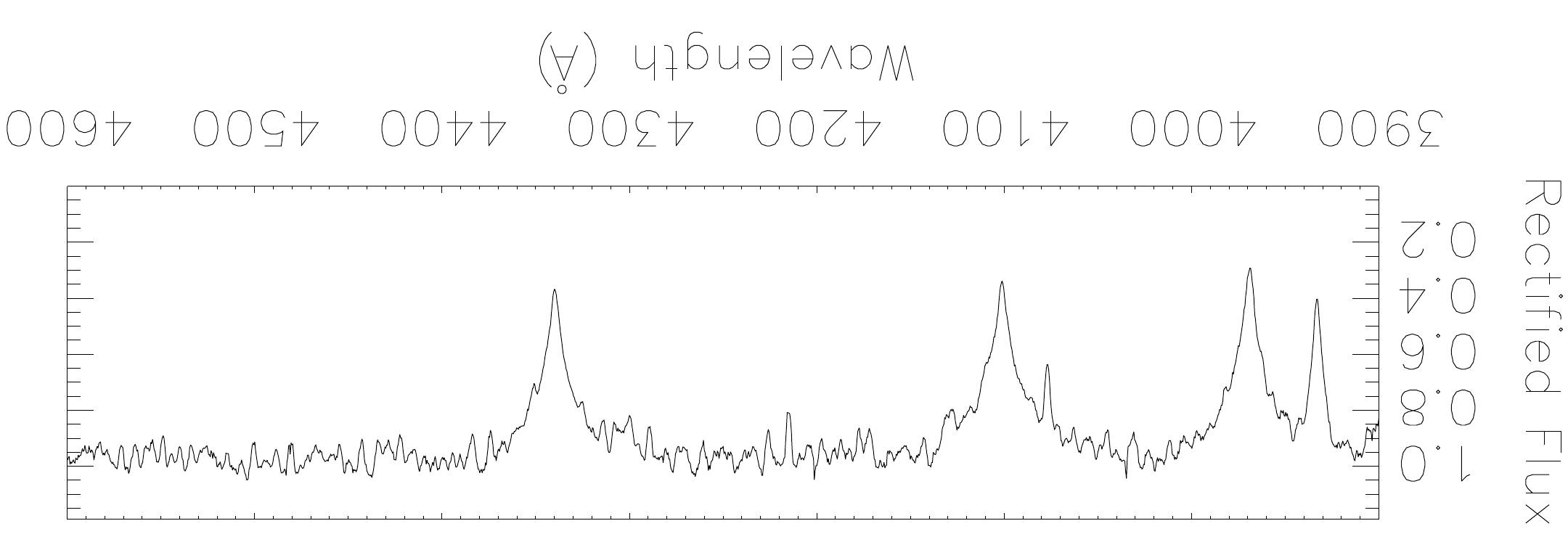}
  \caption{Periodogram and WHT/ISIS spectrum of J$1640$.}
  \label{fig:J1640}
\end{figure}

\setcounter{figure}{11}
\begin{figure}
  \centering
  \includegraphics[angle=180,width=80mm, trim= 23mm 21mm 37mm 110mm,clip]{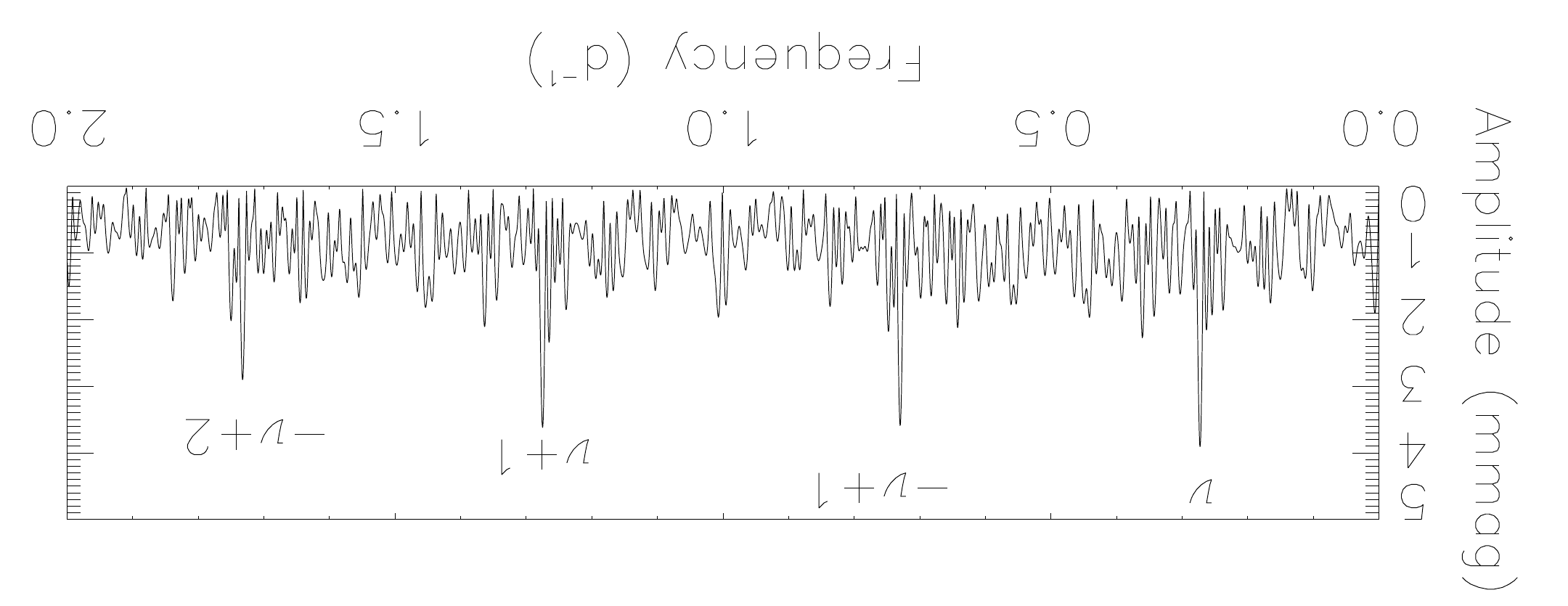}
  \includegraphics[angle=180,width=80mm, trim= 23mm 30mm 27mm 110mm,clip]{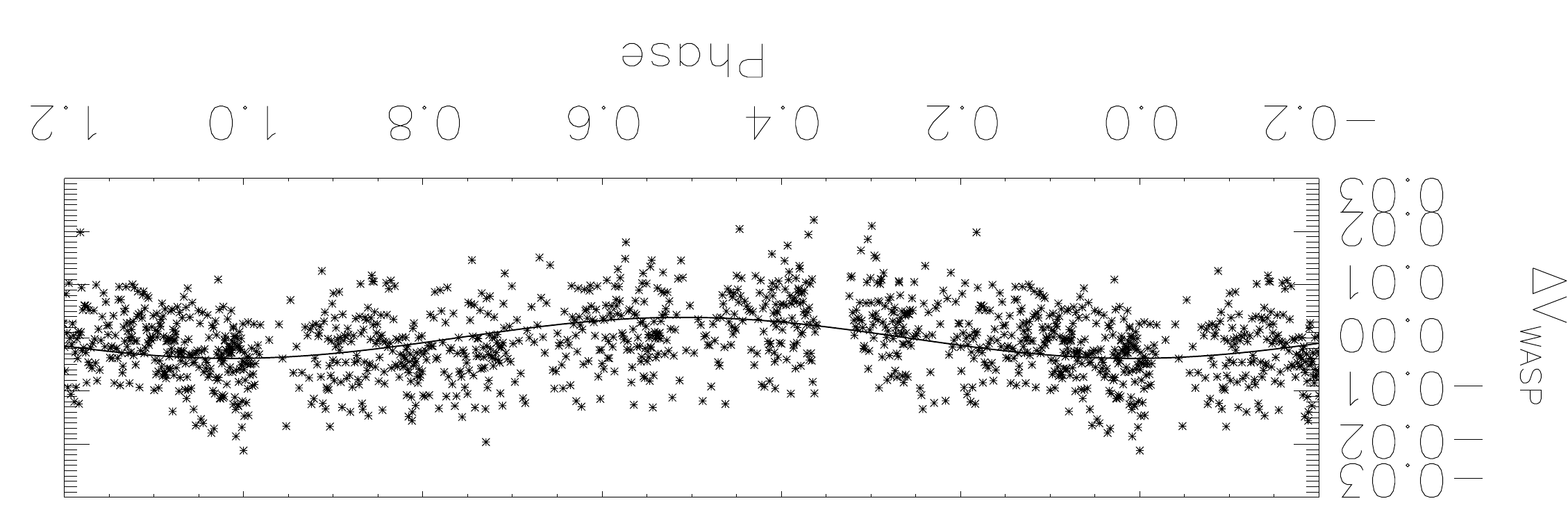}
  \caption{Low frequency periodogram and phase folded lightcurve of J$1640$. The data are folded on a period of $3.67$ d and shown in phase bins of $0.001$. The labelled peaks are the true frequency ($\nu$) and the positive and negative aliases.}
  \label{fig:J1640-lc}
\end{figure}

\subsubsection{J$1844$}

J$1844$~shows a low amplitude pulsation at $181.73$ d$^{-1}$ (Fig. \ref{fig:J1844}). We classify this star as A$7$p with a $T_{\rm eff}$~of about $7000$~K. The spectra for this target were obtained using the Hamilton Echelle Spectrometer mounted on the Shane $3$-m telescope at Lick observatory (Fig. \ref{fig:J1844}). We utilised this instrument to gain a high resolution spectrum to be able to perform a full abundance analysis on J$1844$ as it lies in the {\textit{Kepler}} field. Identified as KIC~$7582608$, the target has been observed in LC mode for the duration of the mission. Analysis of both the \kepler\, data and the HamSpec spectrum is under-way (Holdsworth et al., in preparation).

We observe J$1844$ to have a rotationally modulated lightcurve with a period of $20$ d (Fig. \ref{fig:J1844-lc}), a signature which is also present in the \kepler~data. This is the longest rotation period roAp star we present. The mono-periodic pulsations are similar to those of HD~$12098$ which pulsates at $189.22$ d$^{-1}$ with a blue-band amplitude of $3$ mmag \citep{martinez00}.

\setcounter{figure}{12}
\begin{figure}
  \centering
  \includegraphics[angle=180,width=80mm, trim= 23mm 21mm 37mm 110mm,clip]{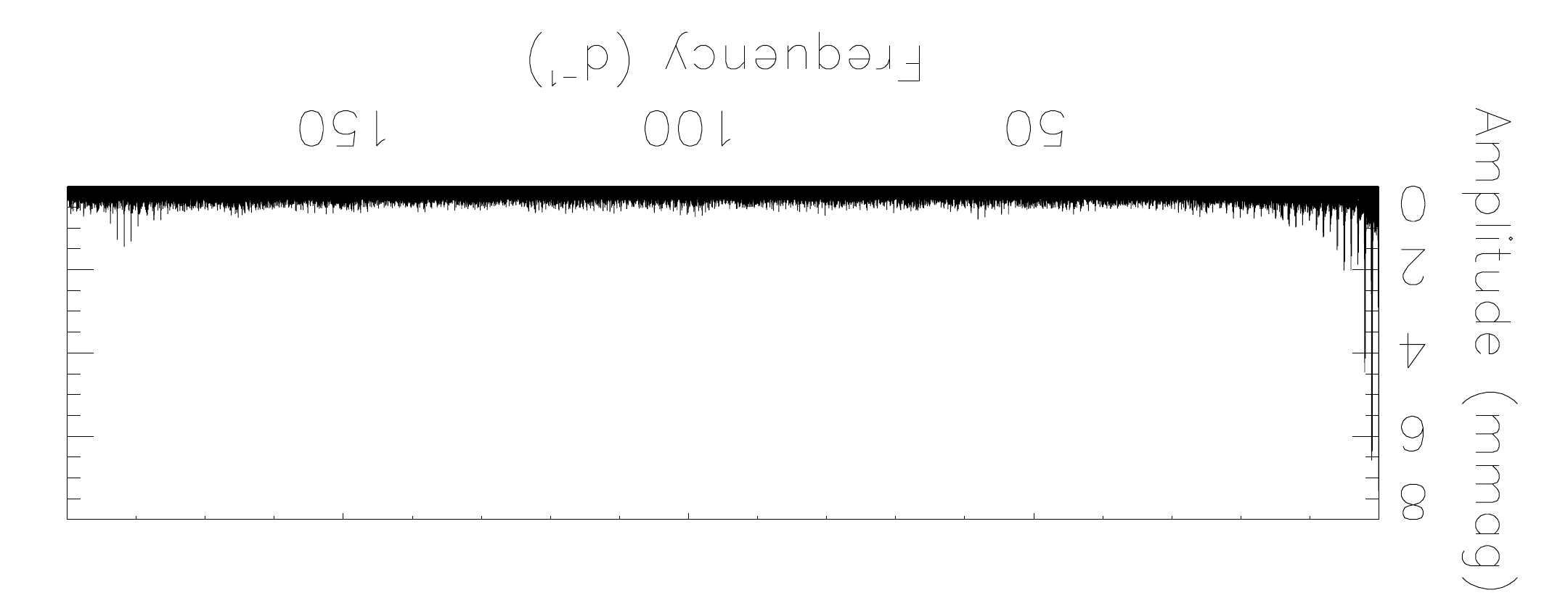}
  \includegraphics[angle=180,width=80mm, trim= 23mm 30mm 37mm 110mm,clip]{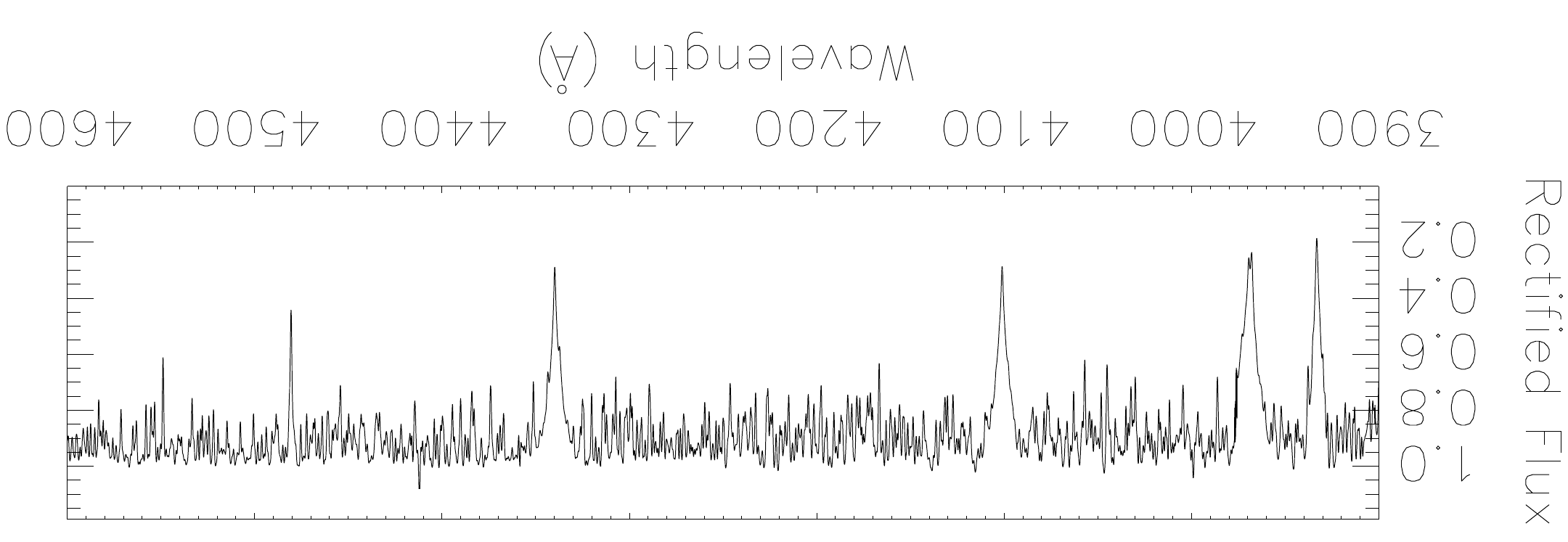}
  \caption{Periodogram and Shane/HamSpec spectrum of J$1844$.}
  \label{fig:J1844}
\end{figure}

\setcounter{figure}{13}
\begin{figure}
  \centering
  \includegraphics[angle=180,width=80mm, trim= 23mm 21mm 37mm 110mm,clip]{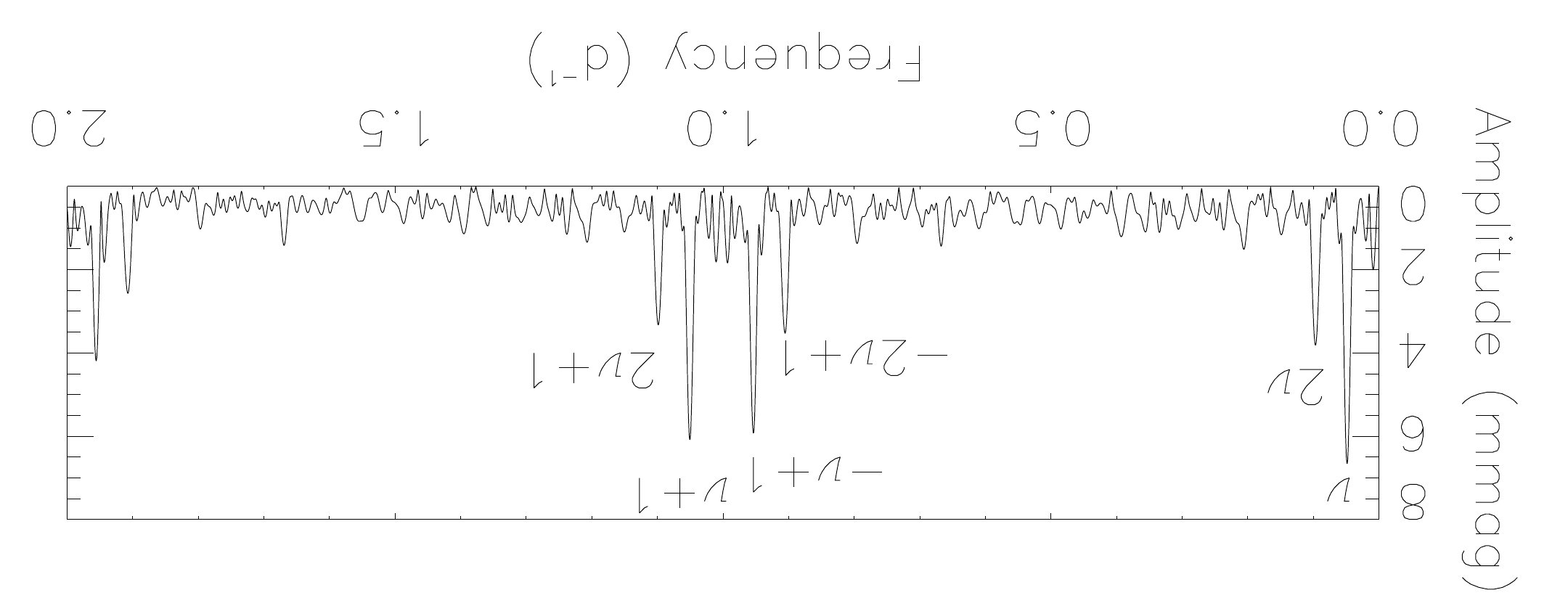}
  \includegraphics[angle=180,width=80mm, trim= 23mm 30mm 27mm 110mm,clip]{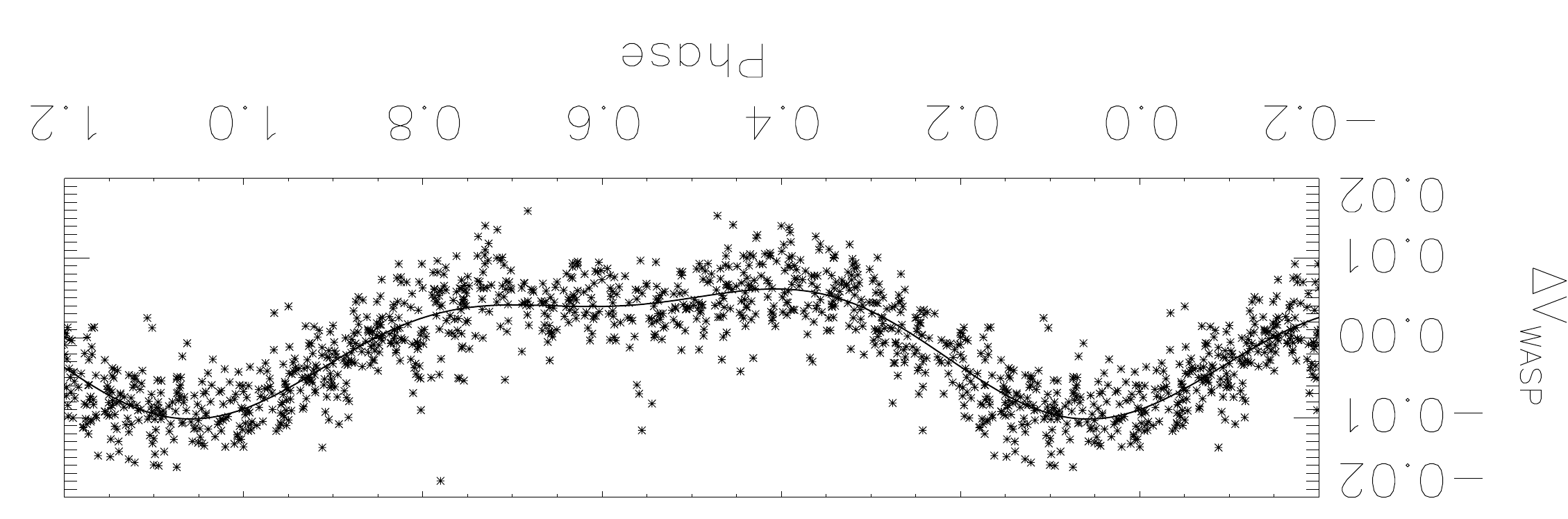}
  \caption{Low frequency periodogram and phase folded lightcurve of J$1844$. The data are folded on a period of $20.2$ d, and are shown in phase bins of $0.001$. The labelled peaks are the true frequency ($\nu$) and its harmonic and their positive and negative aliases.}
  \label{fig:J1844-lc}
\end{figure}

\subsubsection{J$1940$}
\label{sec:1940}

J$1940$ shows pulsations at $176.39$ d$^{-1}$, and is the highest amplitude roAp star discovered by SuperWASP at $4.2$ mmag (Fig. \ref{fig:J1940}). Given the effects of amplitude reduction of pulsations in the WASP data, J$1940$ may be the highest amplitude roAp star known. The SALT/RSS classification spectrum was obtained at a low S/N due to the faintness of the target. After smoothing the spectrum (by convolving it with a Gaussian profile), we deduce that J$1940$~is an F$2$p star with enhancements of Eu~{\sc{ii}} (Fig. \ref{fig:J1940}). We also detect a low-frequency signal in the photometry with a period of $9.58$ d (Fig. \ref{fig:J1940-lc}). J$1940$~is a prime example of the large pixel sizes of the WASP cameras, and the possible blending that this introduces. Fig. \ref{fig:J1940_app} shows an DSS image of the target star (centre) and surrounding objects. The over-plotted circle represents the photometric aperture of WASP (with a radius of $48\arcsec$). To confirm the source of the high-frequency pulsation, we used the TRAPPIST telescope \citep{jehin11} to observe the target for $3$ h. TRAPPIST is a $0.6$-m robotic telescope situated at the ESO La Silla Observatory. Backed by a CCD camera of $2048\times2048$ pixels, the instrument achieves a plate scale of $0.6$\arcsec per pixel. This vastly superior plate scale compared to WASP, enabled us to confirm that J$1940$~is the source of the pulsation (Fig. \ref{fig:TRAPPIST}). We were, however, unable to confirm the origin of the low frequency signature which may originate on one of the other objects in the aperture.

\setcounter{figure}{14}
\begin{figure}
  \centering
  \includegraphics[angle=180,width=80mm, trim= 23mm 21mm 37mm 110mm,clip]{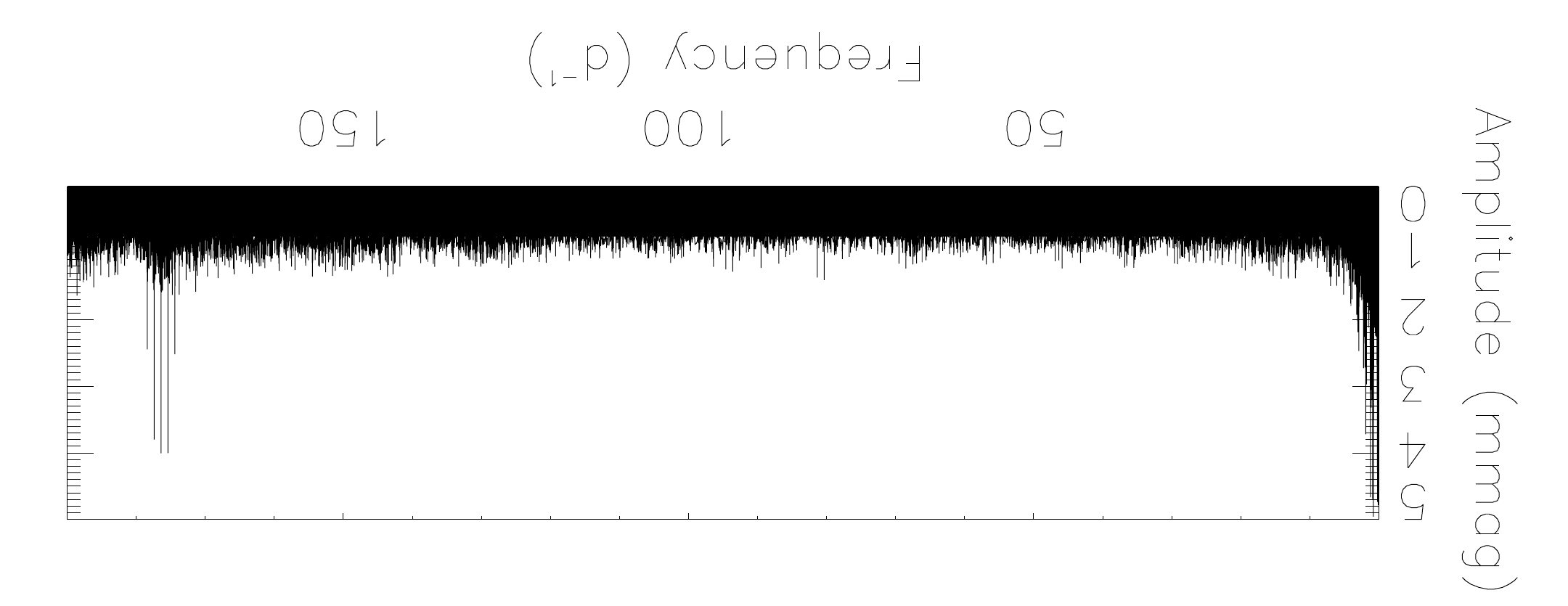}
  \includegraphics[angle=180,width=80mm, trim= 23mm 30mm 37mm 110mm,clip]{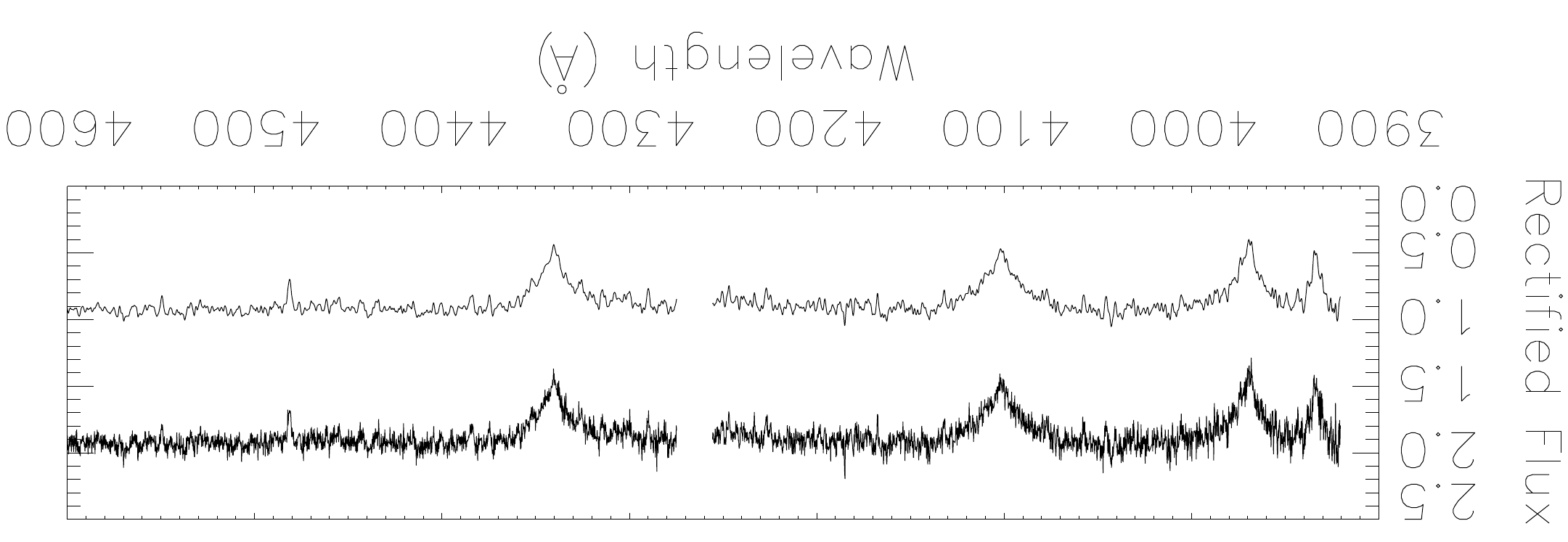}
  \caption{Periodogram and SALT/RSS spectrum of star J$1940$.}
  \label{fig:J1940}
\end{figure}

\setcounter{figure}{15}
\begin{figure}
  \centering
  \includegraphics[angle=180,width=80mm, trim= 23mm 21mm 37mm 110mm,clip]{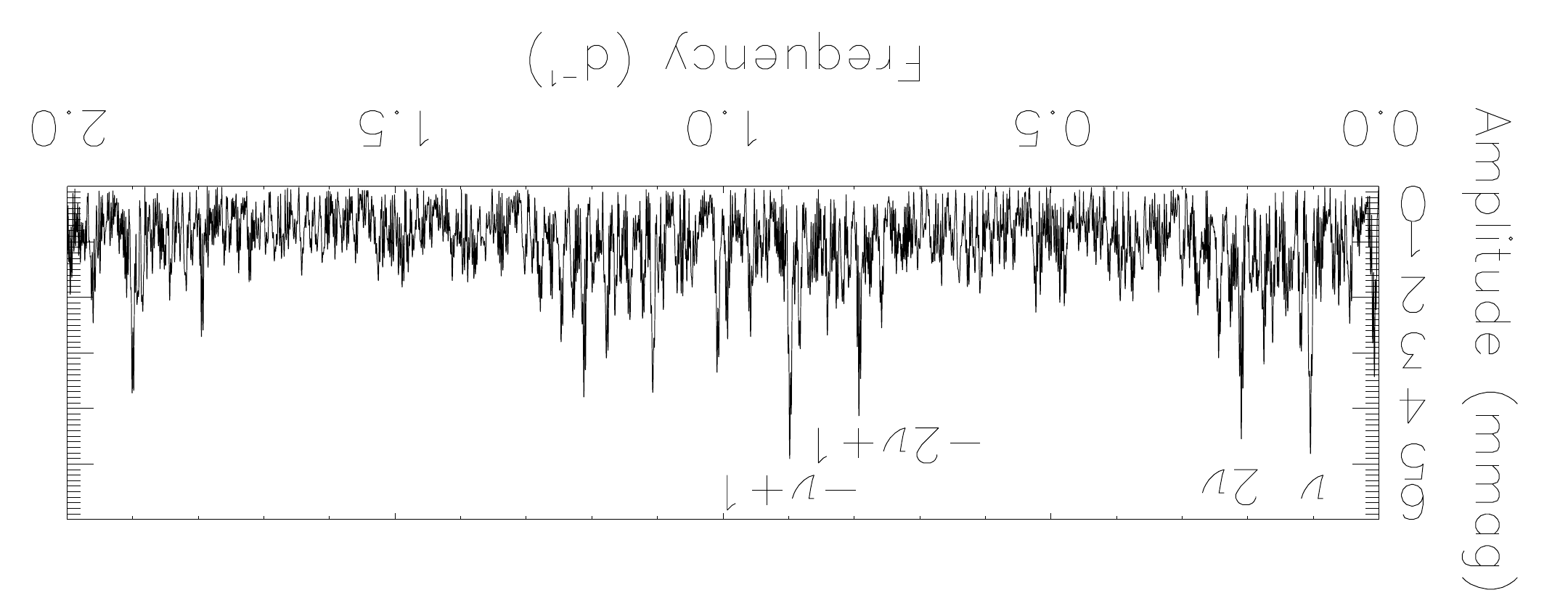}
  \includegraphics[angle=180,width=80mm, trim= 23mm 30mm 27mm 110mm,clip]{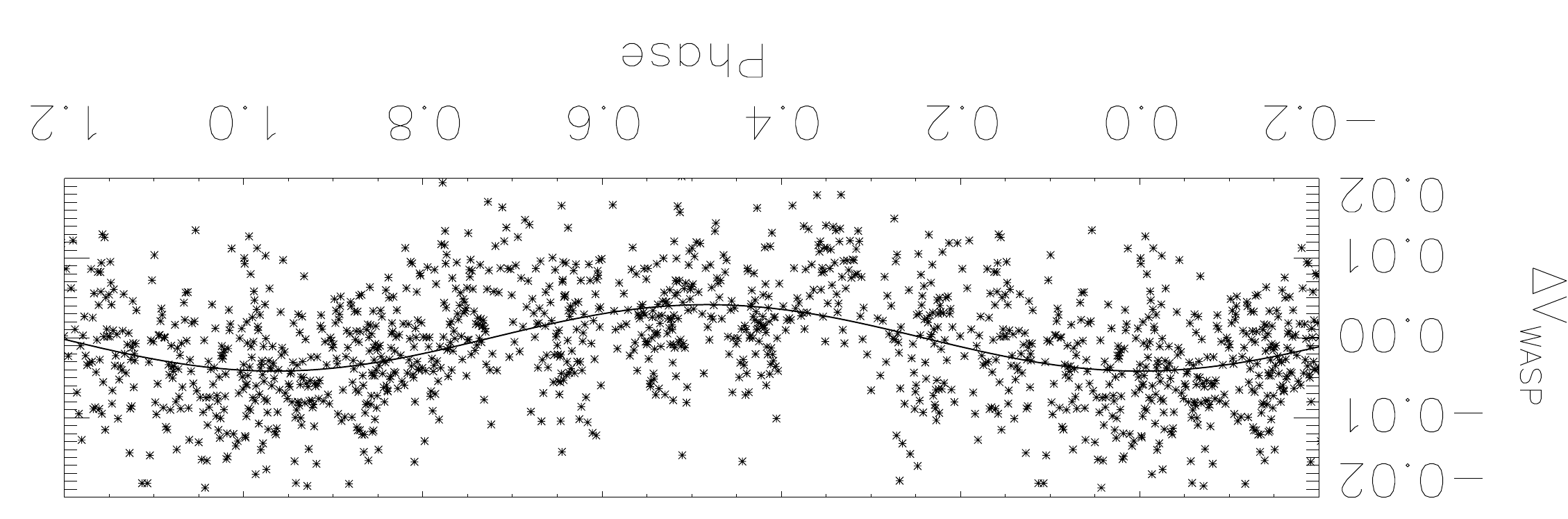}
  \caption{Low frequency periodogram and phase folded lightcurve of J$1940$. The data are folded on a period of $9.58$ d and shown in phase bins of $0.001$. The labelled peaks are the true frequency ($\nu$) and its harmonic and their positive and negative aliases.}
  \label{fig:J1940-lc}
\end{figure}

\setcounter{figure}{16}
\begin{figure}
  \centering
  \includegraphics[width=60mm,trim=27mm 140mm 105mm 58mm,clip]{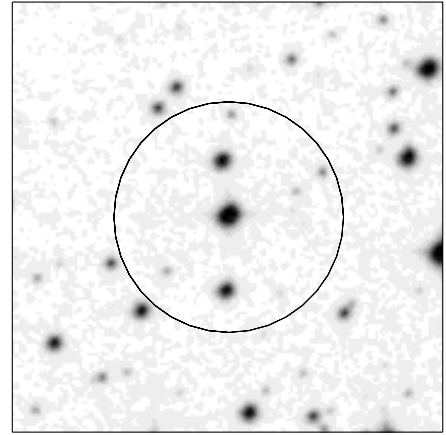}
  \caption{Photometric aperture (dark circle) for J$1940$~showing how multiple stars may fall into the aperture. Such targets require follow-up photometry to confirm which object is varying. For this example, TRAPPIST photometry was obtained, confirming the origin of the variations to be the target (central) star. Image from DSS.}
  \label{fig:J1940_app}
\end{figure}

\setcounter{figure}{17}
\begin{figure}
  \includegraphics[angle=180,width=80mm,trim=25mm 22mm 30mm 23mm,clip]{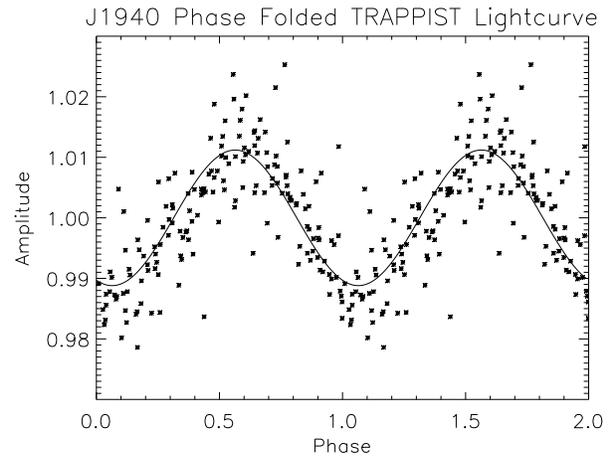}
  \caption{Phase folded TRAPPIST lightcurve for J1940. Obtained to confirm the origin of the variability seen in the WASP data. Data folded on the principal frequency of $176.39$ d$^{-1}$.}
  \label{fig:TRAPPIST}
\end{figure}

\subsection{Other pulsating stars}

As we have a lower boundary of $50$ d$^{-1}$ as our search criterion, we also anticipated the detection of fast $\delta$~Sct systems. In total, we detect $375$~objects which have periods less than $30$ min. We chose, for follow-up spectroscopic observations, objects which appear to us as either possibly relatively slow roAp stars, with the slowest known being HD~$17765$~\citep[$61$ d$^{-1}$;][]{alentiev12}, or multi-periodic $\delta$~Sct stars. 

\setcounter{table}{4}
\begin{table*}
  \centering
  \begin{minipage}{155mm}
    \centering
    \caption{Abridged version of the photometric information for the $\delta$~Sct pulsators. In the full table, available on-line$^{a}$, columns $6-15$~give the first five frequencies, if present, above $50$ d$^{-1}$.}
    \label{tab:online}
    \begin{tabular}{cccccccc}
      
      \hline
      \hline
      
      \multicolumn{1}{c}{\multirow{1}{*}{WASP}} & 
      \multicolumn{1}{c}{\multirow{1}{*}{Other}} & 
      \multicolumn{1}{c}{\multirow{2}{*}{{\textit{V}}}}&
      \multicolumn{1}{c}{\multirow{1}{*}{{$\nu_1$}}}&
      \multicolumn{1}{c}{\multirow{1}{*}{A$_1$}}&
      \multicolumn{1}{c}{\multirow{2}{*}{{\ldots}}}&
      \multicolumn{1}{c}{\multirow{1}{*}{{$\nu_5$}}}&
      \multicolumn{1}{c}{\multirow{1}{*}{A$_5$}}\\

      \multicolumn{1}{c}{\multirow{1}{*}{ID}} & 
      \multicolumn{1}{c}{\multirow{1}{*}{ID}} & 
      \multicolumn{1}{c}{} & 
      \multicolumn{1}{c}{\multirow{1}{*}{(d$^{-1}$)}} &
      \multicolumn{1}{c}{\multirow{1}{*}{(mmag)}} &
      \multicolumn{1}{c}{\multirow{1}{*}{}} &
      \multicolumn{1}{c}{\multirow{1}{*}{(d$^{-1}$)}} &
      \multicolumn{1}{c}{\multirow{1}{*}{(mmag)}} \\
      \hline
      
      1SWASPJ000415.12-172529.6 & HD 225186               & 9.05  & 60.08  & 3.40 & {\ldots}  & --  &  --  \\
      1SWASPJ000537.79+313058.8	& TYC 2259-818-1          & 11.70 & 52.92  & 1.47 & {\ldots}  & --  &  --  \\
      1SWASPJ000830.50+042818.1 & TYC 4-562-1             & 10.16 & 150.26 & 0.76 & {\ldots}  & --  &  --  \\
      1SWASPJ000940.84+562218.9	& TYC 3660-1935-1         & 10.34 & 66.37  & 3.15 & {\ldots}  & --  &  --  \\
      1SWASPJ002436.35+165847.3	& HD 2020                 & 8.50  & 54.41  & 3.40 & {\ldots}  & --  &  --  \\

      \hline
      \multicolumn{8}{l}{$^{a}$ Also available at CDS}\\
    \end{tabular}
  \end{minipage}
\end{table*}

Of our spectroscopically observed targets, we classify $13$~stars as new pulsating Am stars, with a frequency range of $65-164$~d$^{-1}$, and temperature range of $7700-8300$~K. However, we note that at classification resolution there is not always a clear distinction between Am and Ap stars. It is also possible that within this group of stars there are Ap stars which show multi-periodicity of the $\delta$~Sct type. Although it was not initially thought that Ap stars pulsate in the $\delta$~Sct range, \citet{kurtz00} proposed a list of likely targets, with the first example of such a system being HD~$21190$ \citep{koen01}. \kepler~observations have also detected this phenomenon in $5$~of the $7$~Ap stars that it has observed \citep[e.g.][]{balona11a}. However, there are currently no known systems which exhibit both high overtone roAp pulsations and low overtone $\delta$~Sct pulsations. Theoretical work by \cite{saio05} suggests that both high and low overtone p-modes cannot co-exist in magnetic Ap stars as $\delta$~Sct pulsations are suppressed by the presence of the magnetic field, whereas roAp pulsations can be enhanced. However, our survey has identified a few targets, such as J$1917$ (Section \ref{sec:1917}), which show both low and high overtone p-modes in a single target. Further observation are needed, however, to eliminate any other explanations such as target blending and binary systems.

We present below four further targets which show pulsations above $80$ d$^{-1}$ for which we have obtained spectra.

The remaining $23$ targets for which we obtained spectra are presented in Appendix \ref{sec:appendix}. We also provide an on-line catalogue of all the variable systems which show periods shorter than $30$ min. Table \ref{tab:online} shows an example of the on-line table format.

\subsubsection{J$1403$}
J$1403$~is an interesting target as our observations show it to pulsate in two distinct frequency ranges (Fig. \ref{fig:J1403} top). We detect $9$~pulsational frequencies between $25$~and $34$ d$^{-1}$ and $4$~between $87$~and $100$ d$^{-1}$. We classify the star as A$9$ (Fig. \ref{fig:J1403} bottom), however we note different classifications recorded in the literature (e.g. A$3$/$5$III; \citealt{houk78}, F$0$V; \citealt{pickles10}, F$8$~(HD)). We detect low-frequency variations in J$1403$~at a period of $1.5053$ d. The phases folded plot (Fig. \ref{fig:J1403-lc}) shows that this target is most likely an ellipsoidal variable. In such a case, we do not expect this target to be a hybrid pulsator, but a pulsating non-eclipsing binary pair. 

\setcounter{figure}{18}
\begin{figure}
  \centering
  \includegraphics[angle=180,width=80mm, trim= 23mm 21mm 37mm 110mm,clip]{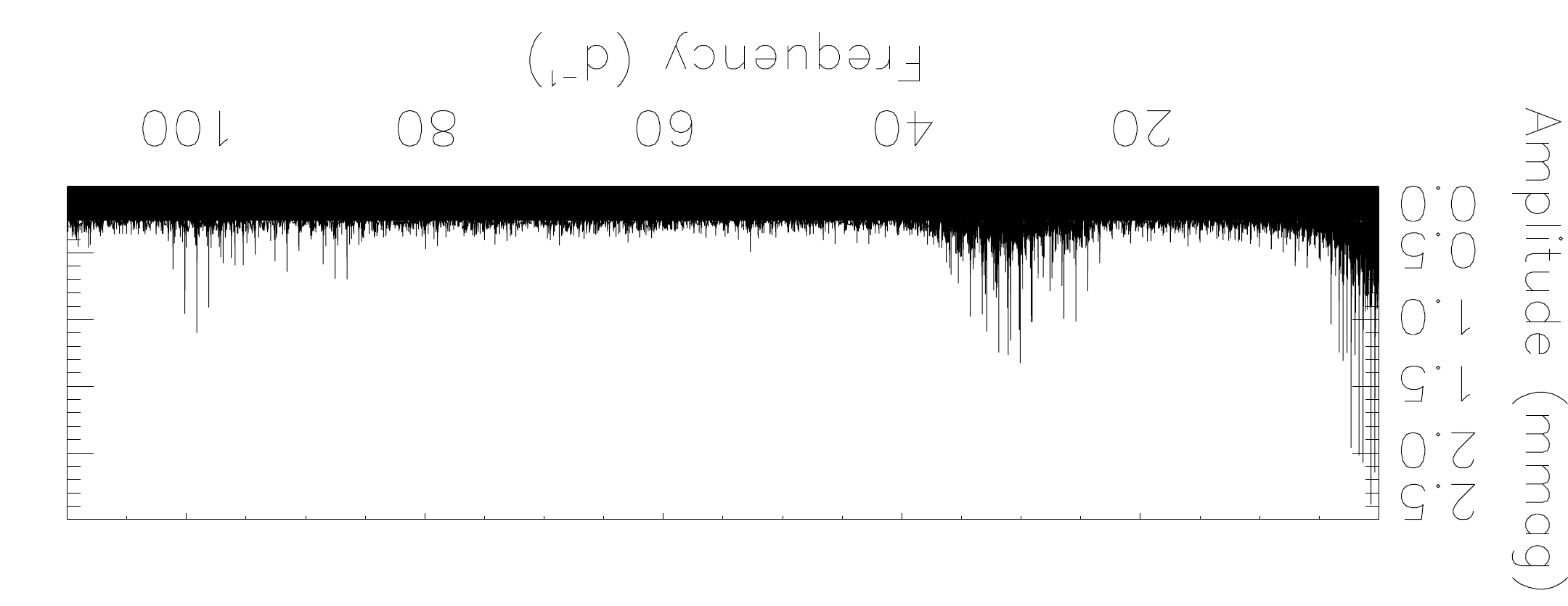}
  \includegraphics[angle=180,width=80mm, trim= 23mm 30mm 37mm 110mm,clip]{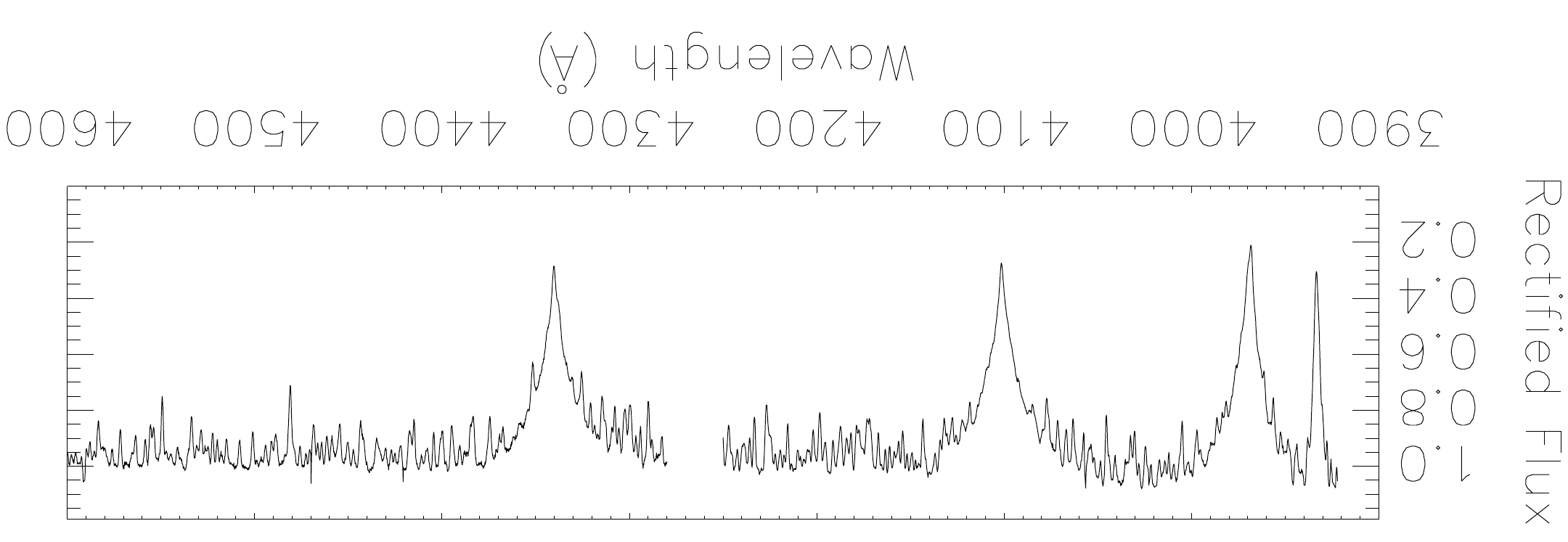}
  \caption{Periodogram and SALT/RSS spectrum of J$1403$.}
  \label{fig:J1403}
\end{figure}

\setcounter{figure}{19}
\begin{figure}
  \centering
  \includegraphics[angle=180,width=80mm, trim= 23mm 21mm 37mm 110mm,clip]{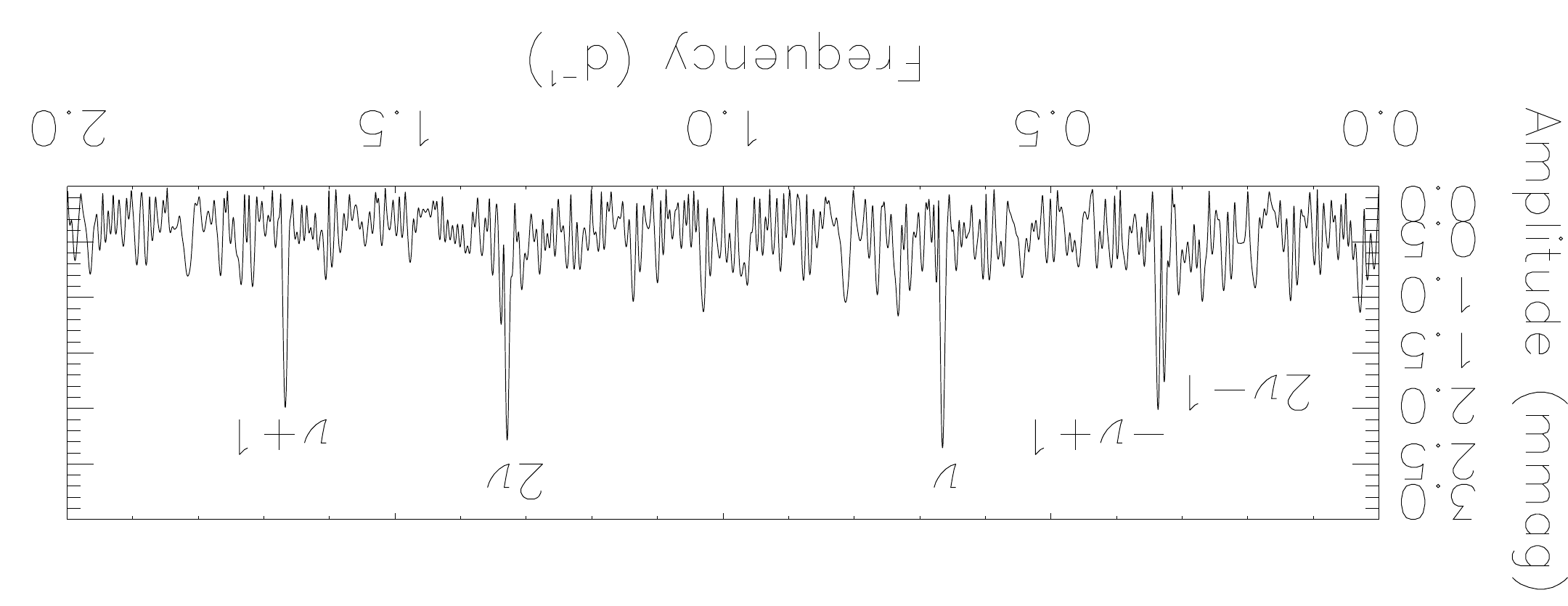}
  \includegraphics[angle=180,width=80mm, trim= 23mm 30mm 22mm 110mm,clip]{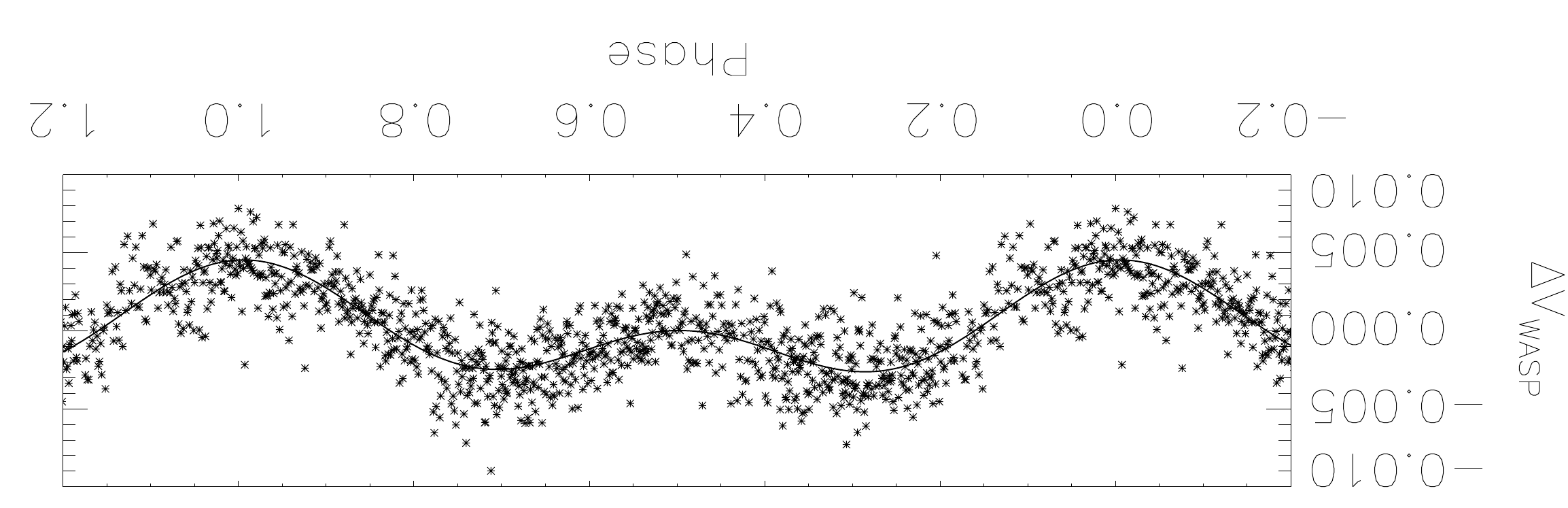}
  \caption{Low frequency periodogram and phase folded lightcurve of J$1403$. The data are folded on a period of $1.5053$ d and shown in phase bins of $0.001$. The frequency on which the data are folded has a FAP of $0.00$. The labelled peaks are the true frequency ($\nu$) and its harmonic and their positive and negative aliases.}
  \label{fig:J1403-lc}
\end{figure}

\subsubsection{J$1917$}
\label{sec:1917}
J$1917$~also shows pulsations in two separate regions. However, here we have pulsations in the $\delta$~Sct range ($41.3$, $45.5$ \& $53.1$ d$^{-1}$) and a single peak in the higher frequency domain ($164$ d$^{-1}$). We obtained a classification spectrum with the RSS spectrograph on SALT, revealing the star to be Am in nature. Similar to the blending problem we saw with J$1940$, J$1917$~also has nearby objects which may have caused this multi-periodic lightcurve (Fig. \ref{fig:J1917_app}). We again used the TRAPPIST telescope to confirm that J$1917$~is the source of the pulsations (see Fig. \ref{fig:J1917-trap-pd}). The star is listed in the Washington Double Star catalogue \citep[WDS;][]{mason01} with a separation of $1.3$\arcsec~based on two observations in the 1930s. We estimate the distance to J$1917$~to be about $400$ pc, which suggests a binary separation of $500$ au. The two components of the visual binary system have magnitudes of $10.8$~and $14$~in the \textit{V}-band \citep{mason01}. If the companion is a main-sequence star, given the spectral type of the primary, we estimate it to be a G-type star. We would expect none of the pulsations to originate from a G star. We propose that either both sets of pulsations originate on the Am star, or that there is an unresolved binary system with two pulsating components. Further observations are required to fully understand the nature of J$1917$.

\setcounter{figure}{20}
\begin{figure}
  \centering
  \includegraphics[width=60mm,trim=27mm 140mm 105mm 58mm,clip]{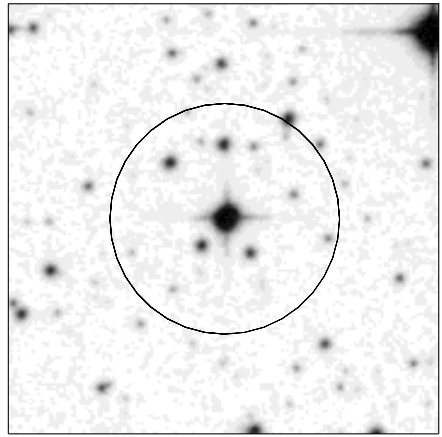}
  \caption{Photometric aperture (dark circle) for J$1917$~showing multiple stars in the WASP aperture. Follow-up observations using the TRAPPIST telescope confirm the pulsations are originating from the central object. Image from DSS.}
  \label{fig:J1917_app}
\end{figure}

\setcounter{figure}{21}
\begin{figure}
  \centering
  \includegraphics[angle=180,width=80mm, trim=23mm 16mm 28mm 110mm,clip]{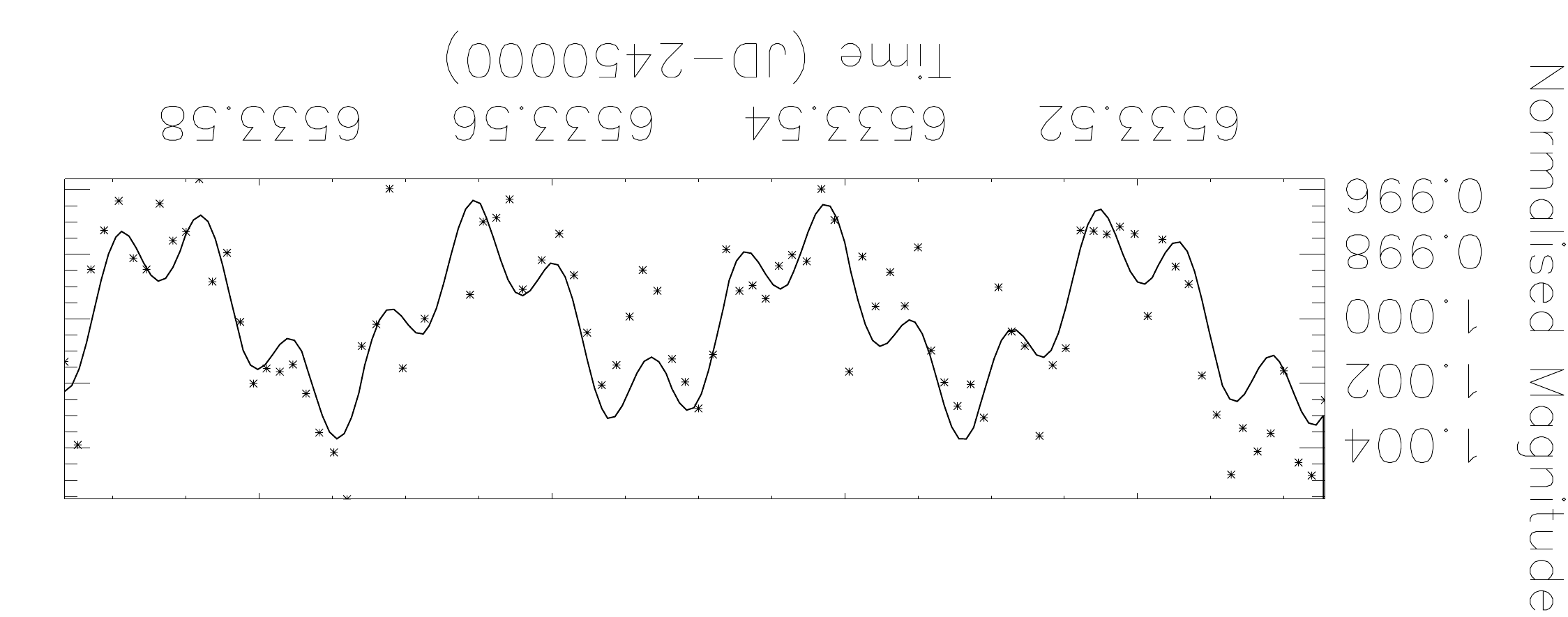}
  \includegraphics[angle=180,width=80mm, trim=23mm 21mm 37mm 110mm,clip]{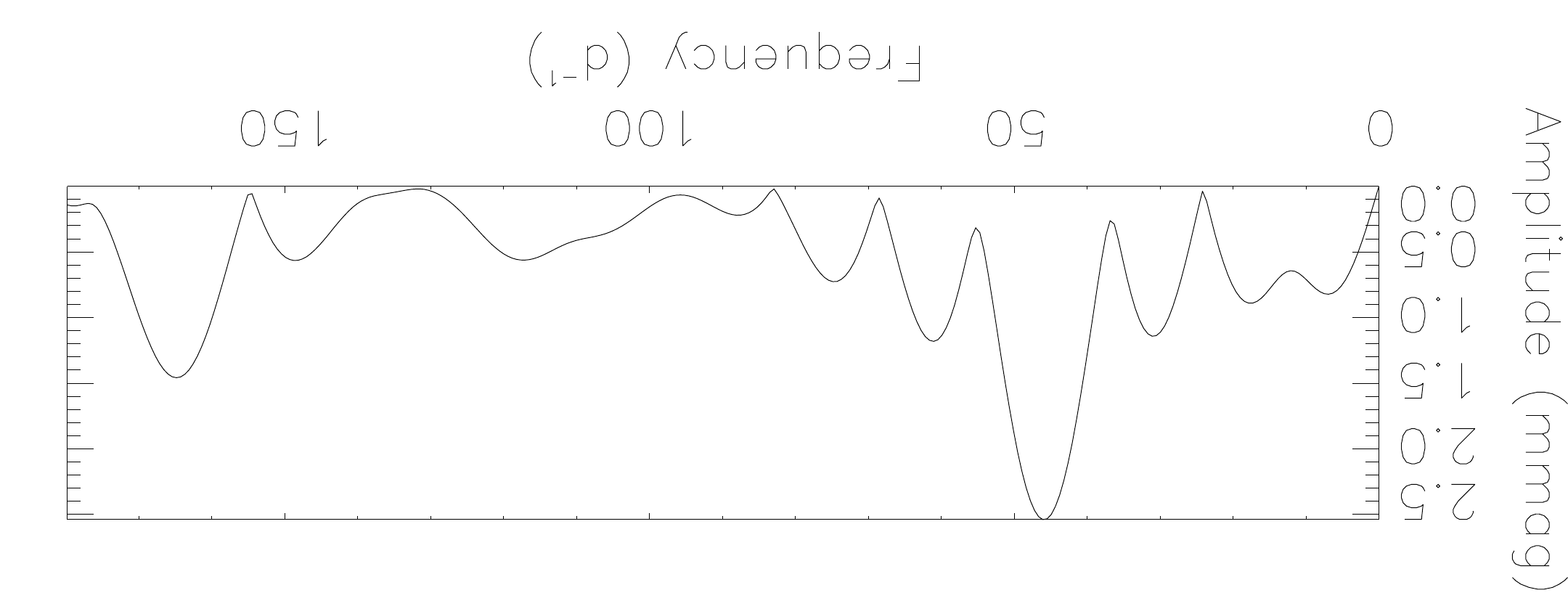}
  \includegraphics[angle=180,width=80mm, trim=23mm 21mm 37mm 110mm,clip]{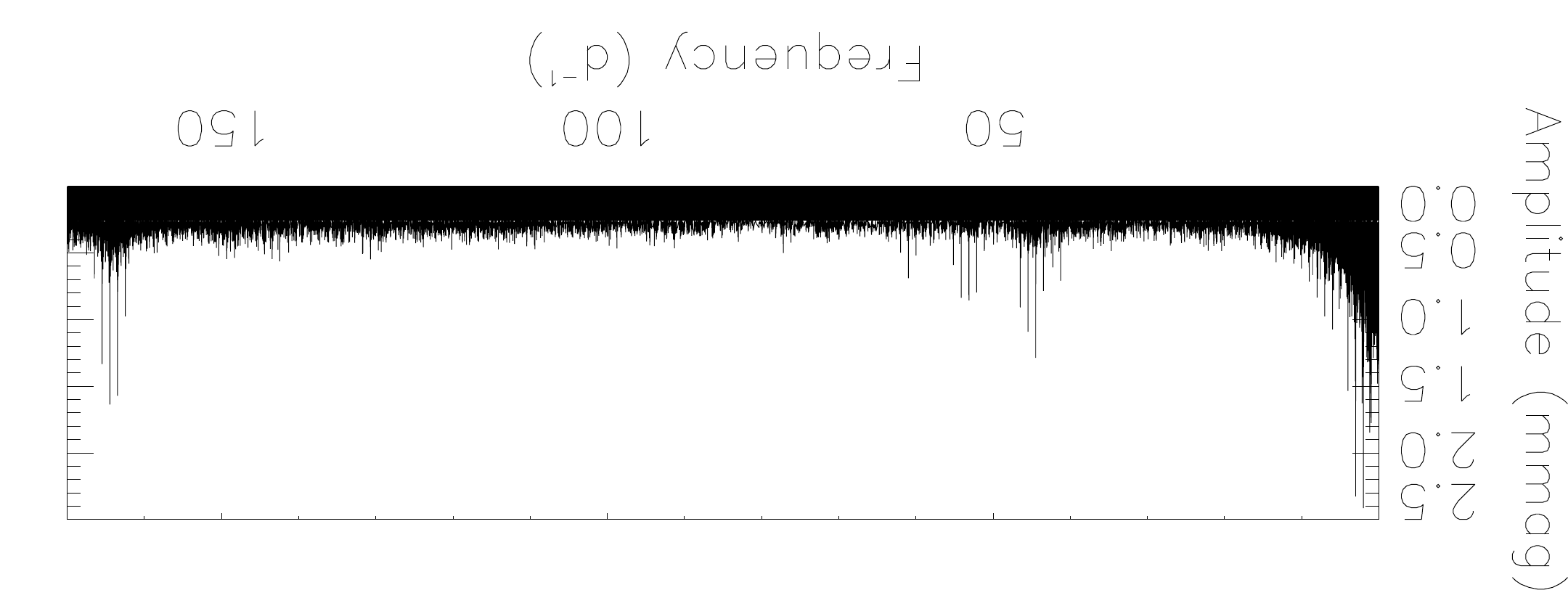}
  \includegraphics[angle=180,width=80mm, trim=23mm 30mm 37mm 110mm,clip]{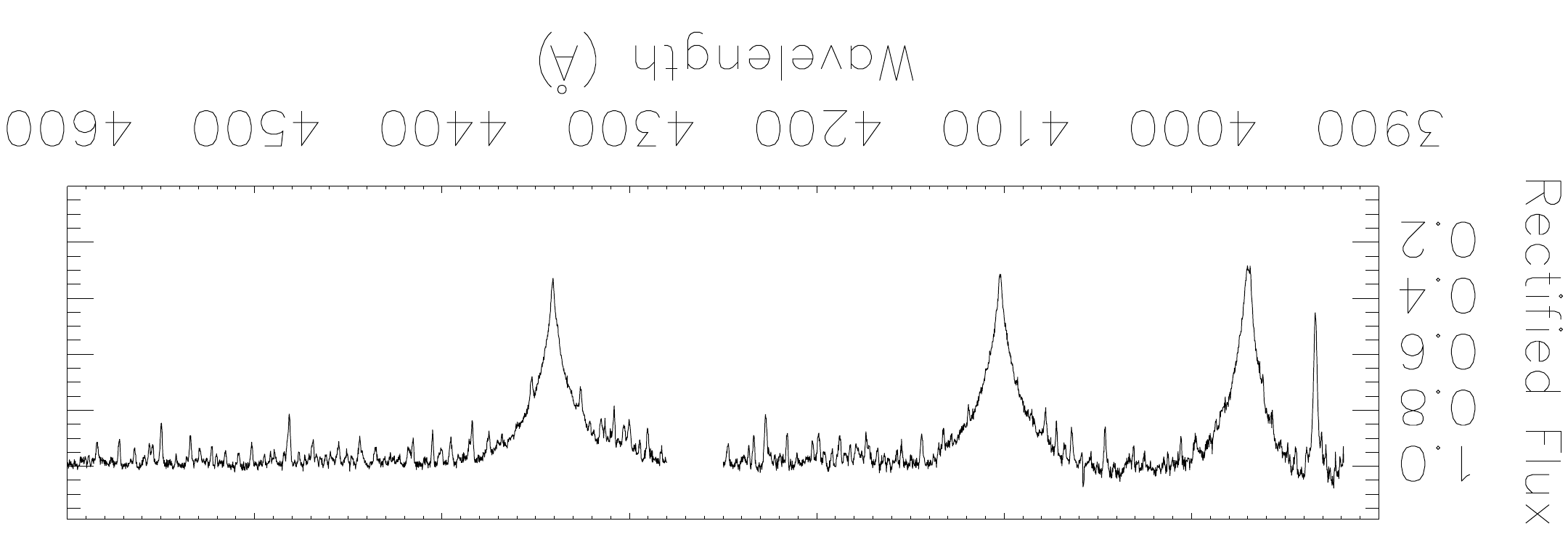}
  \caption{Top: Two hours of TRAPPIST follow-up observations confirm the pulsations originate on J$1917$ (the data have been binned on $\sim1$ min intervals). Second: Periodogram of the TRAPPIST data. The peak at $45$ d$^{-1}$ is stronger than in the WASP periodogram as there are many pulsations contributing power that are not resolved in just $2$ h. The high-frequency is at about the same amplitude as the WASP data. Third: The WASP periodogram of J$1917$~for comparison. Bottom: SALT/RSS spectrum of J$1917$~showing it to be an A7m star.}
  \label{fig:J1917-trap-pd}
\end{figure}

\subsubsection{J$2054$}

J$2054$ shows two pulsations over $100$ d$^{-1}$. The strongest is at $104.86$ d$^{-1}$ with an amplitude of $1.1$ mmag, with the second at $100.44$ d$^{-1}$ with an amplitude of $0.53$ mmag. The Shane/HamSpec spectrum obtained for this target indicates that it is an A$3$m: star based on a weakened Ca~K line and slightly enhanced Sr. We note that there is no clear depletion of Sc {\sc{ii}}. We estimate a $v\sin i$ of $\sim$~50~km~s$^{-1}$ for J$2054$.

\setcounter{figure}{22}
\begin{figure}
  \centering
  \includegraphics[angle=180,width=80mm, trim= 23mm 21mm 37mm 110mm,clip]{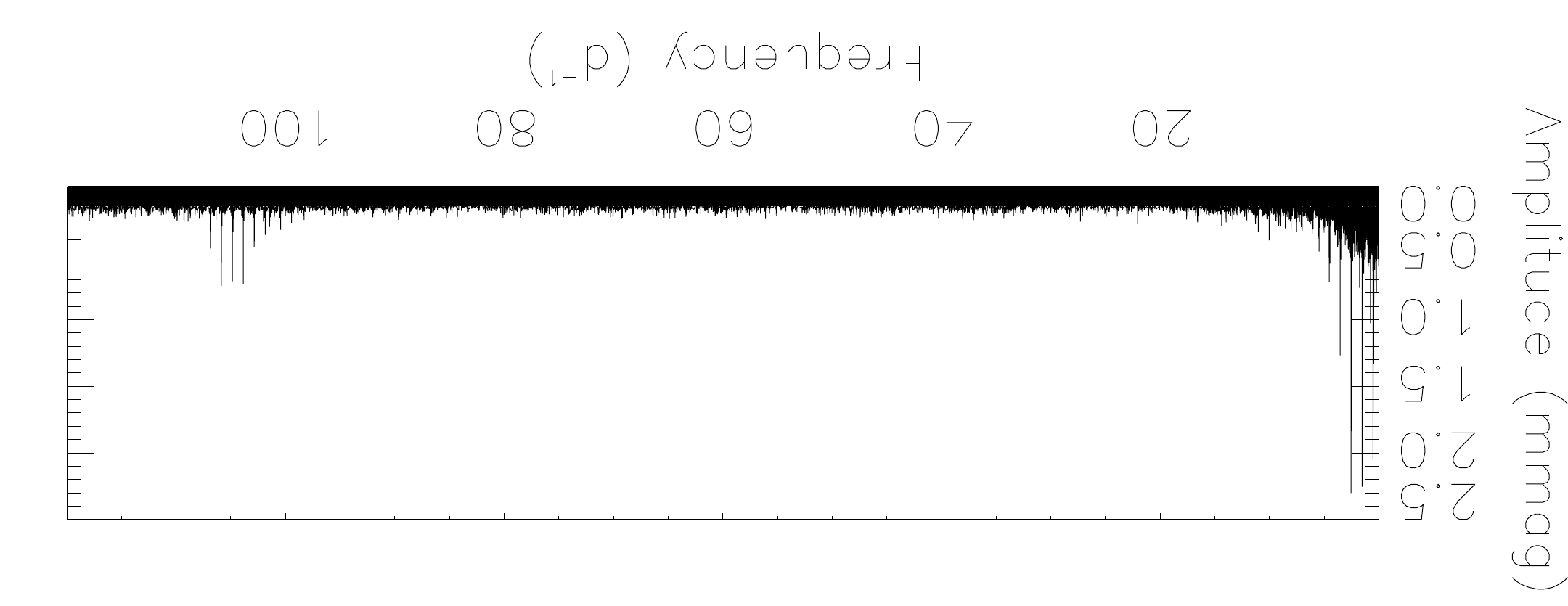}
  \includegraphics[angle=180,width=80mm, trim= 23mm 30mm 37mm 110mm,clip]{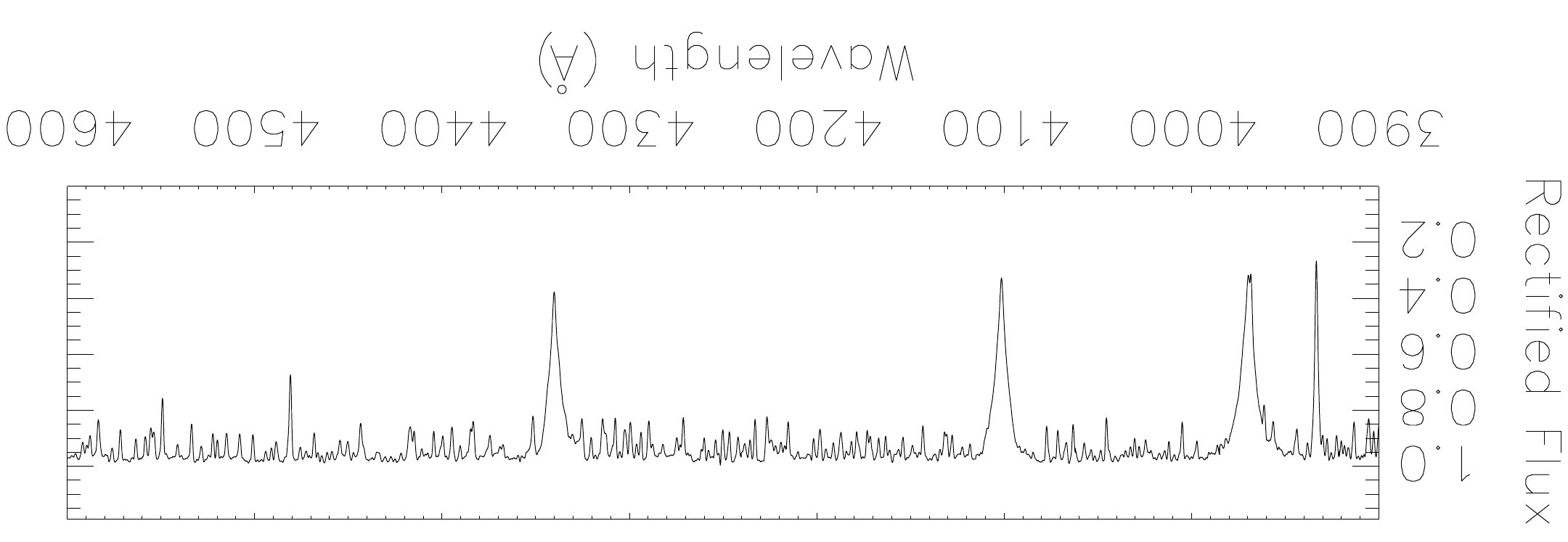}
  \caption{Periodogram and Shane/HamSpec spectrum of J$2054$.}
  \label{fig:J2054}
\end{figure}

Low frequency variations are noted in the photometry, corresponding to a period of $1.3$ d (Fig. \ref{fig:J2054-lc}). When the data are folded on this period, the resulting phase diagram shows two maxima with unequal minima. The lightcurve indicates a binary target, where the pulsations may originate in one or both components. However there is no evidence for this in the single spectrum we obtained.

J$2054$ is the fastest Am star that we have found with SuperWASP, superseding the previous fastest, HD~$108452$~\citep{smalley11} pulsating at $71$ d$^{-1}$. Our results have therefore pushed the boundary of the pulsating Am stars farther into the domain of the roAp stars, further blurring the distinction between these two types of pulsator.

\setcounter{figure}{23}
\begin{figure}
  \centering
  \includegraphics[angle=180,width=80mm, trim= 23mm 21mm 37mm 110mm,clip]{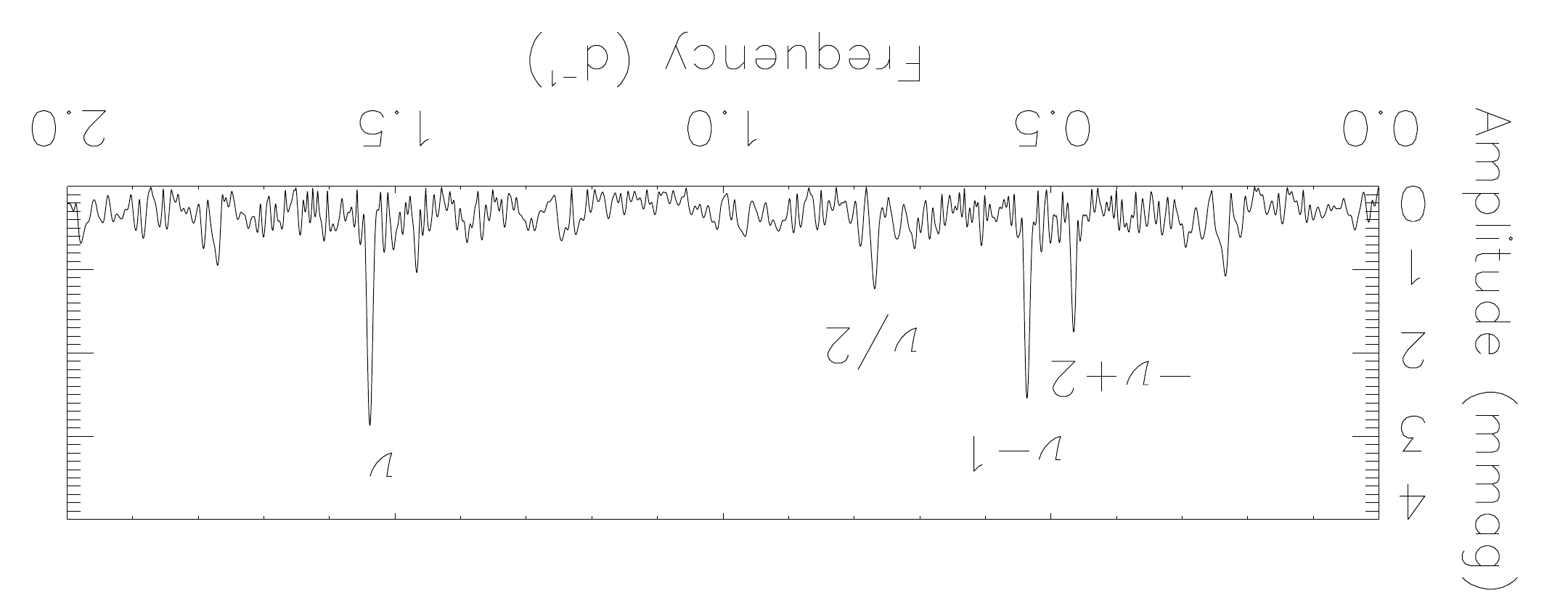}
  \includegraphics[angle=180,width=80mm, trim= 23mm 30mm 22mm 110mm,clip]{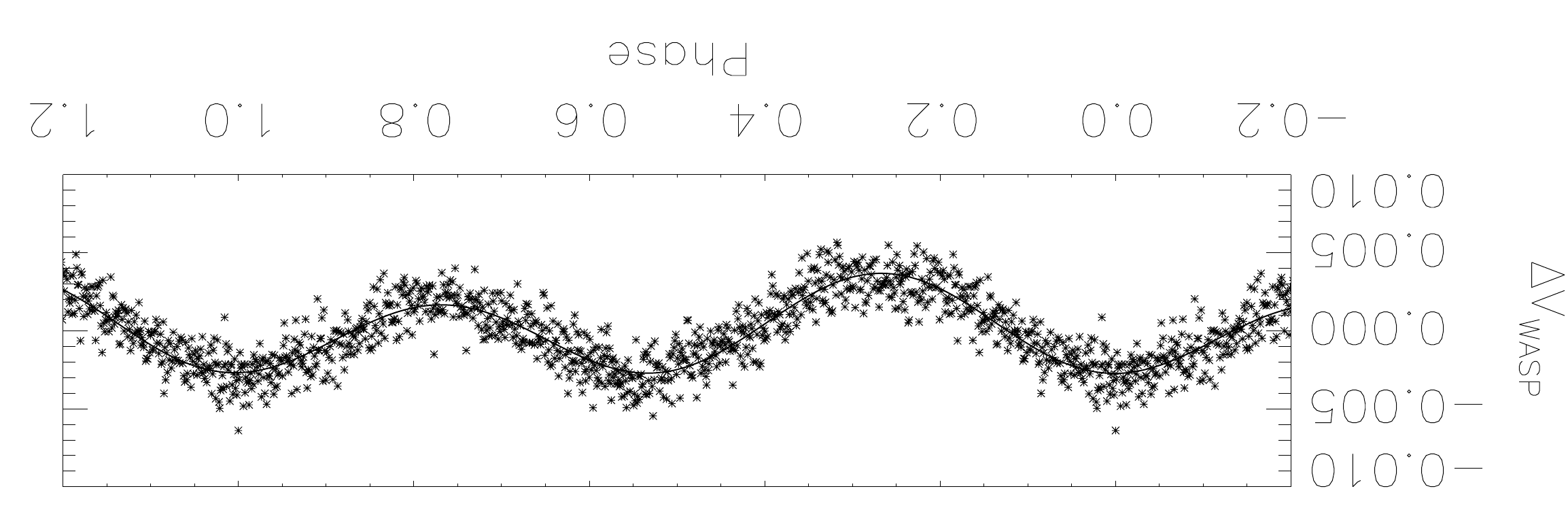}
  \caption{Low frequency periodogram and phase folded lightcurve of J$2054$. The data are folded on a period of $1.3$ d and shown in phase bins of $0.001$. The frequency on which the data are folded has a FAP of $0.00$. The labelled peaks are the true frequency ($\nu$) and its subharmonic and their positive and negative aliases.}
  \label{fig:J2054-lc}
\end{figure}

\subsubsection{J$2305$}

J$2305$~is a double-mode, high-frequency Am pulsating star similar in nature to J$2054$. It pulsates slightly slower than J$2054$ at $92.75$~and $101.68$ d$^{-1}$ (Fig. \ref{fig:J2305} top). We classify J$2035$~as an A$7$m star (Fig. \ref{fig:J2305} bottom). 

We detect no low-frequency variability in the SuperWASP photometry, indicating that both pulsations are most likely originating in J$2305$. Multiple periods have previously been observed in Am stars \citep[e.g.][]{joshi03,balona11b}, however not at the frequencies or amplitudes presented here.

\setcounter{figure}{24}
\begin{figure}
  \centering
  \includegraphics[angle=180,width=80mm, trim= 23mm 21mm 37mm 110mm,clip]{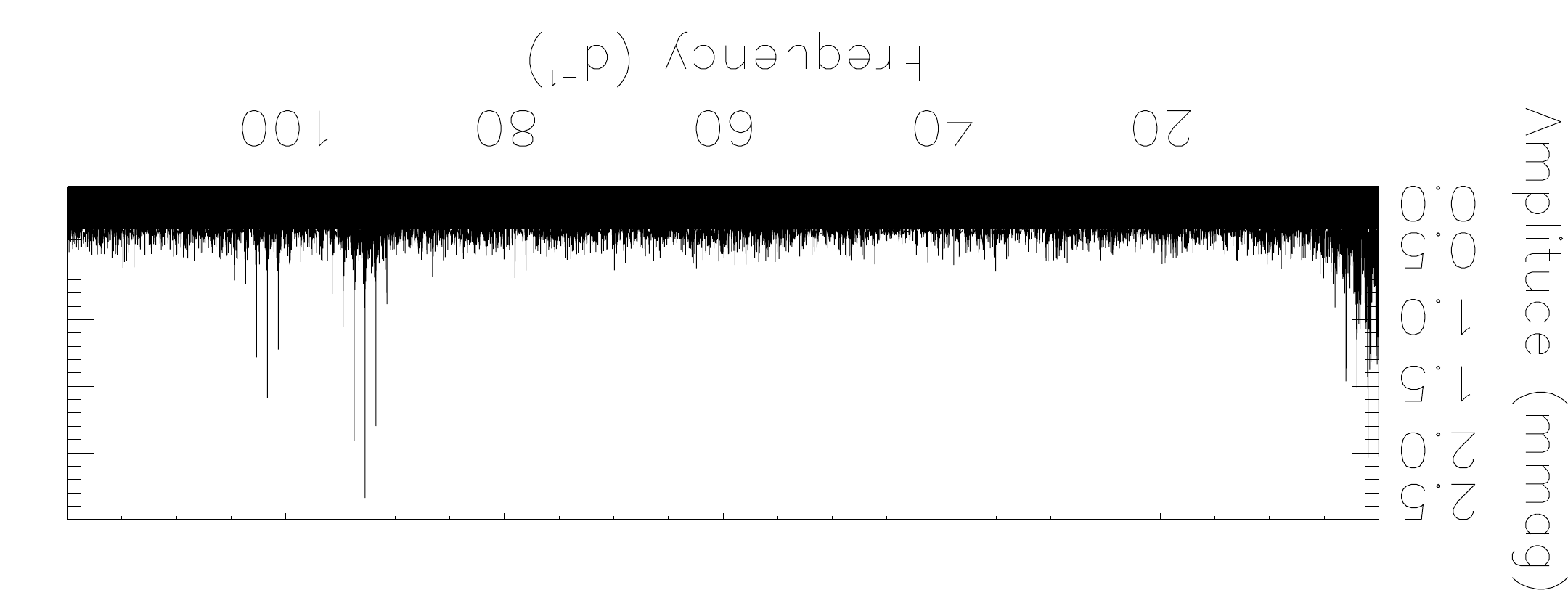}
  \includegraphics[angle=180,width=80mm, trim= 23mm 30mm 37mm 110mm,clip]{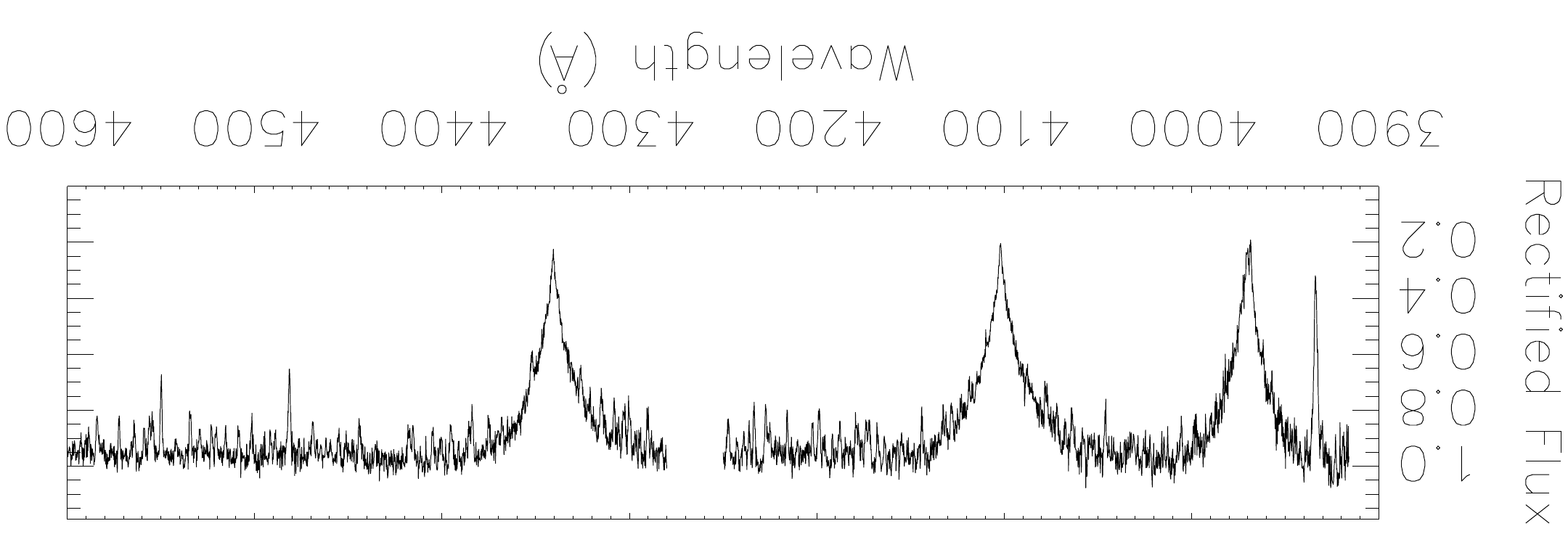}
  \caption{Periodogram and SALT/RSS spectrum of J$2305$.}
  \label{fig:J2305}
\end{figure}

\section{Conclusion}
\label{conc}

We have exploited the SuperWASP archive in the search for rapidly varying F-, A- and B-type stars. Our survey, of over $1.5$~million objects, has resulted in the discovery of $10$~new roAp stars as well as $13$~new pulsating Am stars. Further to this, there are over $350$~systems which show variations on periods less than $30$ min.

The discovery of this number of new roAp stars increases the known stars of this class by $20$~per cent, providing a larger sample for further study. 

This work shows the power of photometric surveys to identify a whole variety of variable stars. There are many more sources of data that can be exploited in a similar way to that which we have presented here for the search for rare and interesting pulsating stars. We have been able to push the previous limits of pulsation frequencies in some types of A stars, leading to a greater frequency overlap between different pulsator classes. 

\section*{Acknowledgements}

DLH acknowledges financial support from the STFC via the Ph.D. studentship programme. The WASP project is funded and maintained by Queen's University Belfast, the Universities of Keele, St. Andrews, Warwick and Leicester, the Open University, the Isaac Newton Group, the Instituto de Astrofisica Canarias, the South African Astronomical Observatory and by the STFC. Some of the observations reported in this paper were obtained with the Southern African Large Telescope (SALT) under programs 2012-1-UKSC-001 (PI: BS), 2012-2-UKSC-001 (PI: DLH) and 2013-1-UKSC-002 (PI: DLH). TRAPPIST is a project funded by the Belgian Fund for Scientific Research (FNRS) with the participation of the Swiss National Science Foundation. The Digitized Sky Surveys were produced at the Space Telescope Science Institute under U.S. Government grant NAG W-2166. The images of these surveys are based on photographic data obtained using the Oschin Schmidt Telescope on Palomar Mountain and the UK Schmidt Telescope. The plates were processed into the present compressed digital form with the permission of these institutions. This publication makes use of data products from the Two Micron All Sky Survey, which is a joint project of the University of Massachusetts and the Infrared Processing and Analysis Center/California Institute of Technology, funded by the National Aeronautics and Space Administration and the National Science Foundation. We thank the referee, Donald Kurtz, for useful comments and suggestions.

\appendix
\section{Spectroscopically observed targets}
\label{sec:appendix}

Below we present the periodograms and spectra for the remaining $23$ spectroscopically observed targets.

\setcounter{figure}{0}
\begin{figure*}
  \centering
  \begin{minipage}{\textwidth}
    \centering
    
    \includegraphics[angle=180,width=80mm, trim= 23mm 21mm 37mm 110mm,clip]{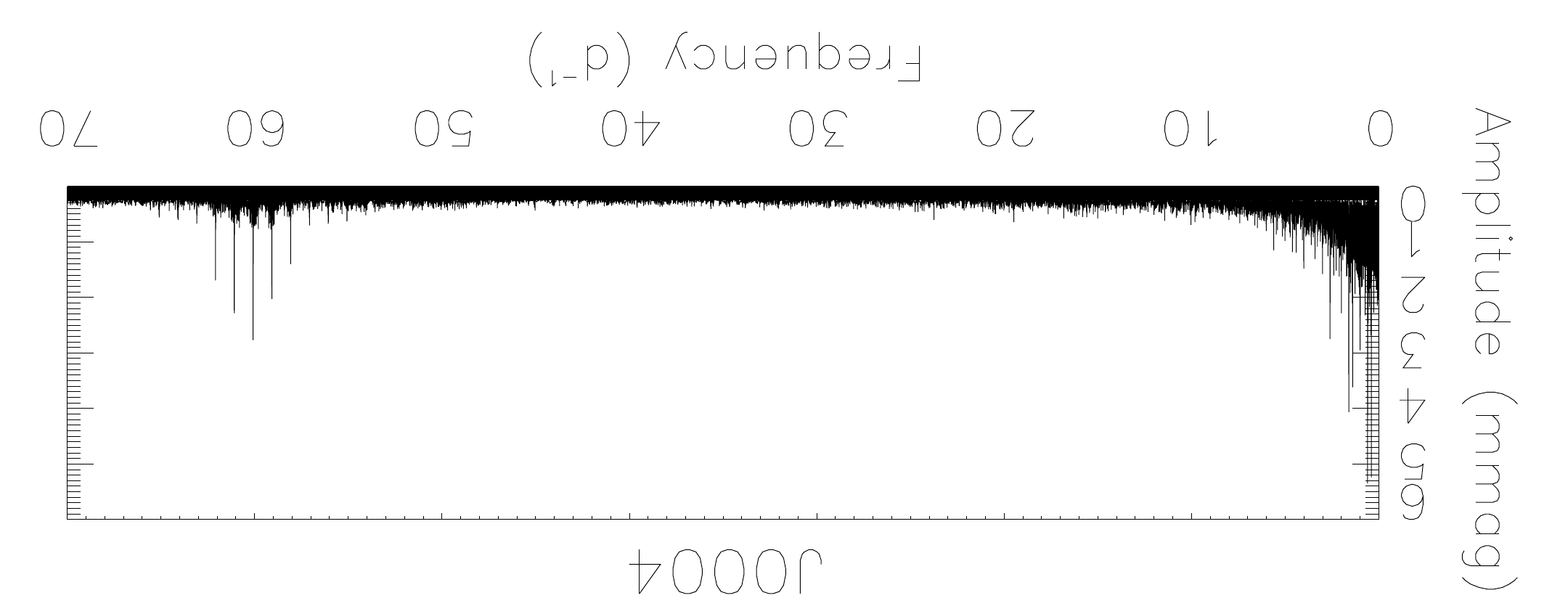}\hfill
    \includegraphics[angle=180,width=80mm, trim= 23mm 21mm 37mm 110mm,clip]{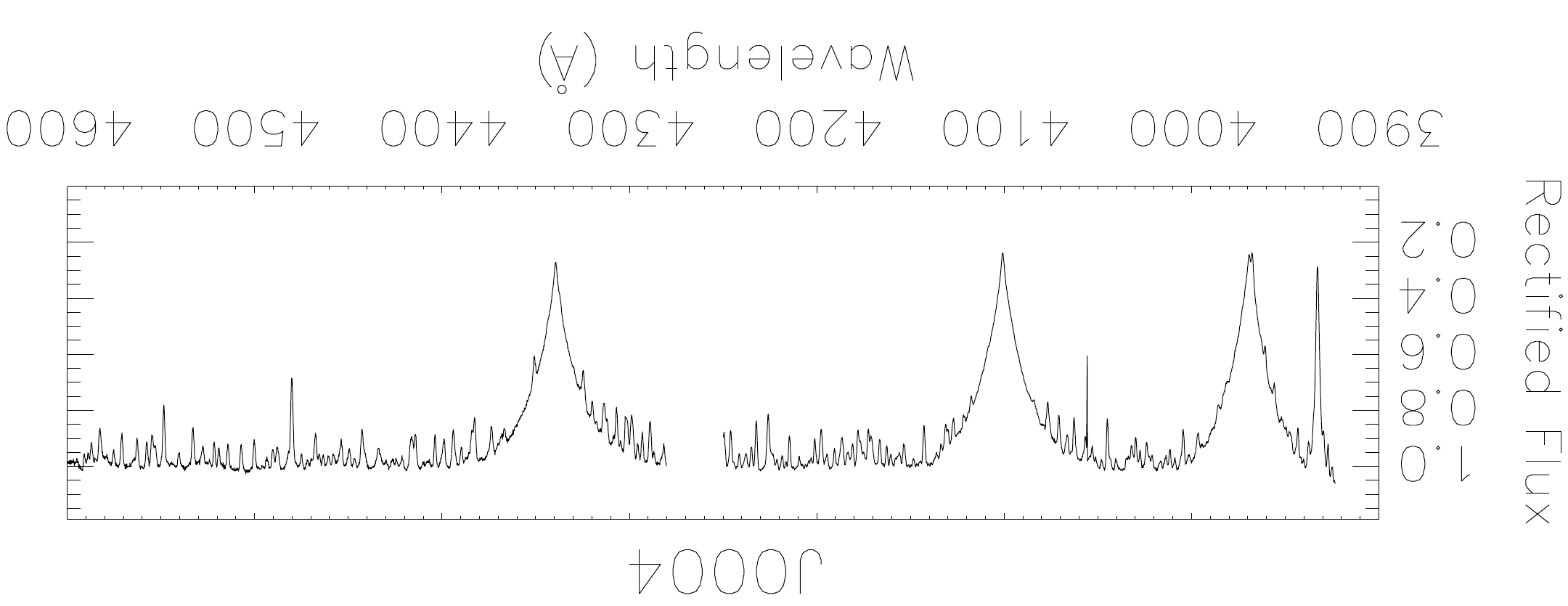}\hfill

    \includegraphics[angle=180,width=80mm, trim= 23mm 21mm 37mm 110mm,clip]{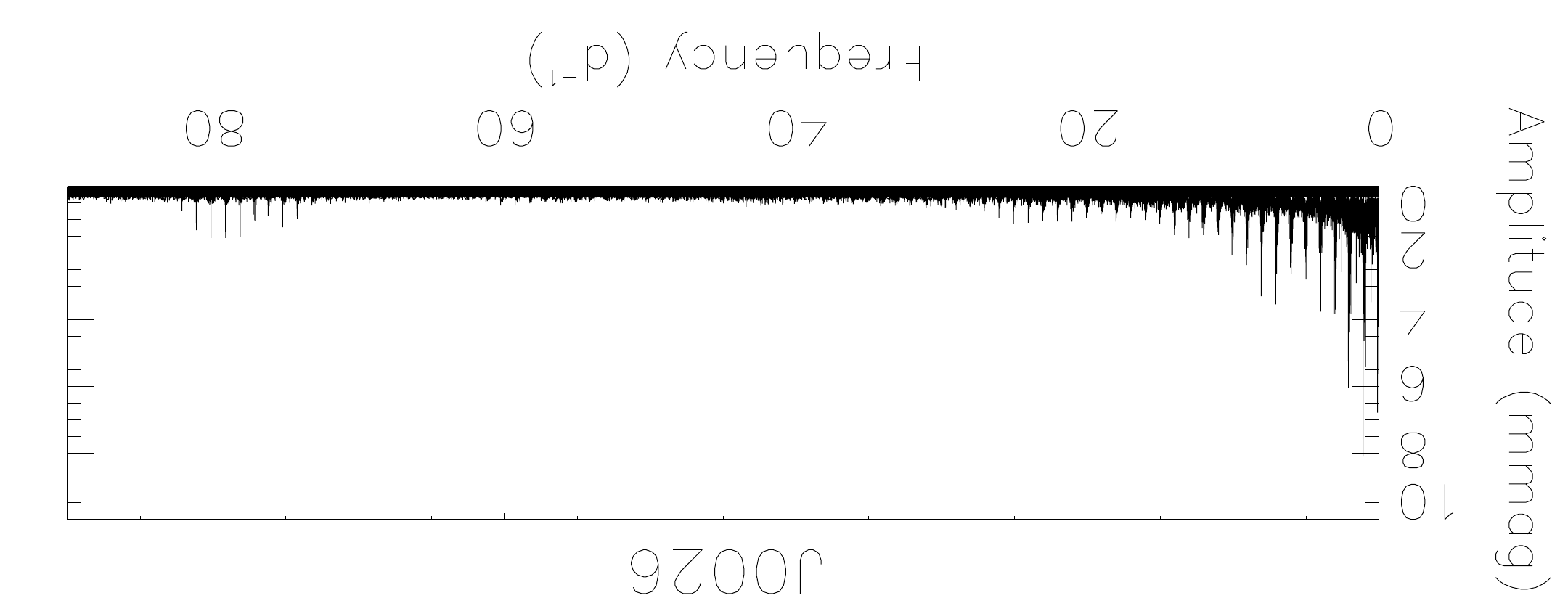}\hfill
    \includegraphics[angle=180,width=80mm, trim= 23mm 21mm 37mm 110mm,clip]{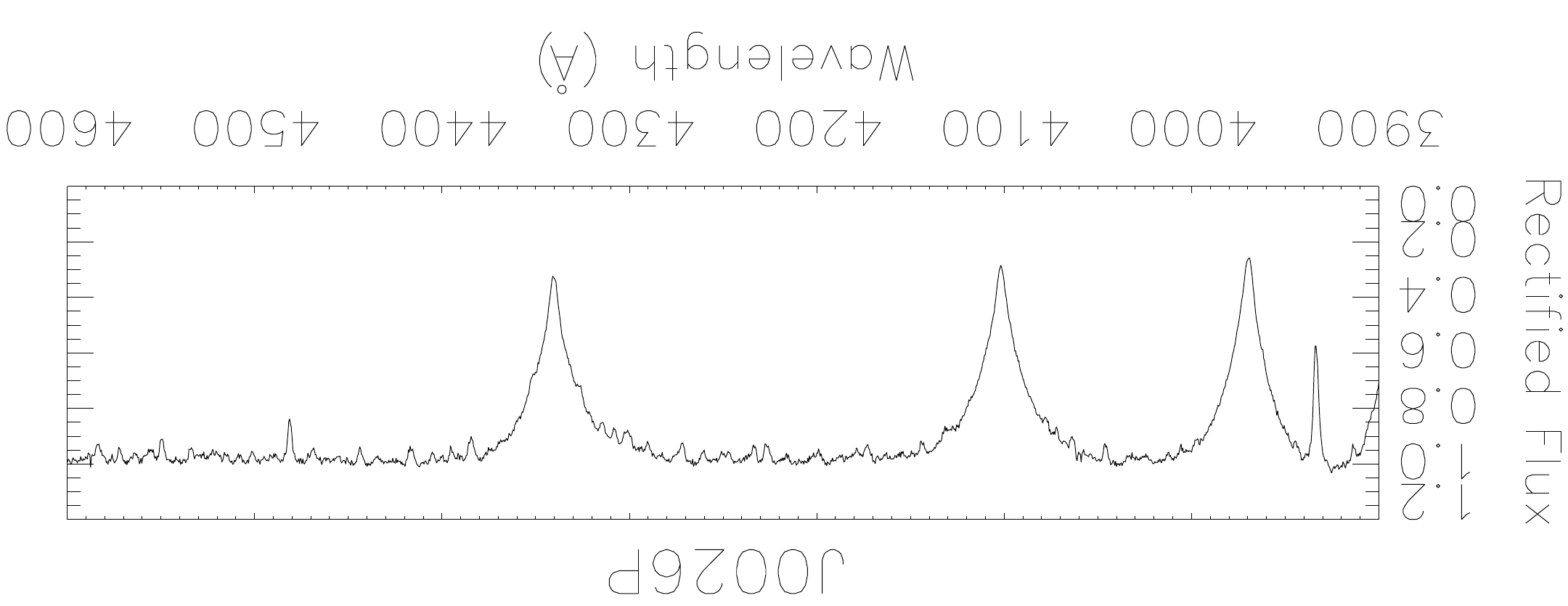}\hfill

    \includegraphics[angle=180,width=80mm, trim= 23mm 21mm 37mm 110mm,clip]{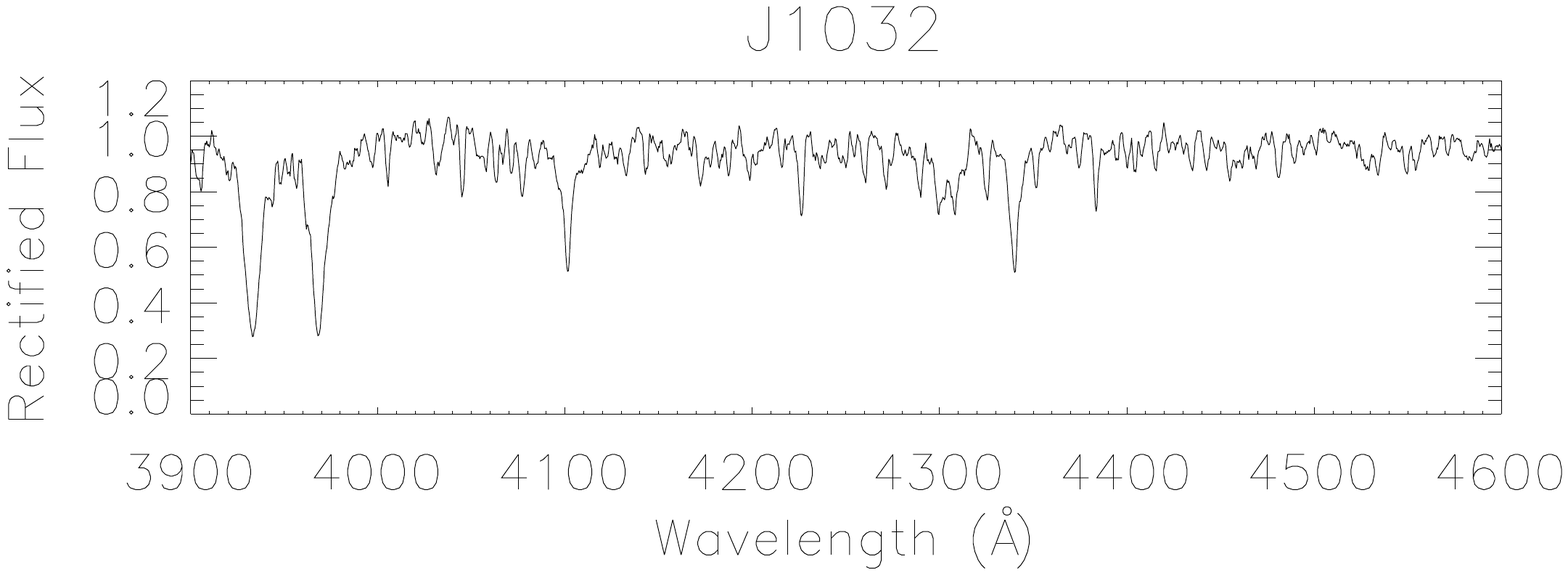}\hfill
    \includegraphics[angle=180,width=80mm, trim= 23mm 21mm 37mm 110mm,clip]{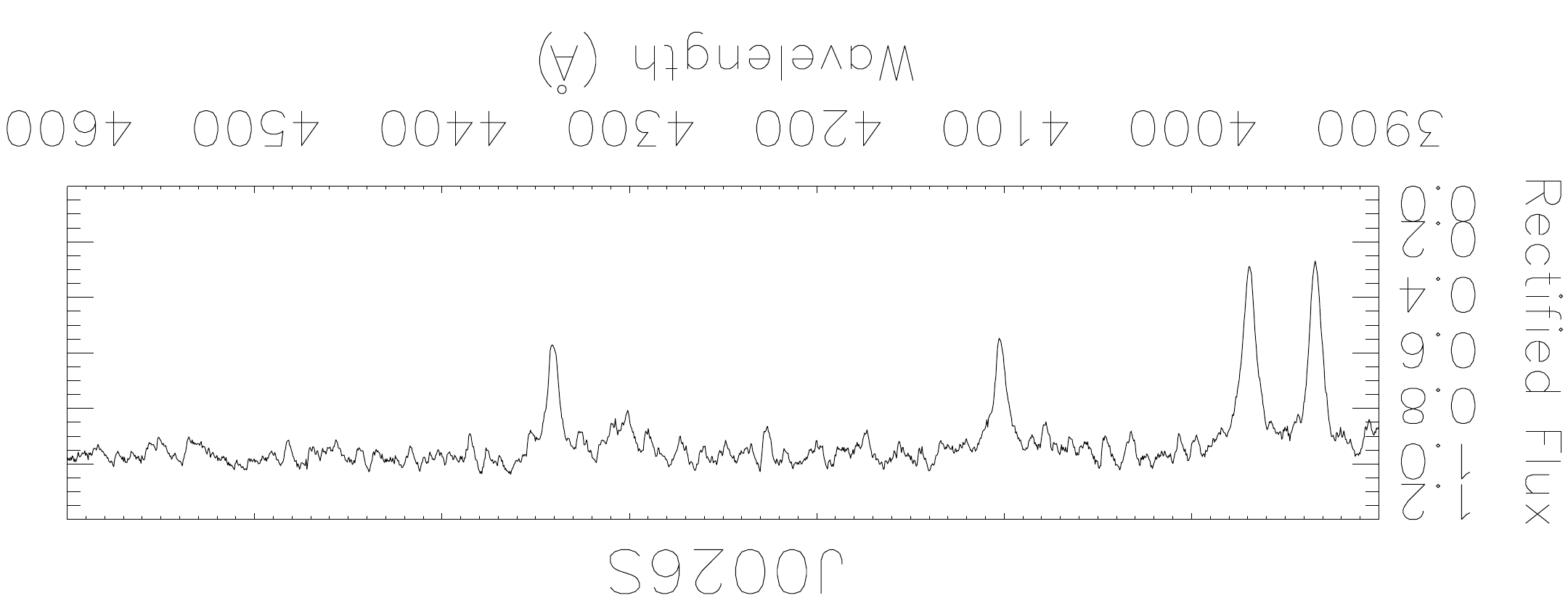}\hfill

    \includegraphics[angle=180,width=80mm, trim= 23mm 21mm 37mm 110mm,clip]{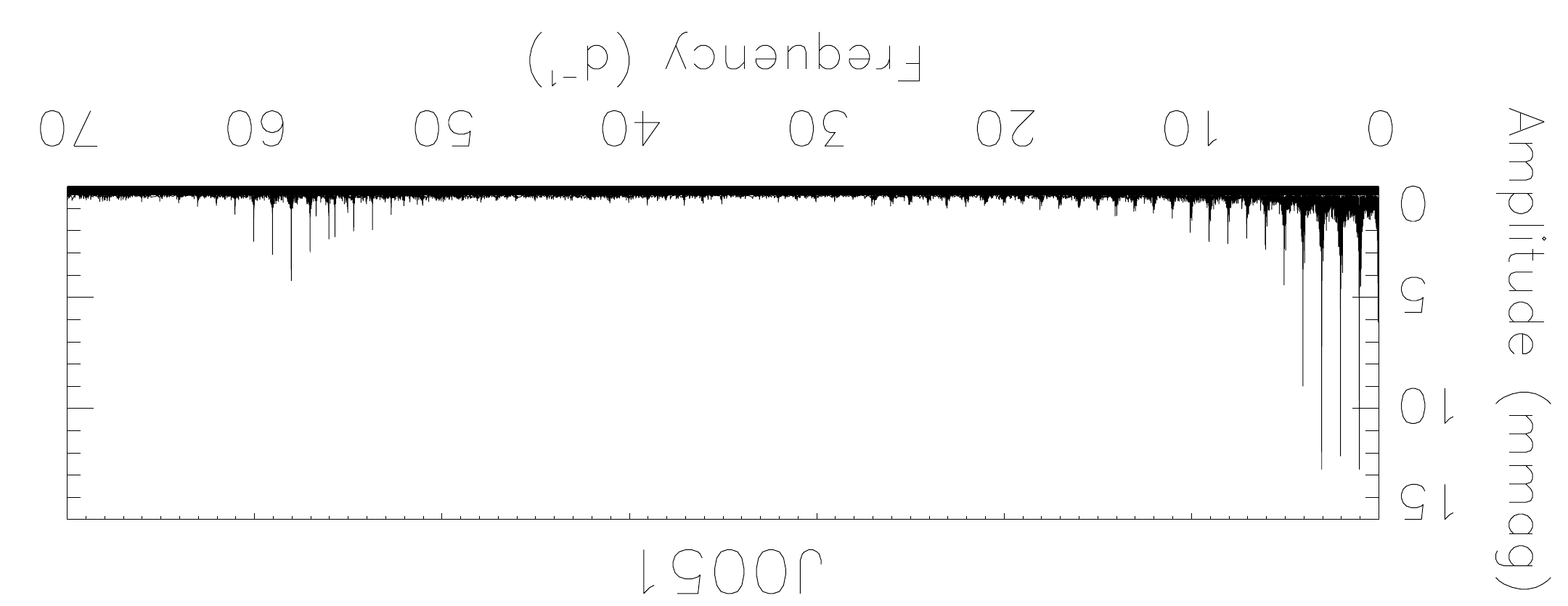}\hfill
    \includegraphics[angle=180,width=80mm, trim= 23mm 21mm 37mm 110mm,clip]{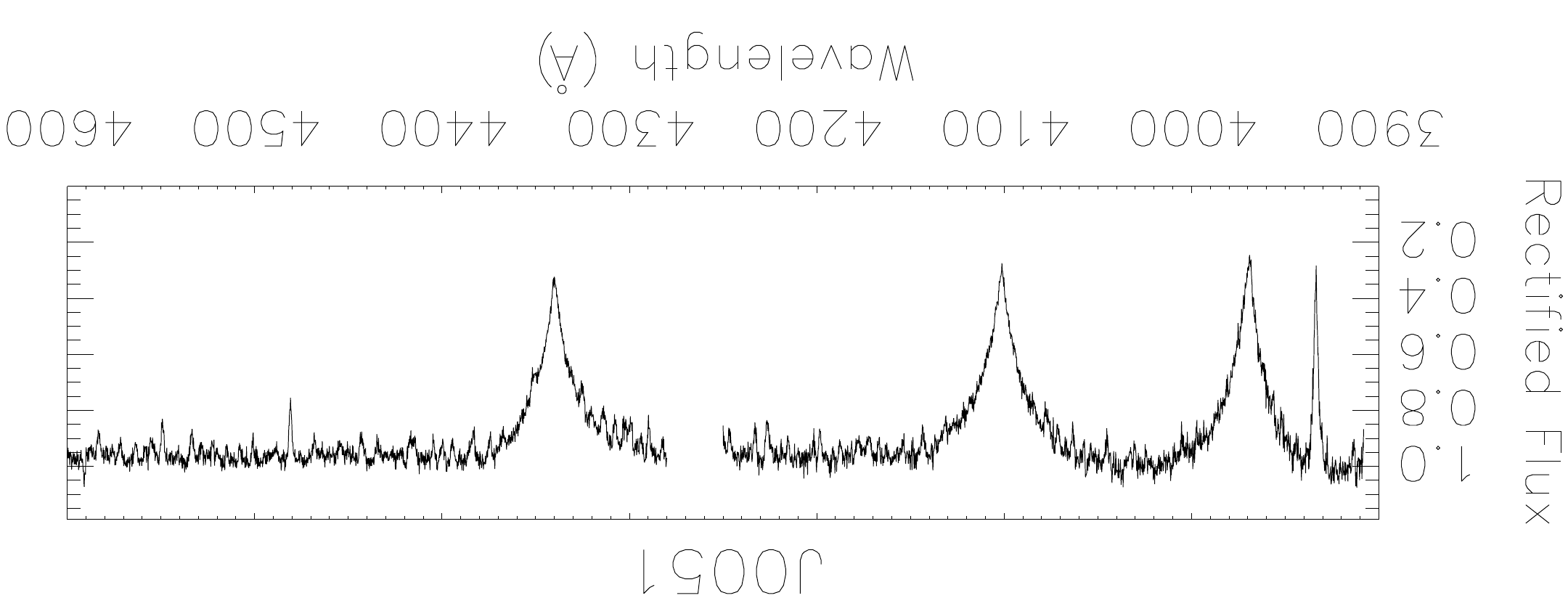}\hfill

    \includegraphics[angle=180,width=80mm, trim= 23mm 21mm 37mm 110mm,clip]{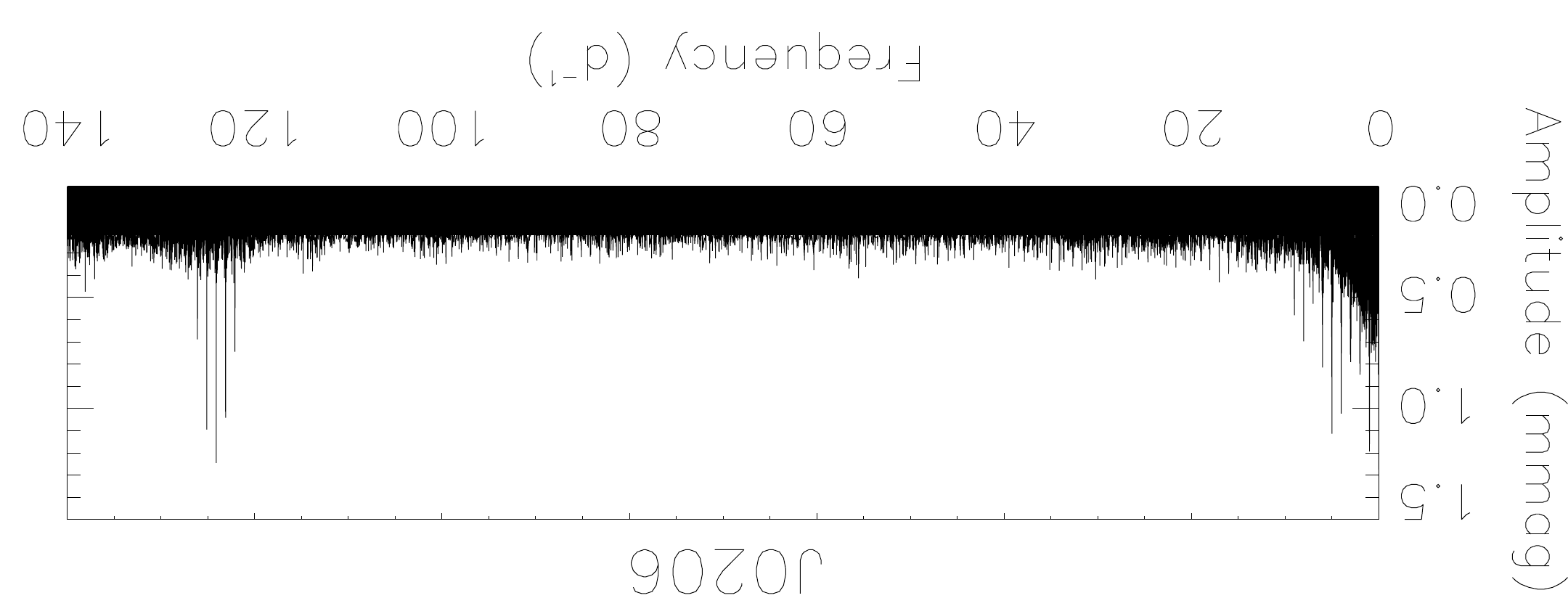}\hfill
    \includegraphics[angle=180,width=80mm, trim= 23mm 21mm 37mm 110mm,clip]{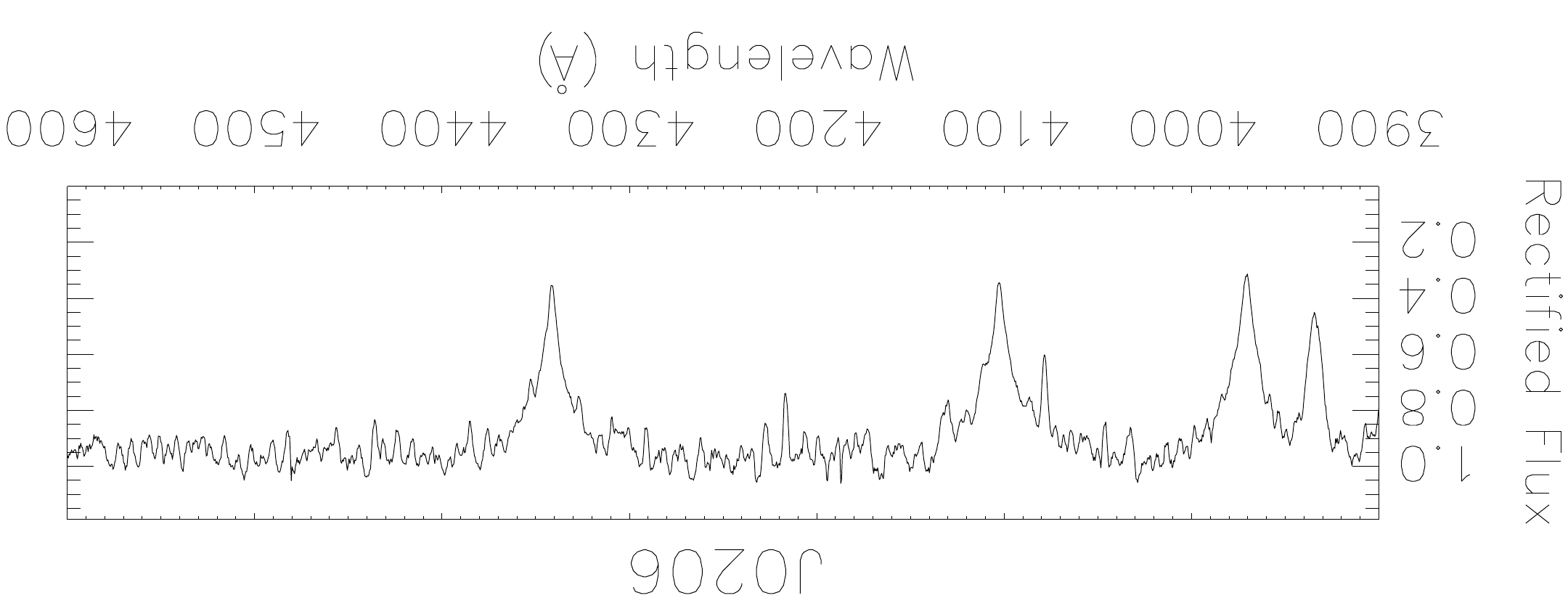}\hfill

    \includegraphics[angle=180,width=80mm, trim= 23mm 21mm 37mm 110mm,clip]{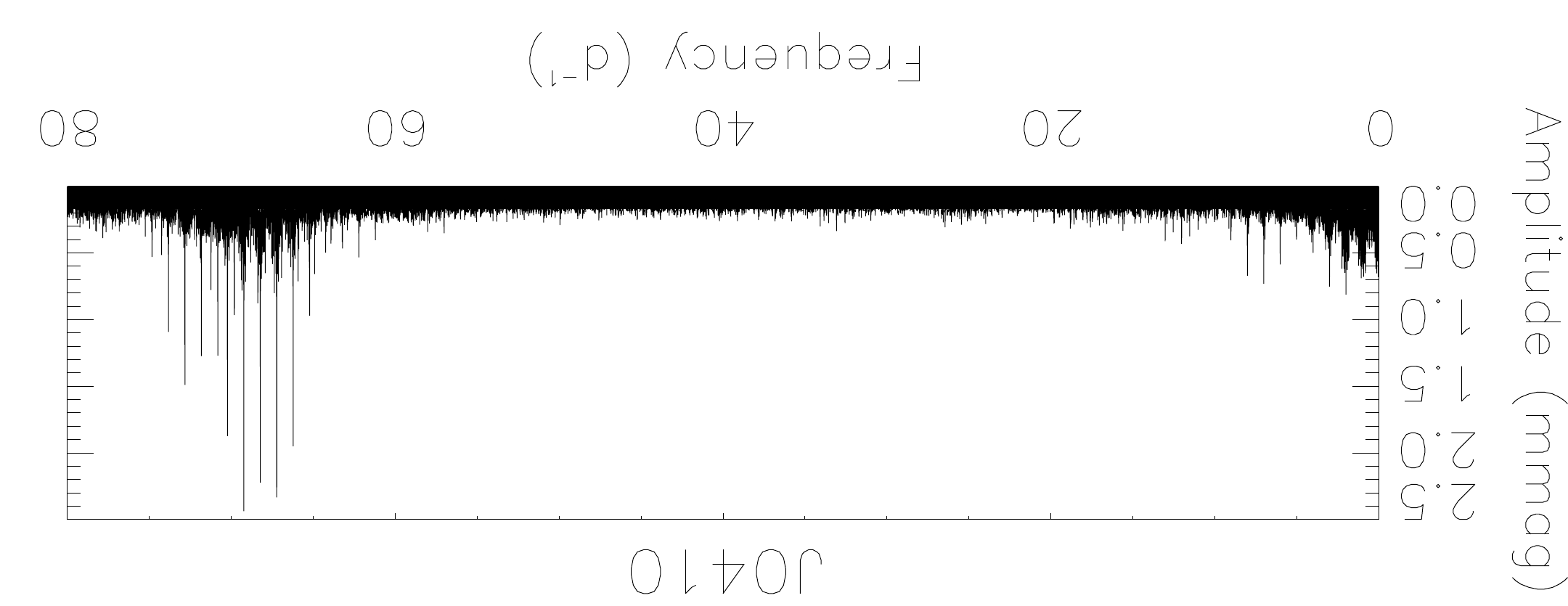}\hfill
    \includegraphics[angle=180,width=80mm, trim= 23mm 21mm 37mm 110mm,clip]{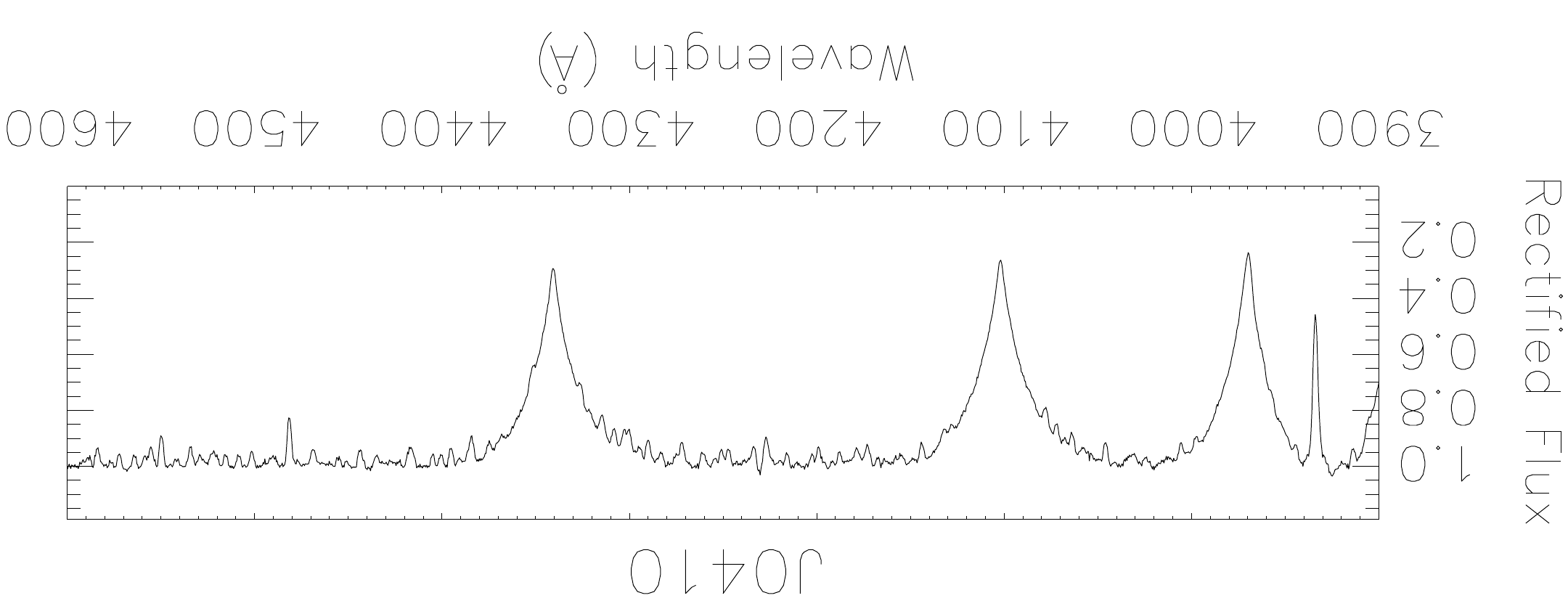}\hfill

    \includegraphics[angle=180,width=80mm, trim= 23mm 21mm 37mm 110mm,clip]{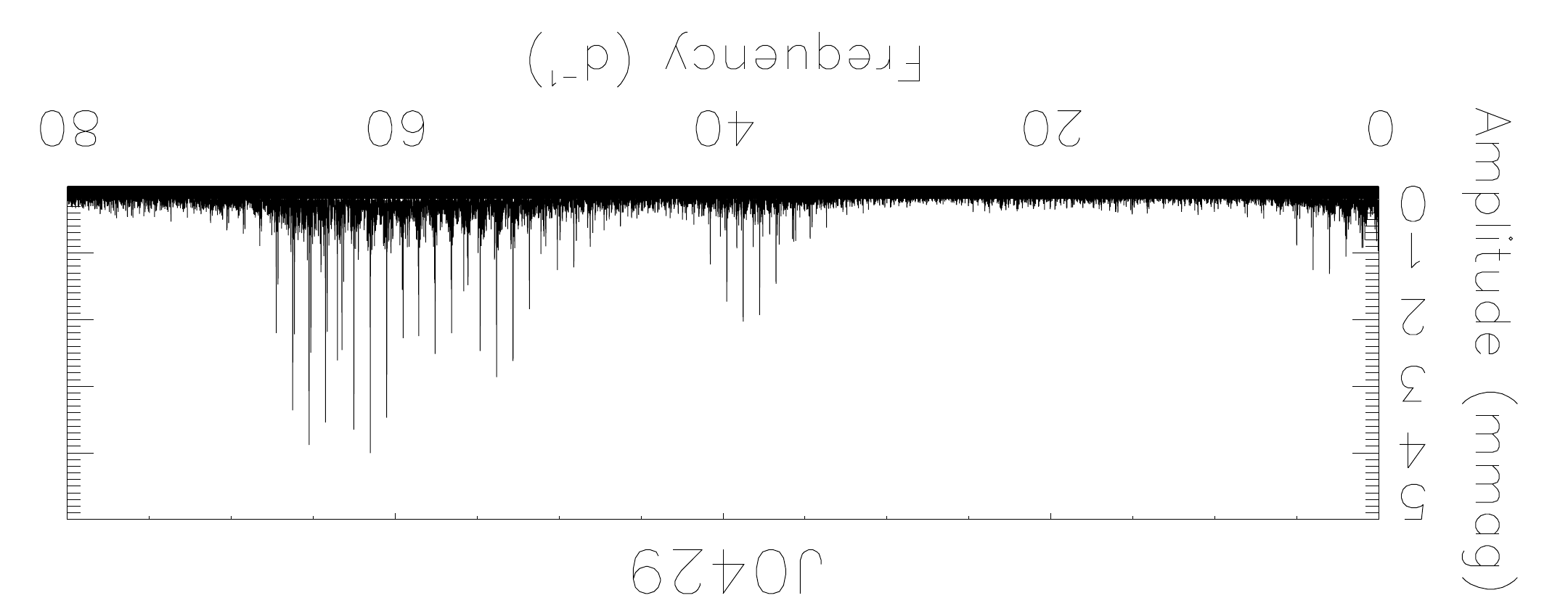}\hfill
    \includegraphics[angle=180,width=80mm, trim= 23mm 21mm 37mm 110mm,clip]{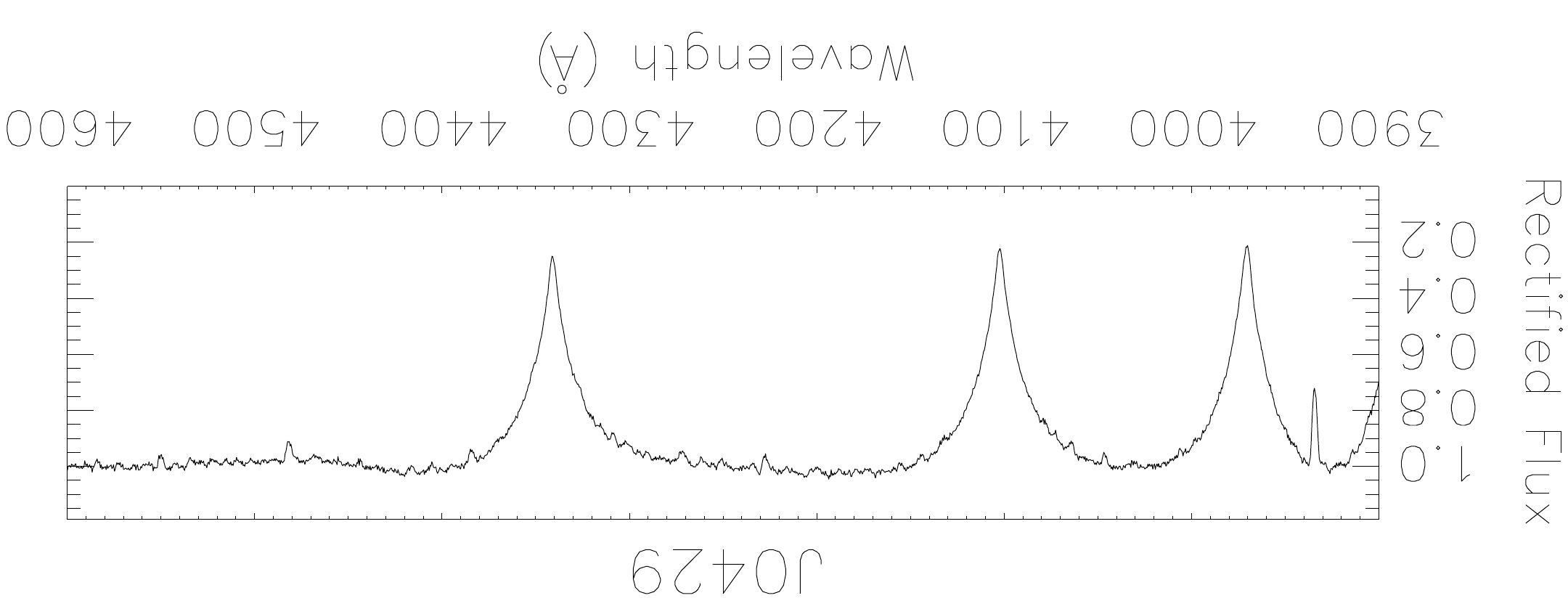}

    \caption{Periodograms and spectra for the remaining spectroscopically observed targets.}

  \end{minipage}
\end{figure*}
\clearpage
\newpage

\setcounter{figure}{0}
\begin{figure*}
  \centering
  \begin{minipage}{\textwidth}
    \centering

    \includegraphics[angle=180,width=80mm, trim= 23mm 21mm 37mm 110mm,clip]{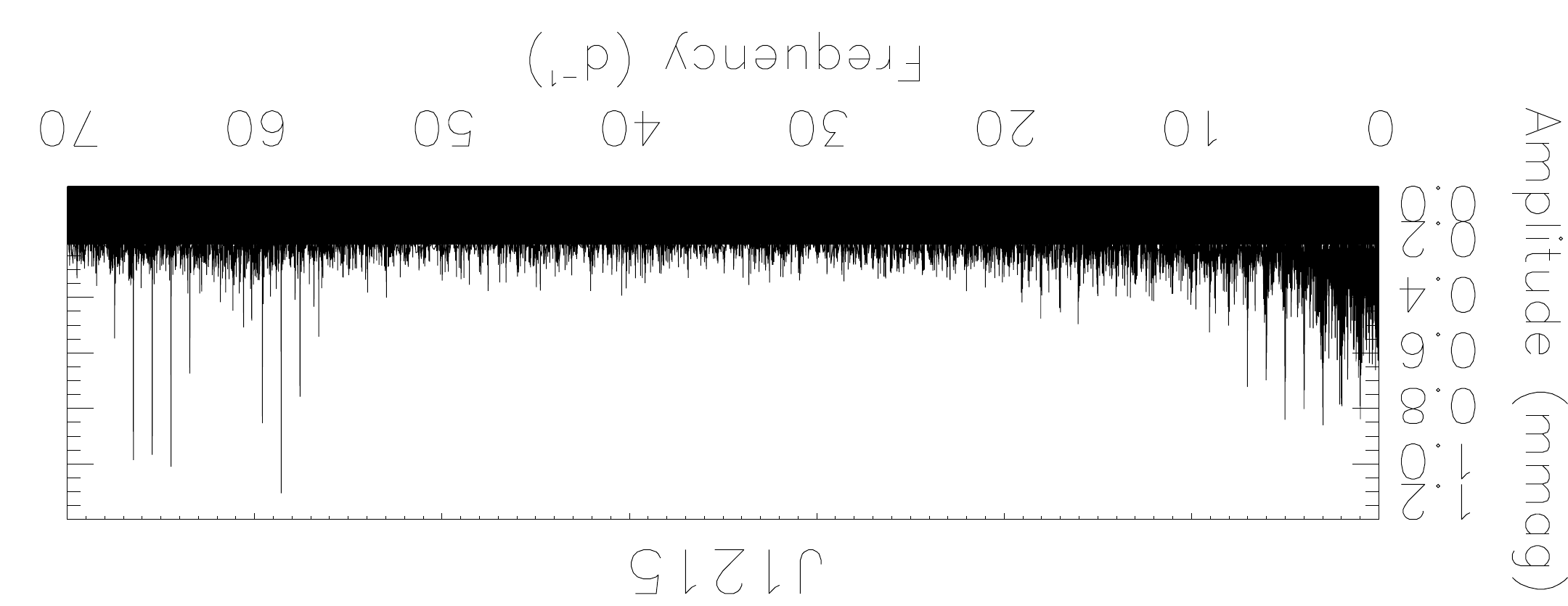}\hfill
    \includegraphics[angle=180,width=80mm, trim= 23mm 21mm 37mm 110mm,clip]{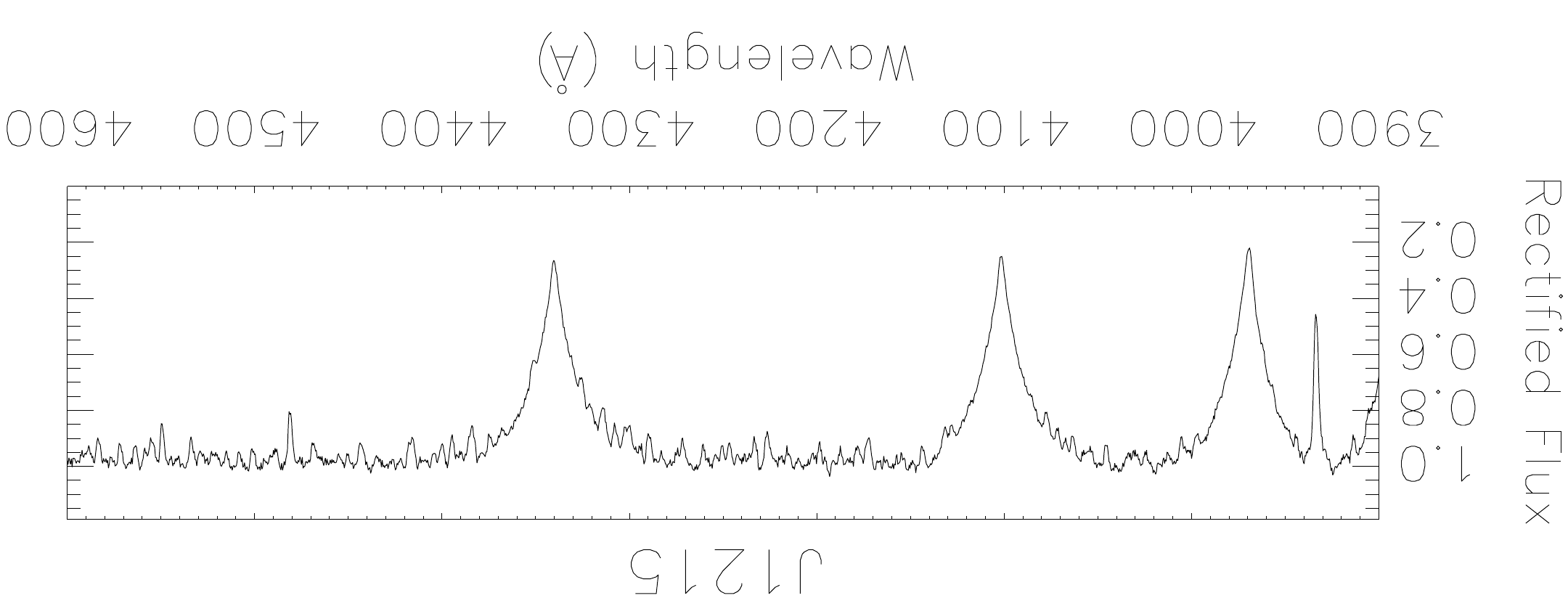}\hfill
    
    \includegraphics[angle=180,width=80mm, trim= 23mm 21mm 37mm 110mm,clip]{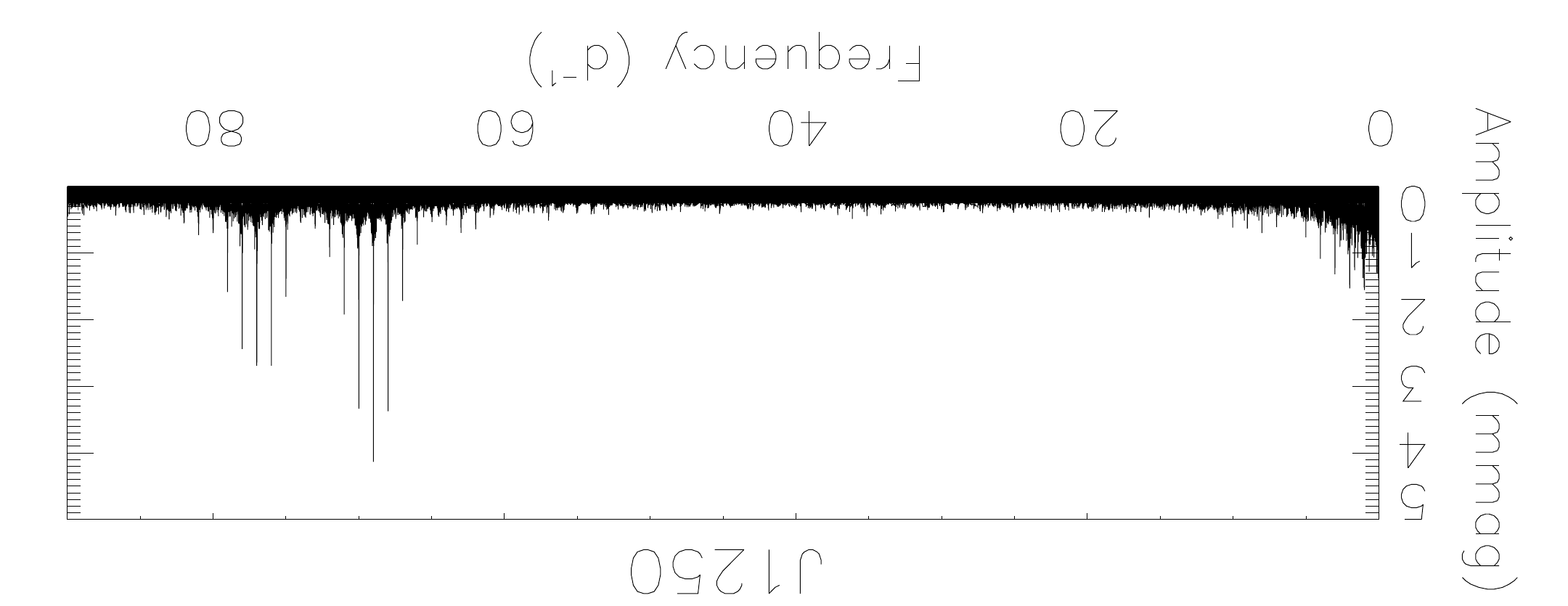}\hfill
    \includegraphics[angle=180,width=80mm, trim= 23mm 21mm 37mm 110mm,clip]{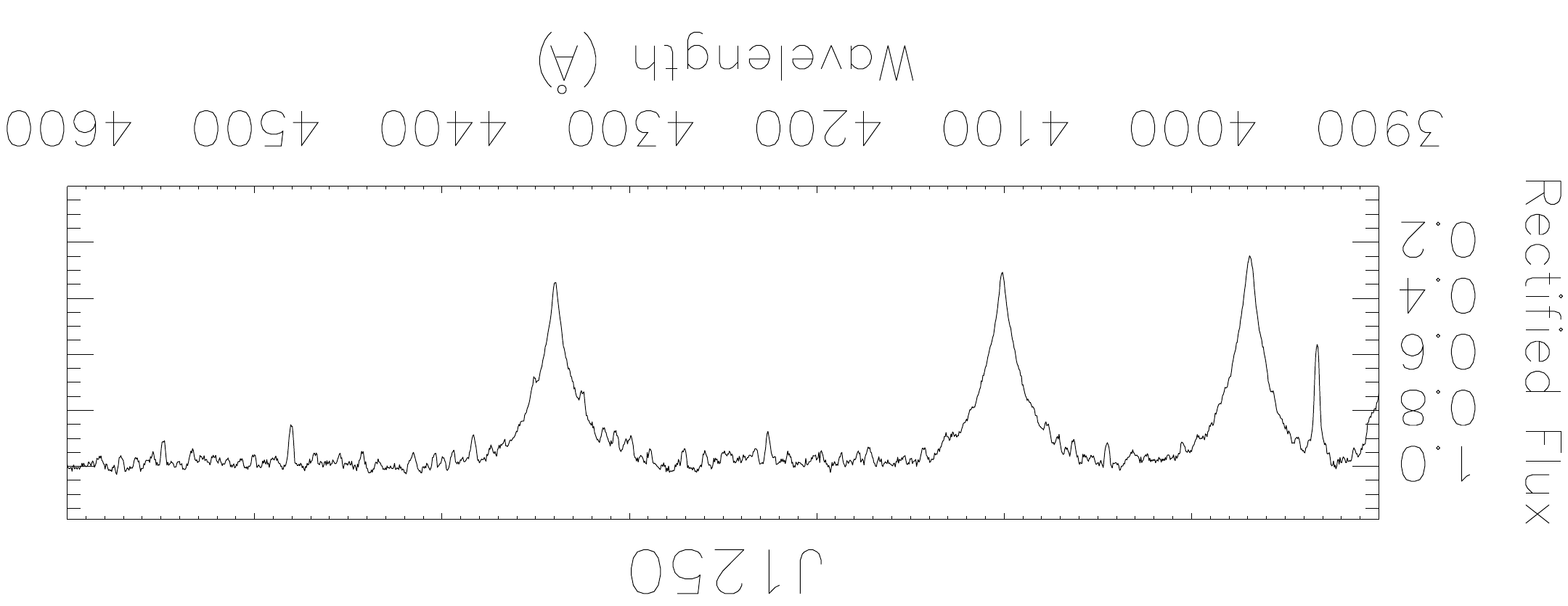}\hfill
    
    \includegraphics[angle=180,width=80mm, trim= 23mm 21mm 37mm 110mm,clip]{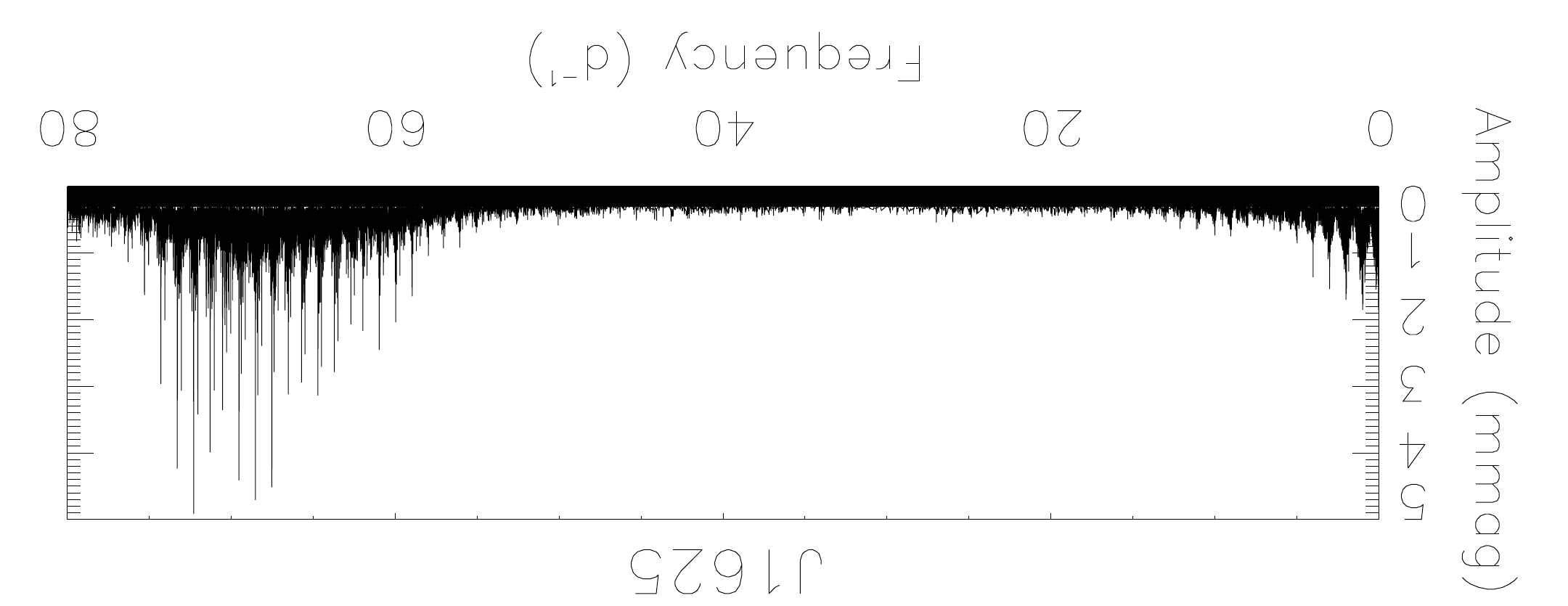}\hfill
    \includegraphics[angle=180,width=80mm, trim= 23mm 21mm 37mm 110mm,clip]{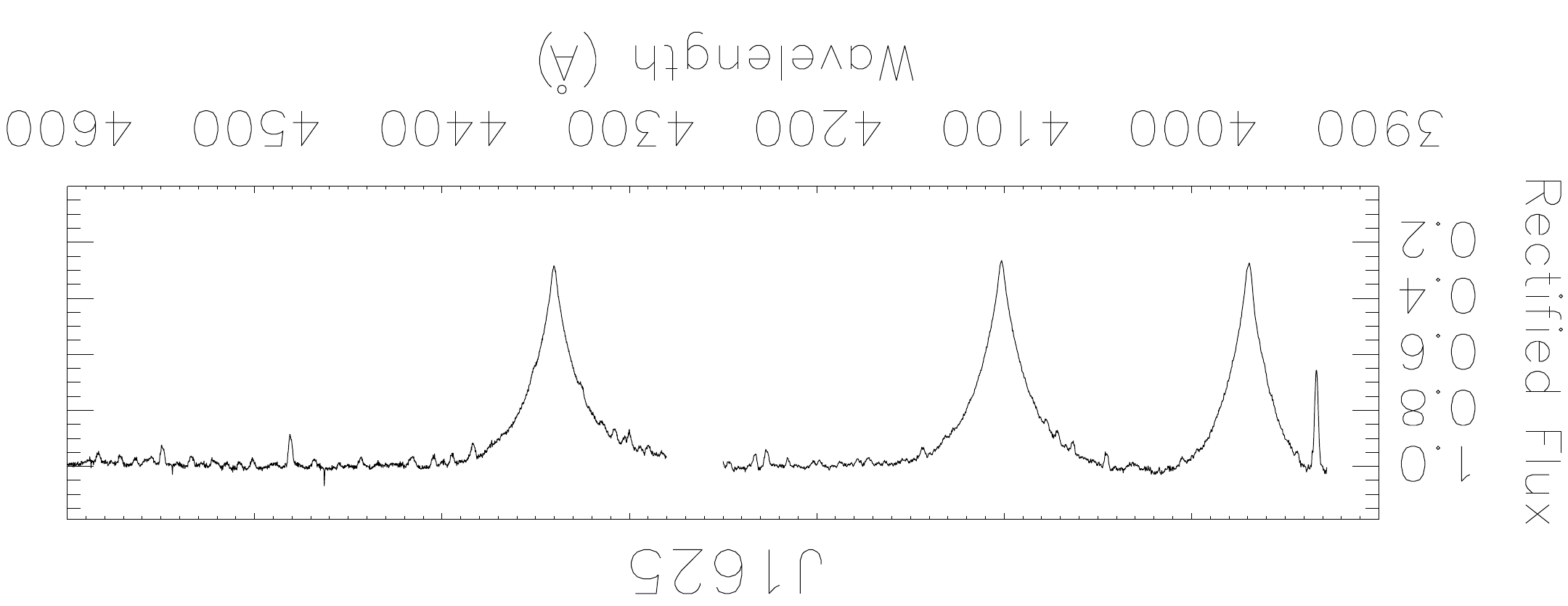}\hfill
    
    \includegraphics[angle=180,width=80mm, trim= 23mm 21mm 37mm 110mm,clip]{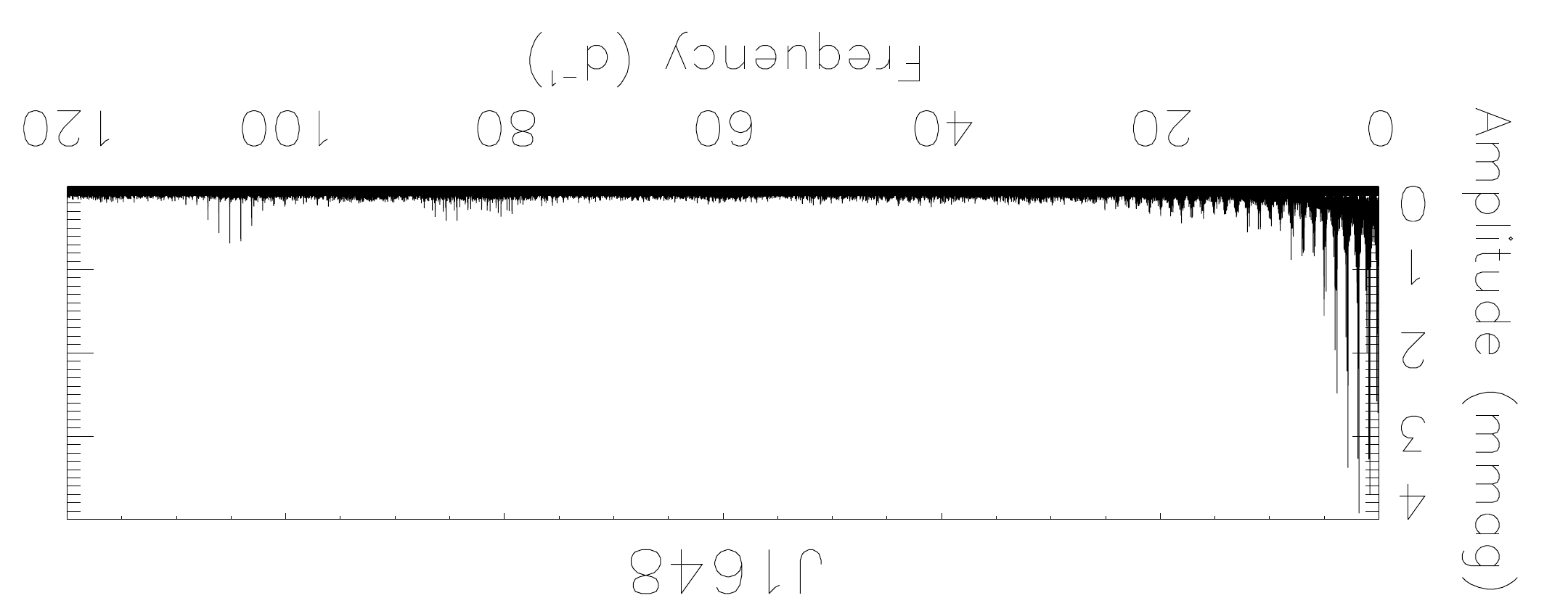}\hfill
    \includegraphics[angle=180,width=80mm, trim= 23mm 21mm 37mm 110mm,clip]{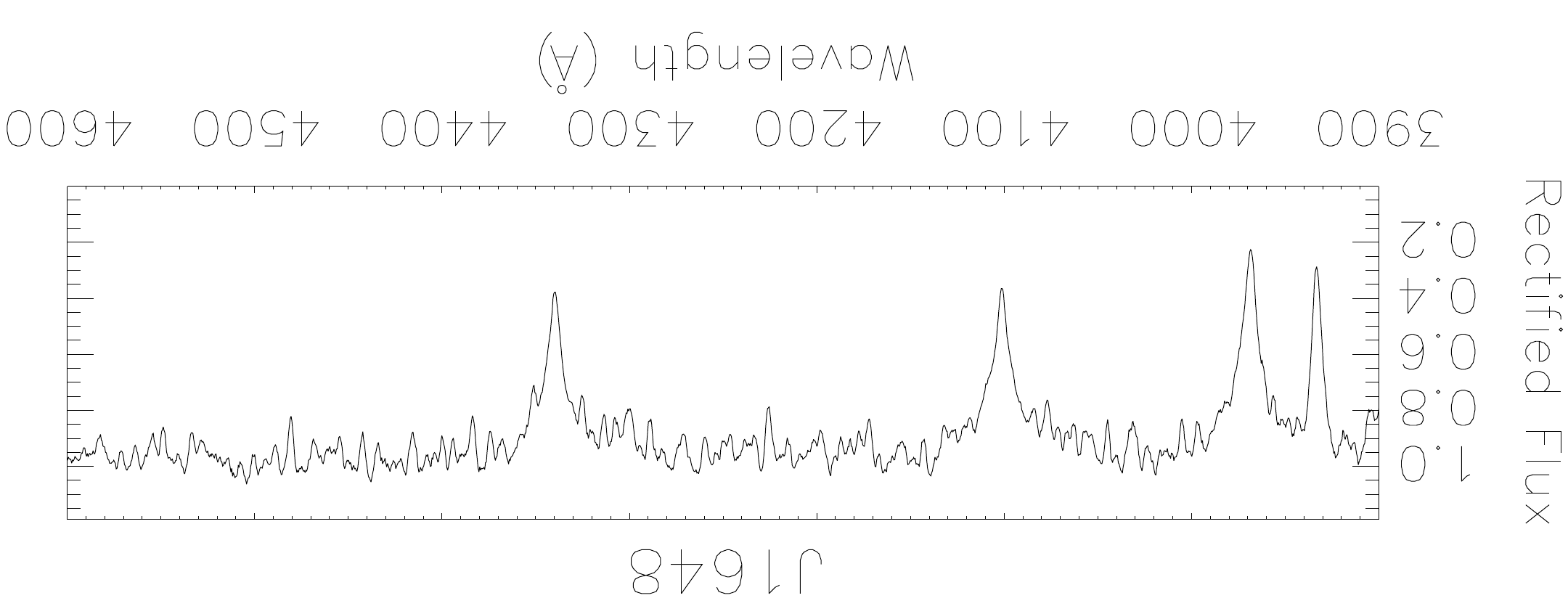}\hfill
    
    \includegraphics[angle=180,width=80mm, trim= 23mm 21mm 37mm 110mm,clip]{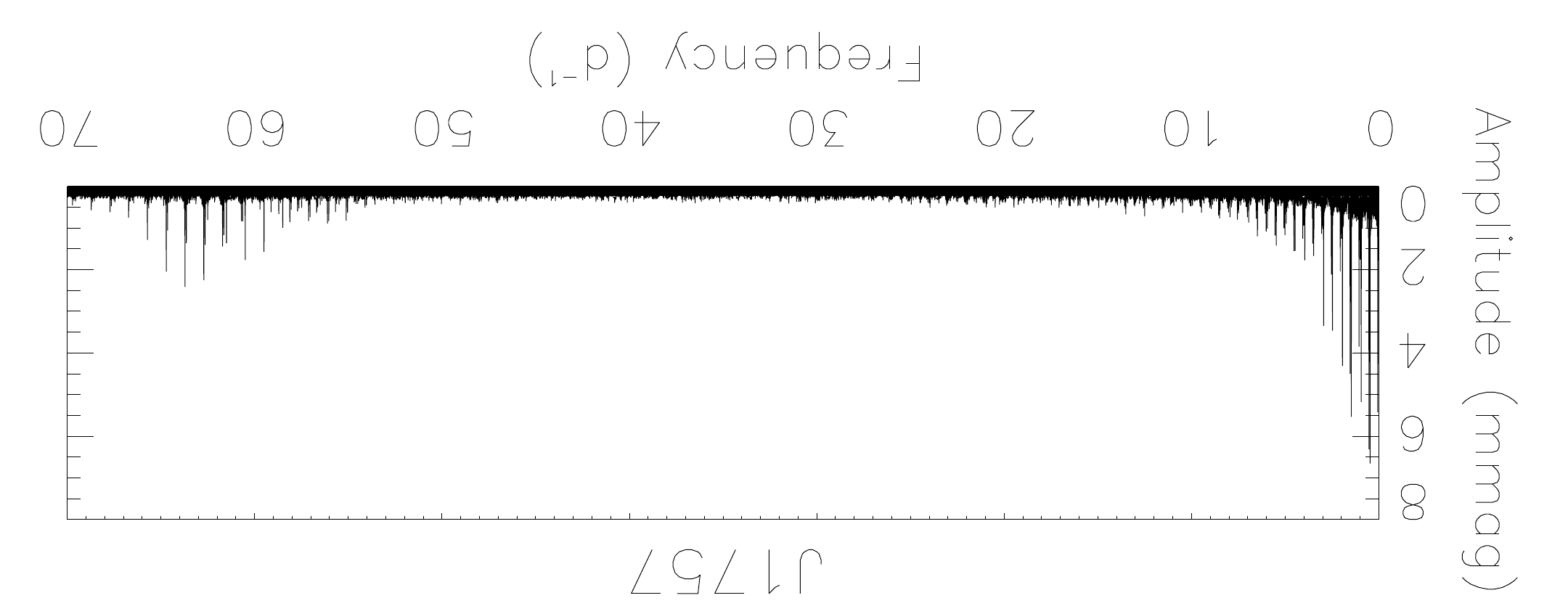}\hfill
    \includegraphics[angle=180,width=80mm, trim= 23mm 21mm 37mm 110mm,clip]{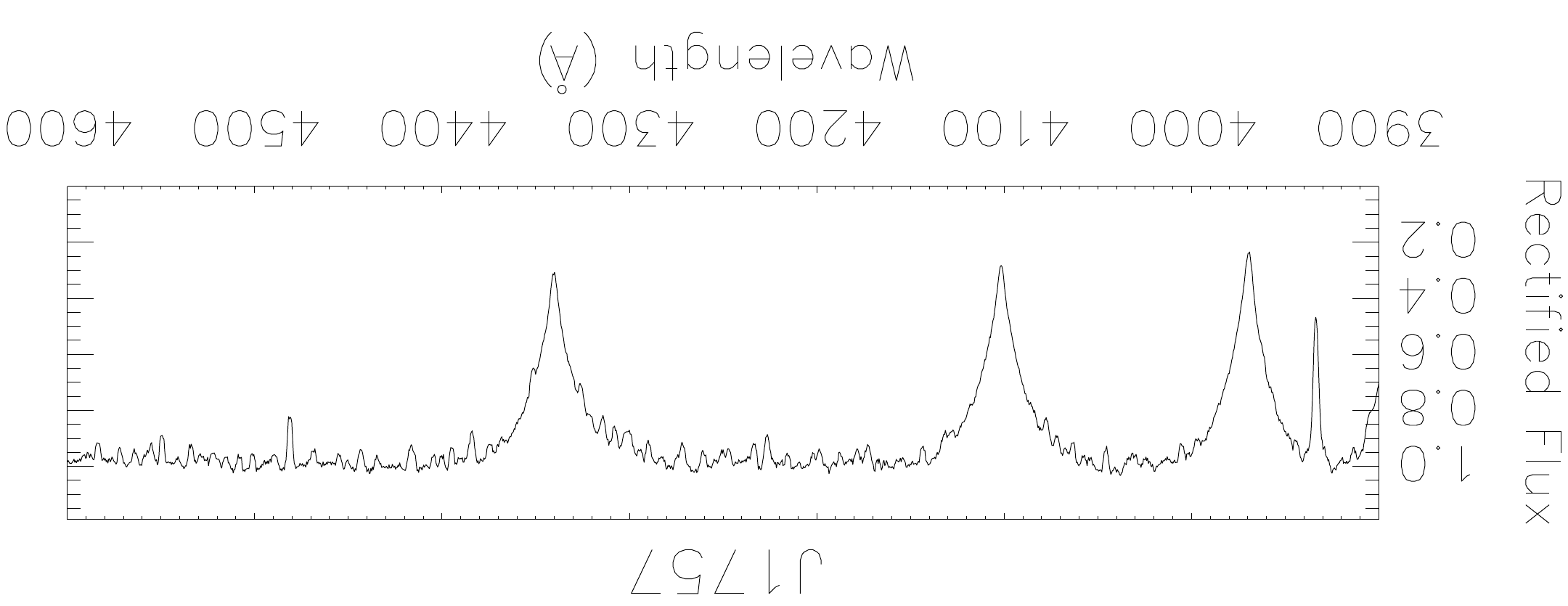}\hfill
    
    \includegraphics[angle=180,width=80mm, trim= 23mm 21mm 37mm 110mm,clip]{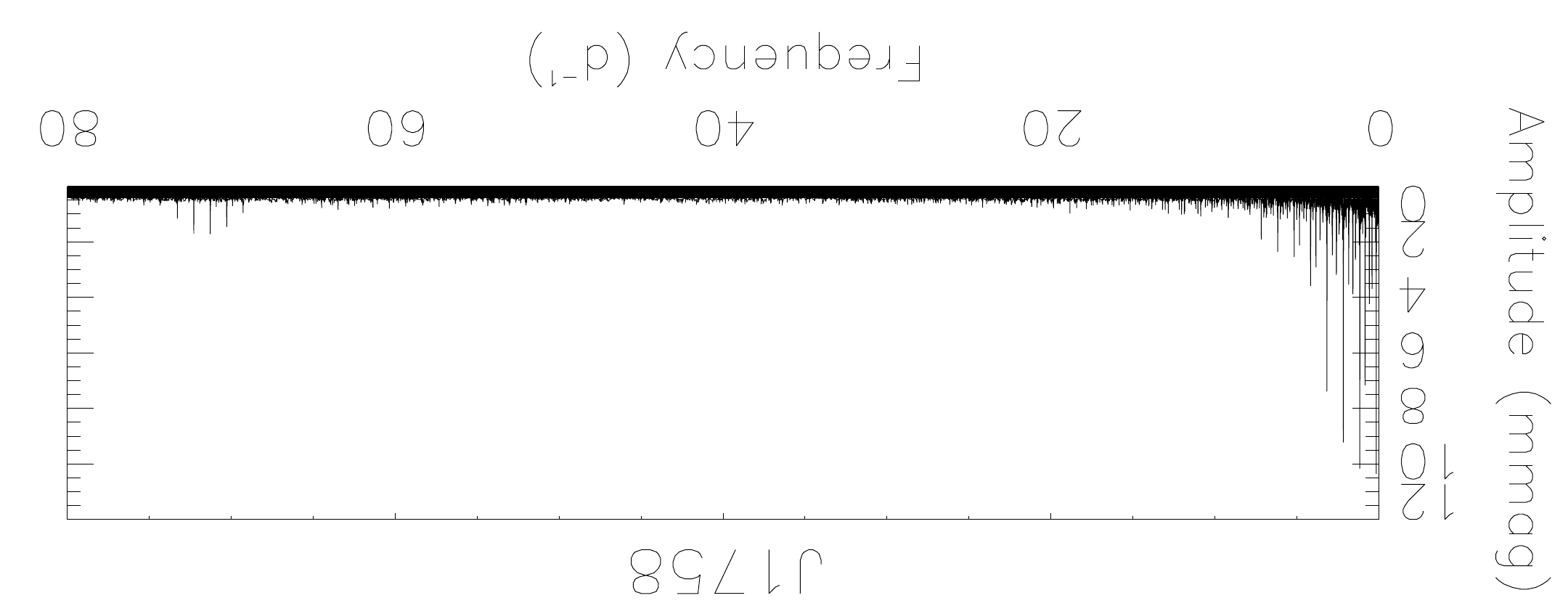}\hfill
    \includegraphics[angle=180,width=80mm, trim= 23mm 21mm 37mm 110mm,clip]{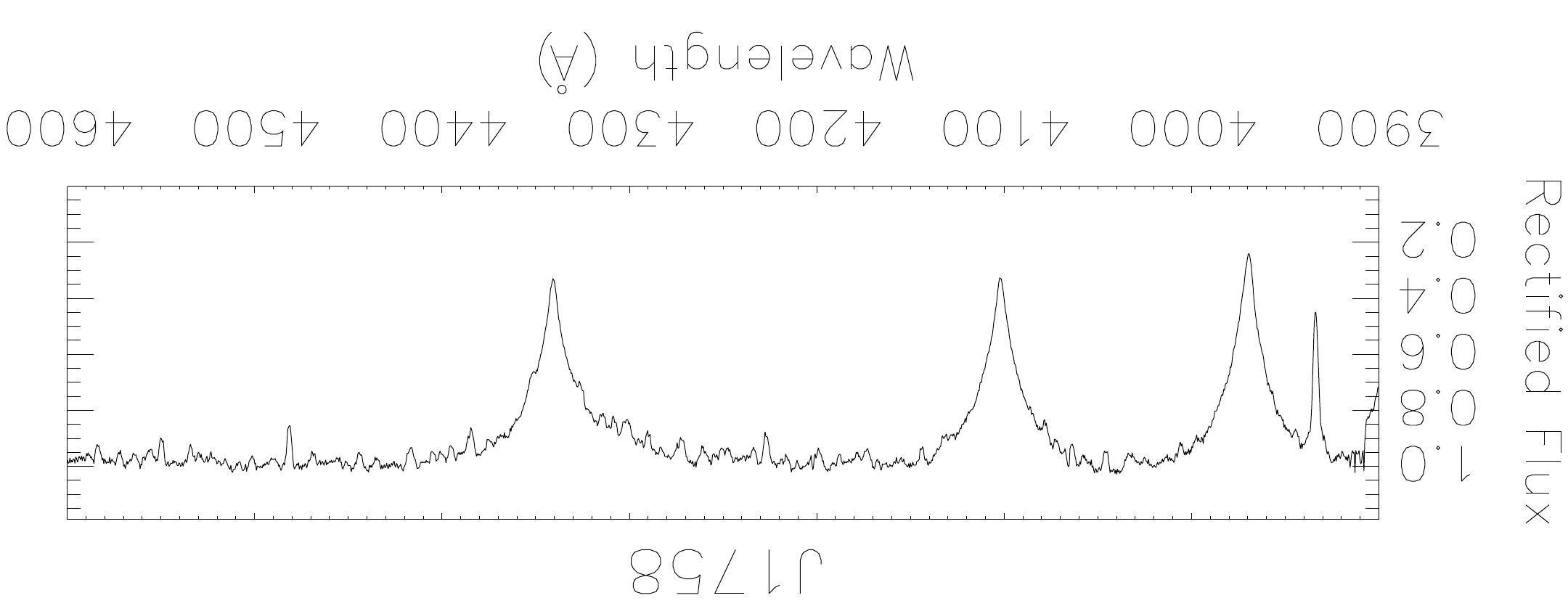}\hfill
    
    \includegraphics[angle=180,width=80mm, trim= 23mm 21mm 37mm 110mm,clip]{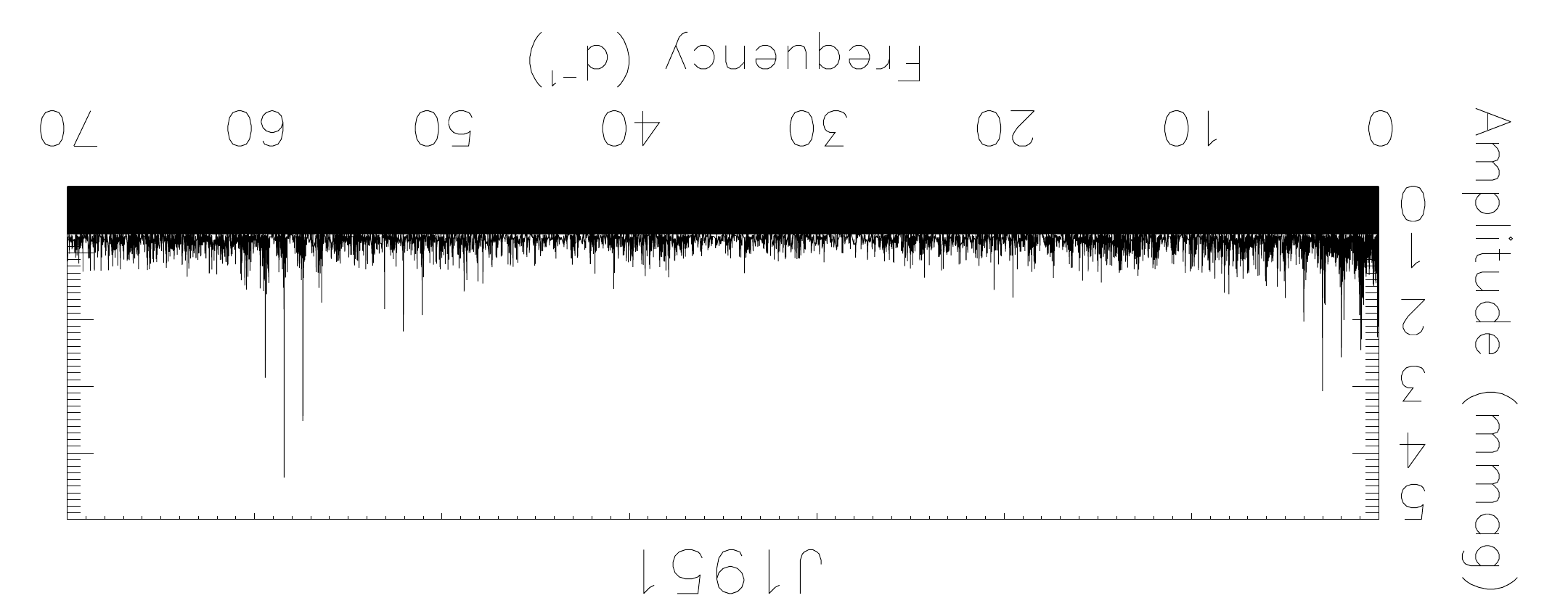}\hfill
    \includegraphics[angle=180,width=80mm, trim= 23mm 21mm 37mm 110mm,clip]{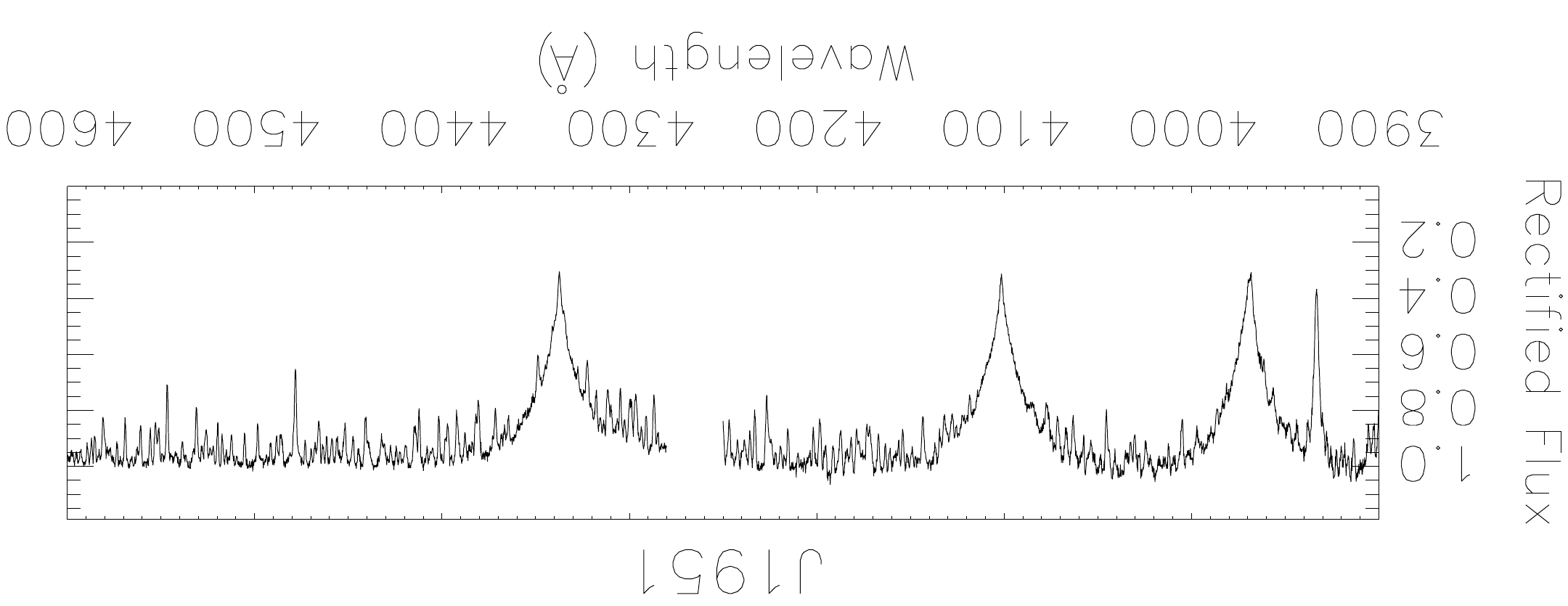}\hfill

    \caption{Continued.}
  \end{minipage}
\end{figure*}

\clearpage
\newpage

\setcounter{figure}{0}
\begin{figure*}
  \centering
  \begin{minipage}{\textwidth}
    \centering

      \includegraphics[angle=180,width=80mm, trim= 23mm 21mm 37mm 110mm,clip]{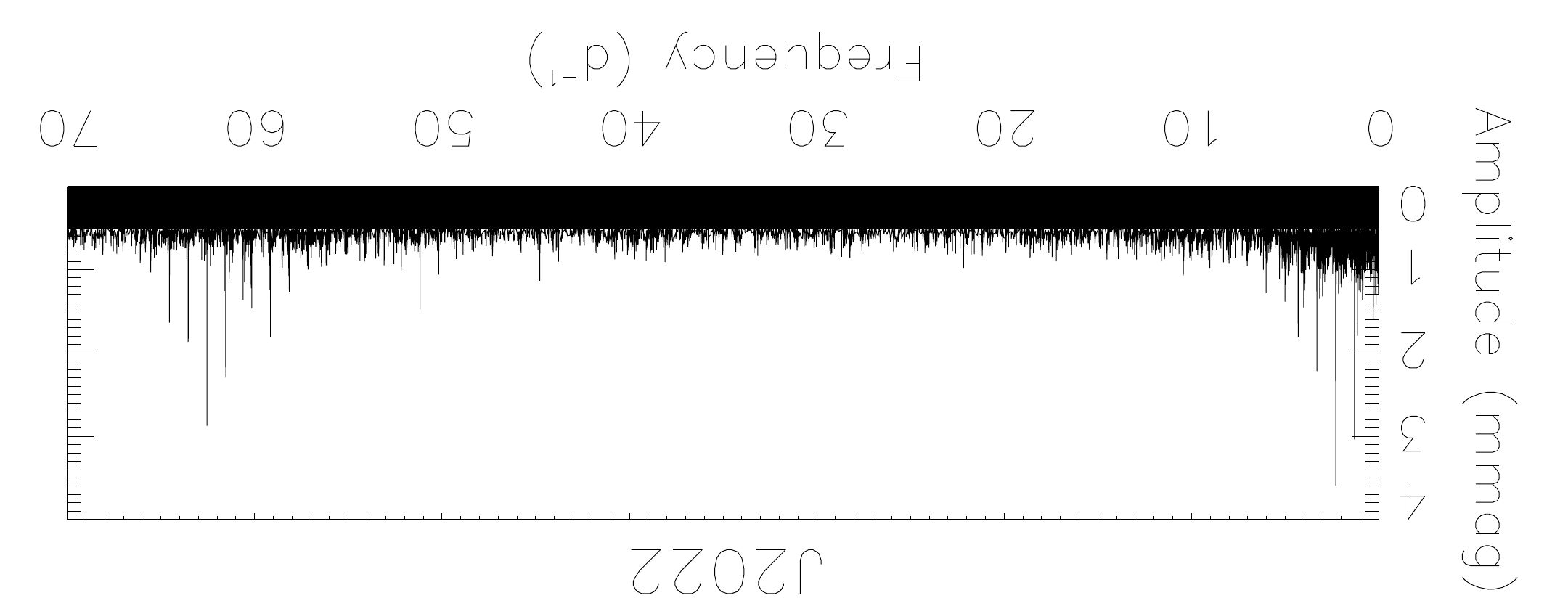}\hfill
      \includegraphics[angle=180,width=80mm, trim= 23mm 21mm 37mm 110mm,clip]{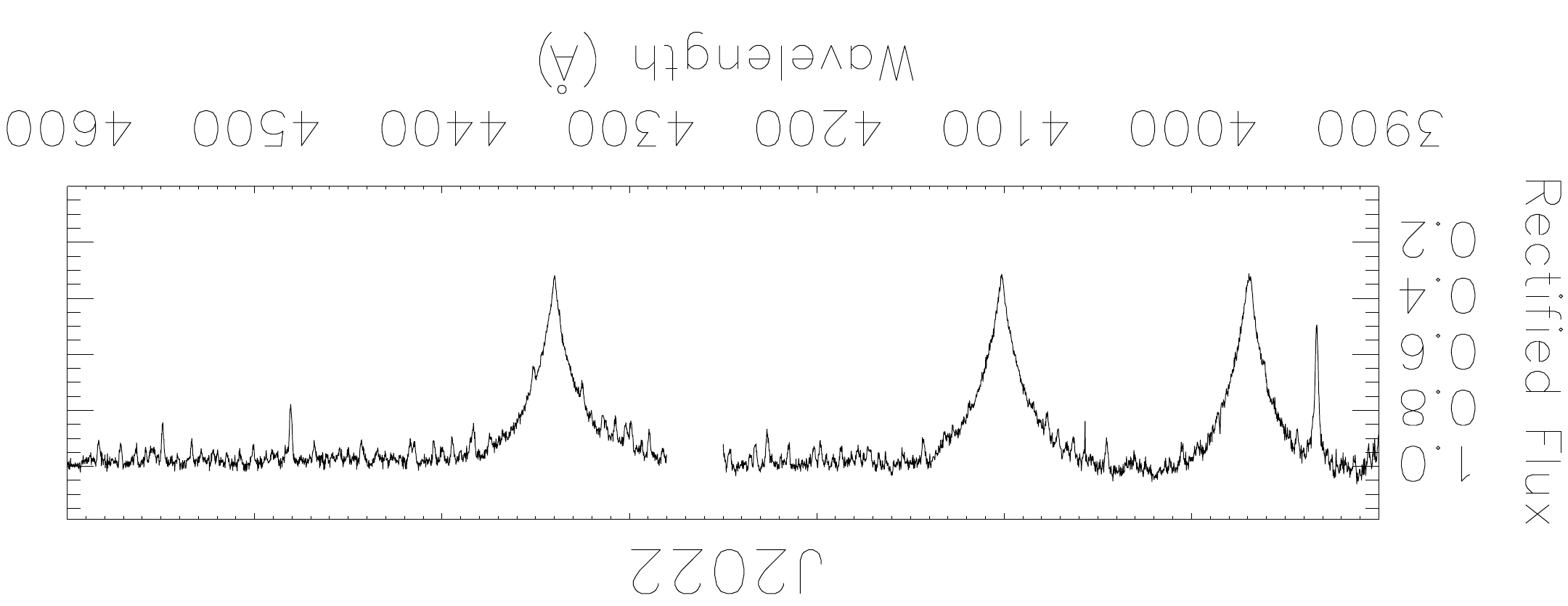}\hfill

      \includegraphics[angle=180,width=80mm, trim= 23mm 21mm 37mm 110mm,clip]{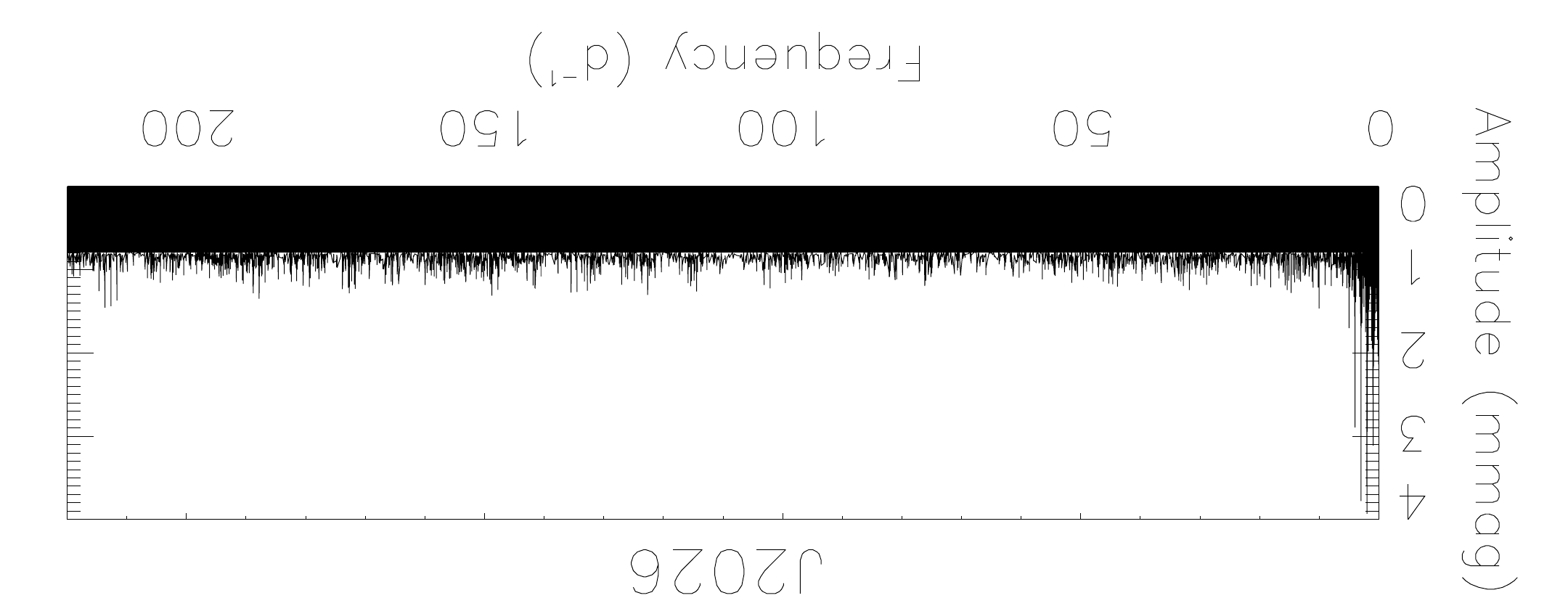}\hfill
      \includegraphics[angle=180,width=80mm, trim= 23mm 21mm 37mm 110mm,clip]{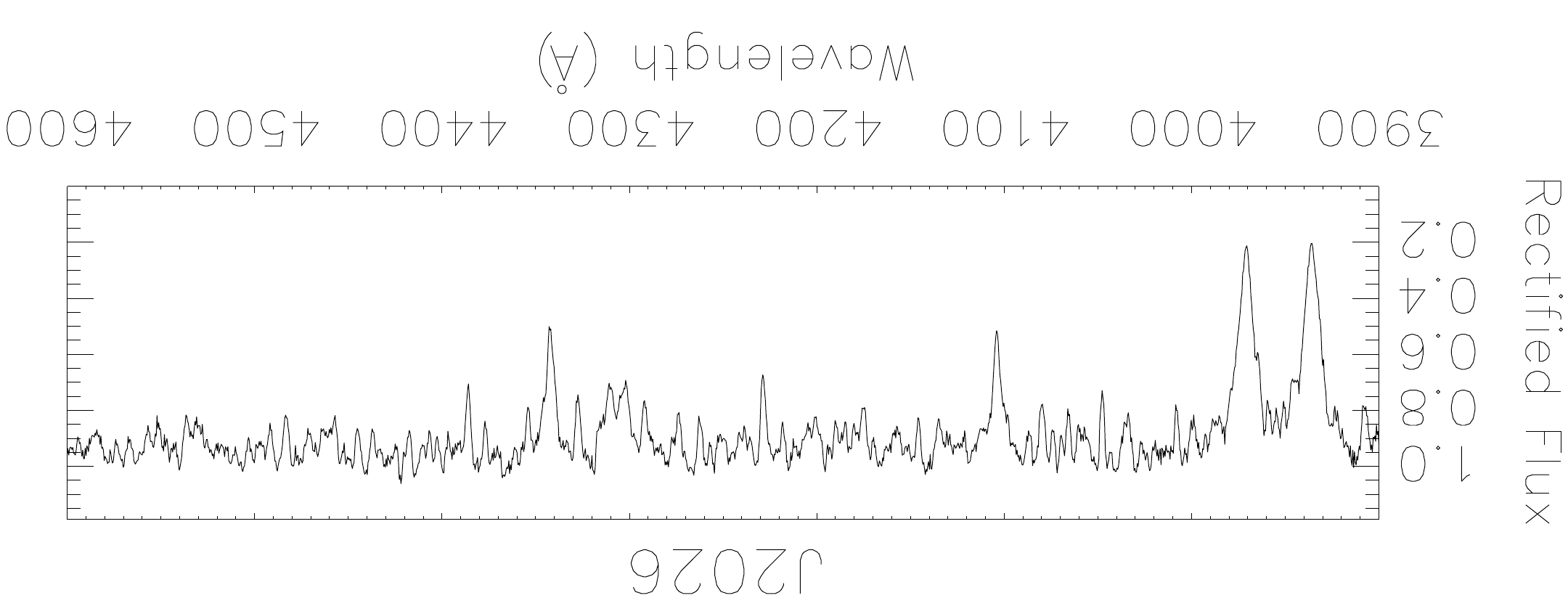}\hfill

      \includegraphics[angle=180,width=80mm, trim= 23mm 21mm 37mm 110mm,clip]{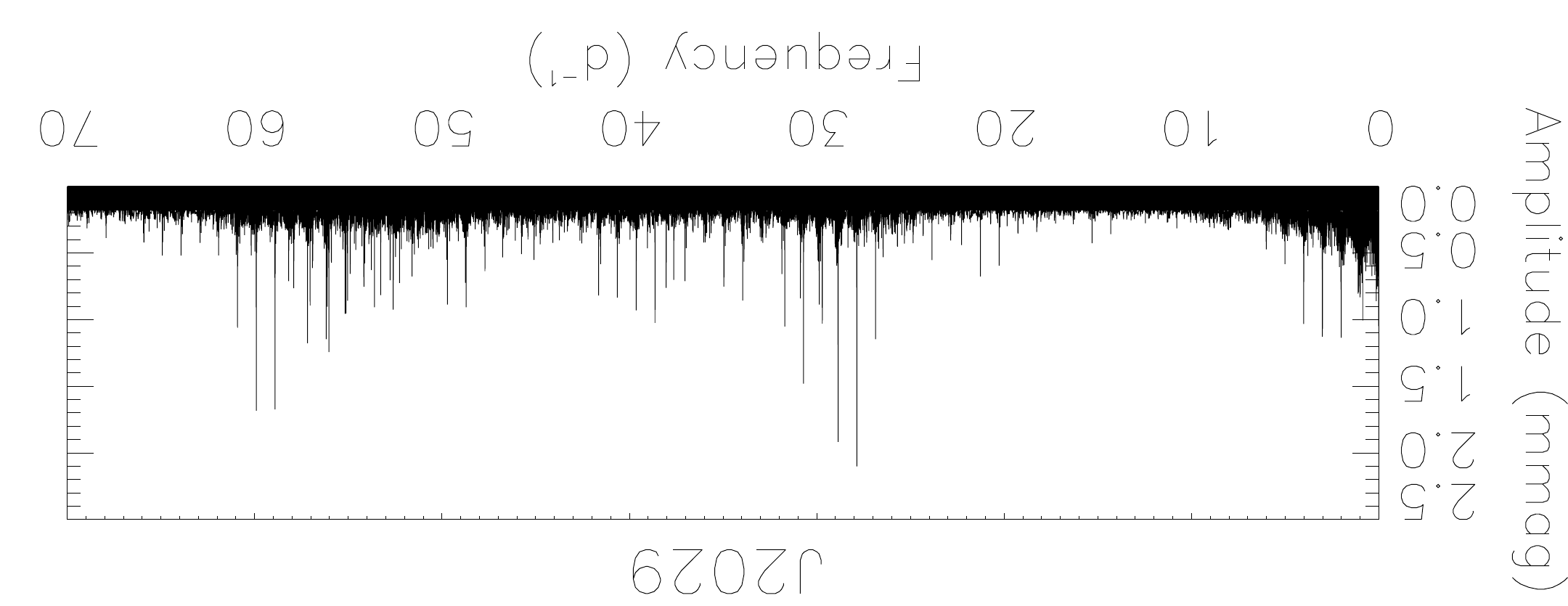}\hfill
      \includegraphics[angle=180,width=80mm, trim= 23mm 21mm 37mm 110mm,clip]{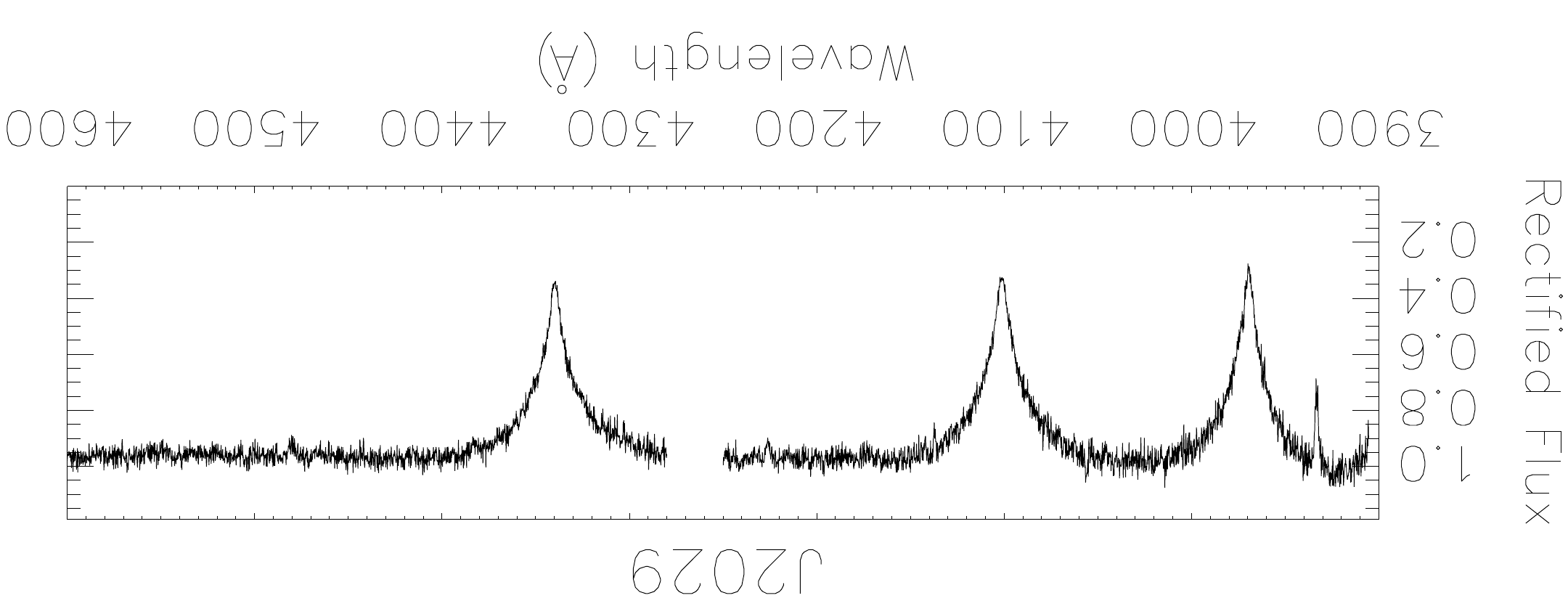}\hfill

      \includegraphics[angle=180,width=80mm, trim= 23mm 21mm 37mm 110mm,clip]{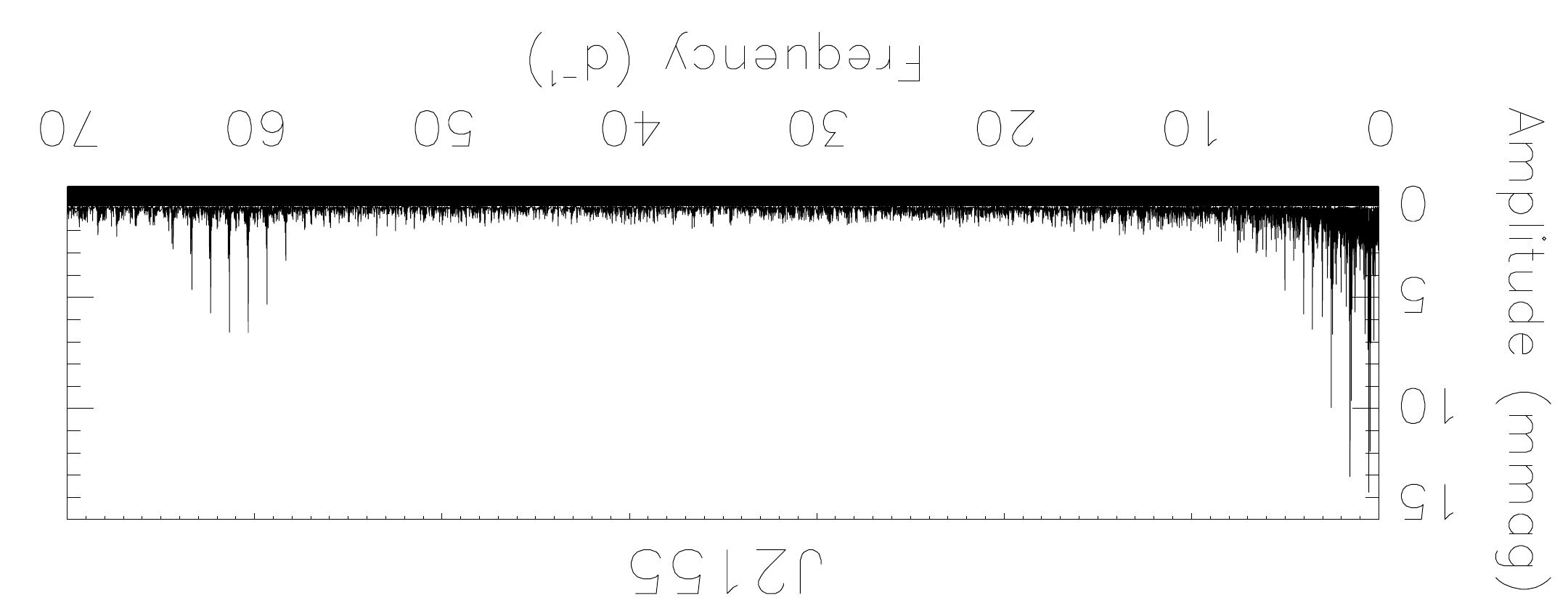}\hfill
      \includegraphics[angle=180,width=80mm, trim= 23mm 21mm 37mm 110mm,clip]{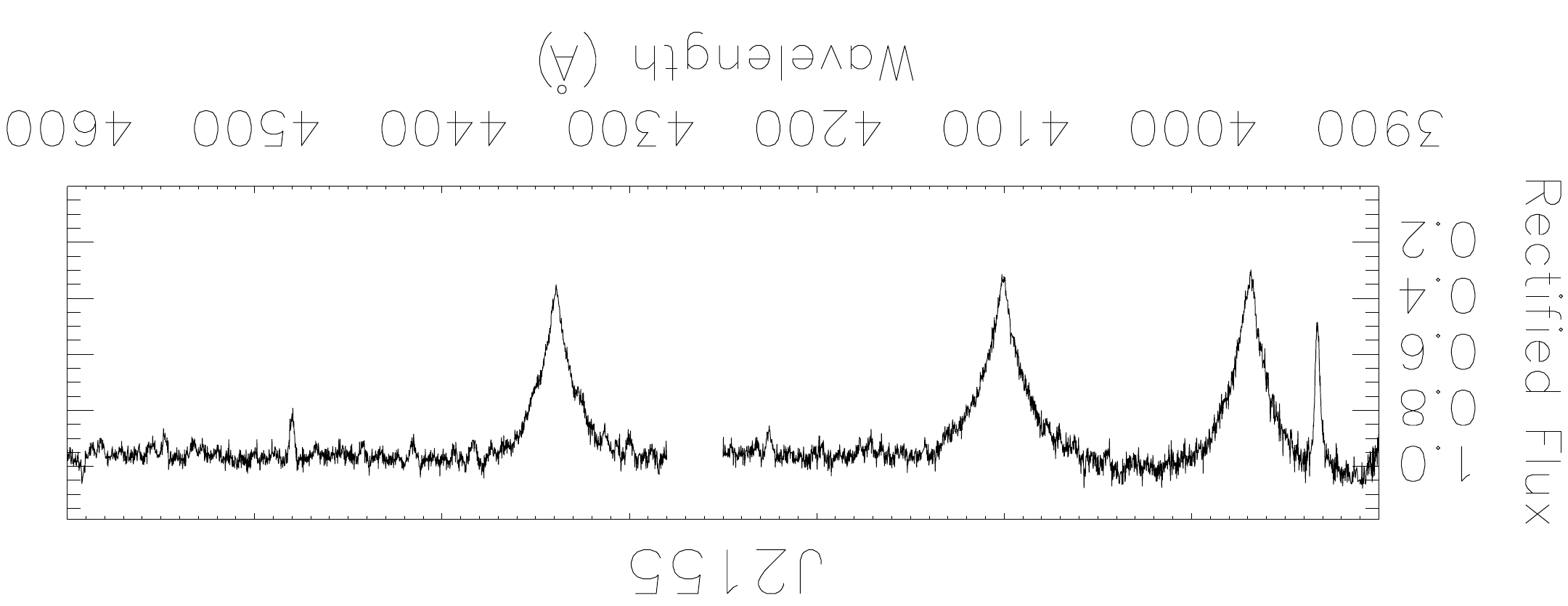}\hfill

      \includegraphics[angle=180,width=80mm, trim= 23mm 21mm 37mm 110mm,clip]{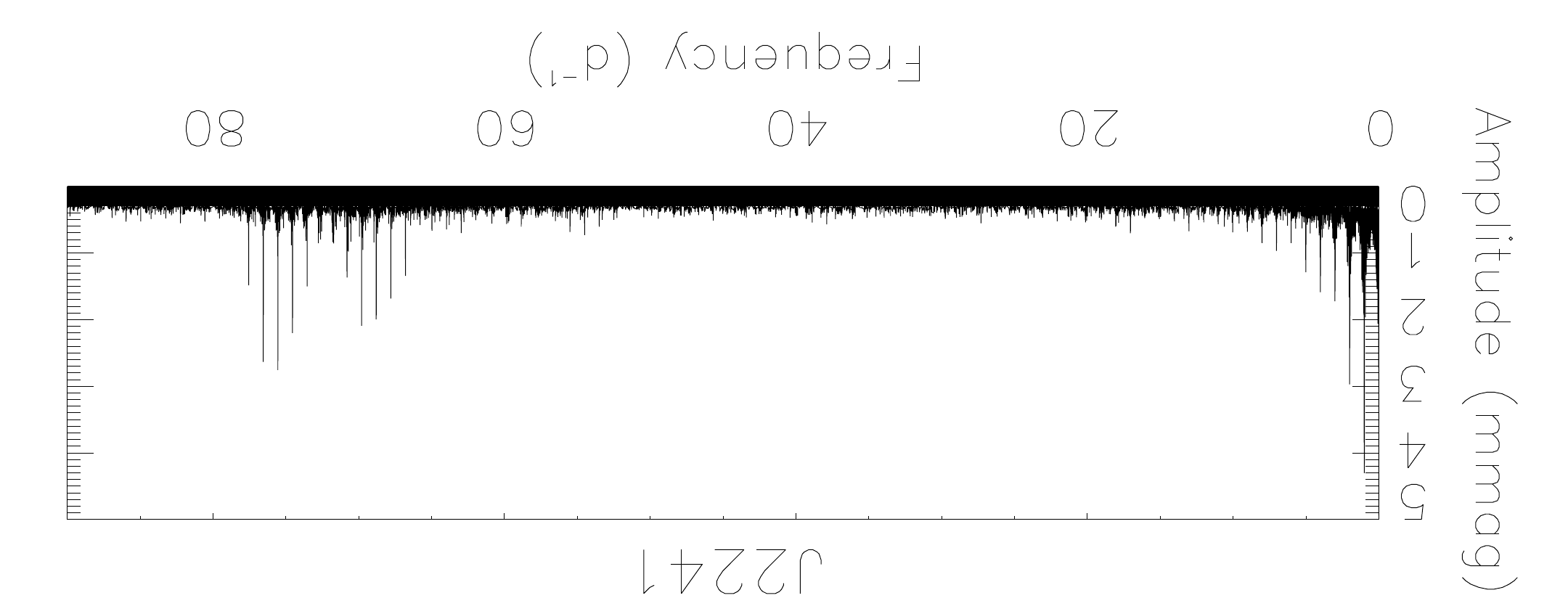}\hfill
      \includegraphics[angle=180,width=80mm, trim= 23mm 21mm 37mm 110mm,clip]{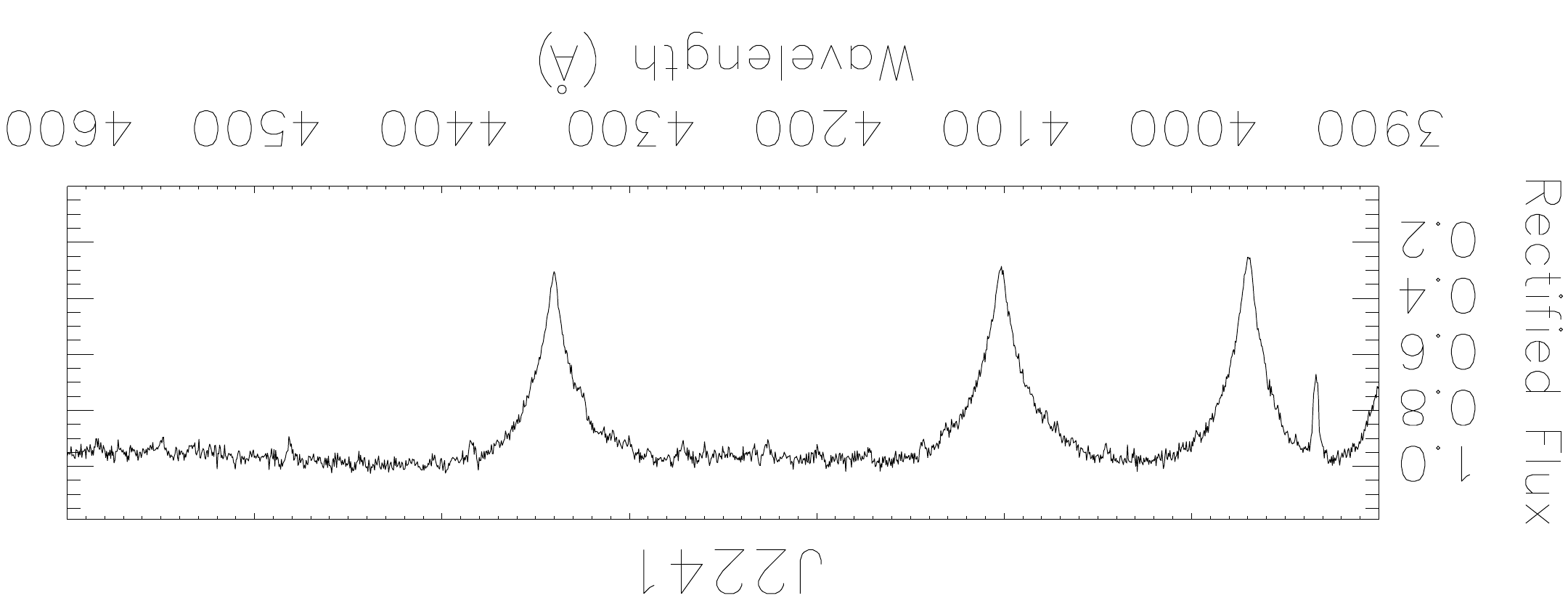}\hfill

      \includegraphics[angle=180,width=80mm, trim= 23mm 21mm 37mm 110mm,clip]{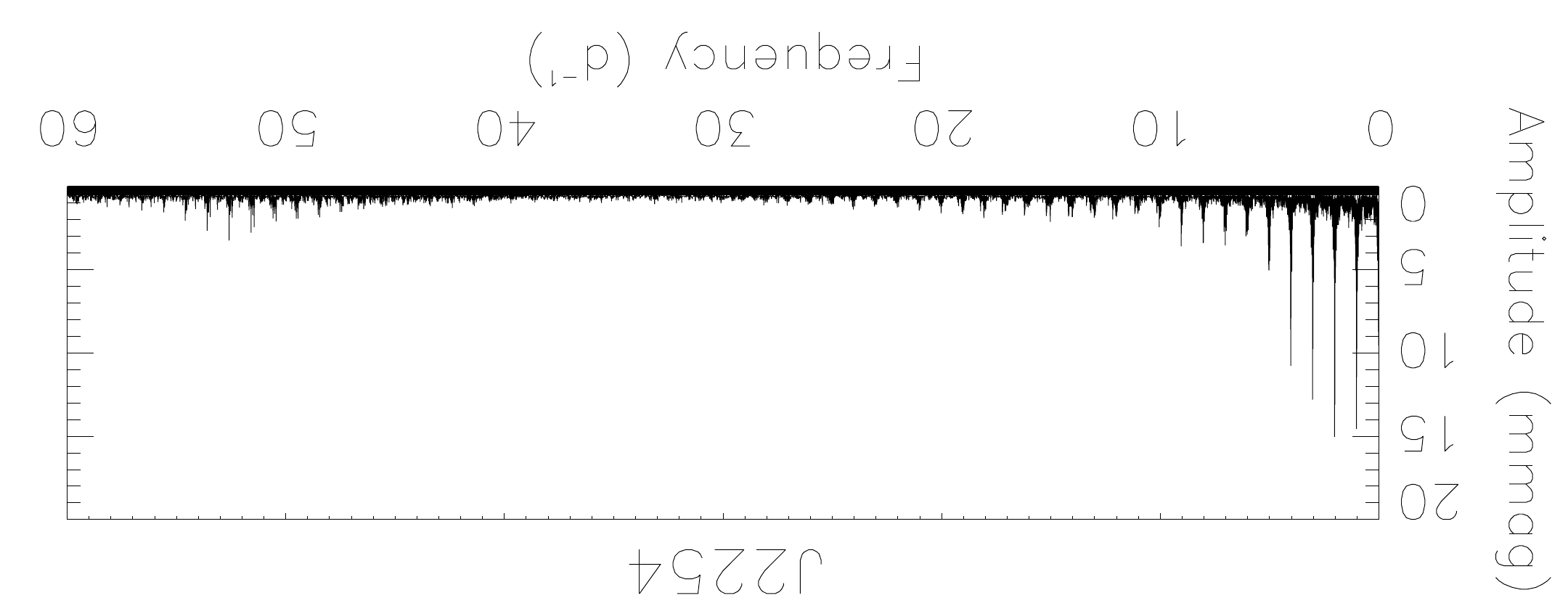}\hfill
      \includegraphics[angle=180,width=80mm, trim= 23mm 21mm 37mm 110mm,clip]{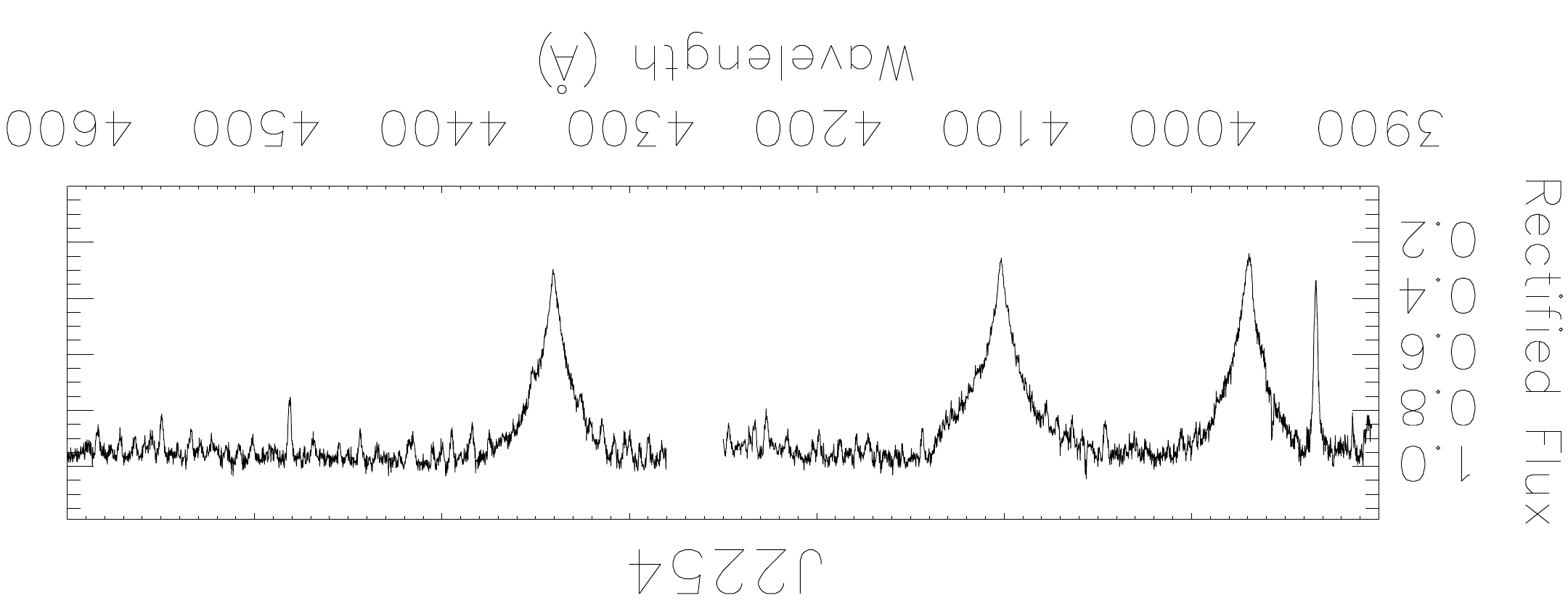}\hfill

      \includegraphics[angle=180,width=80mm, trim= 23mm 21mm 37mm 110mm,clip]{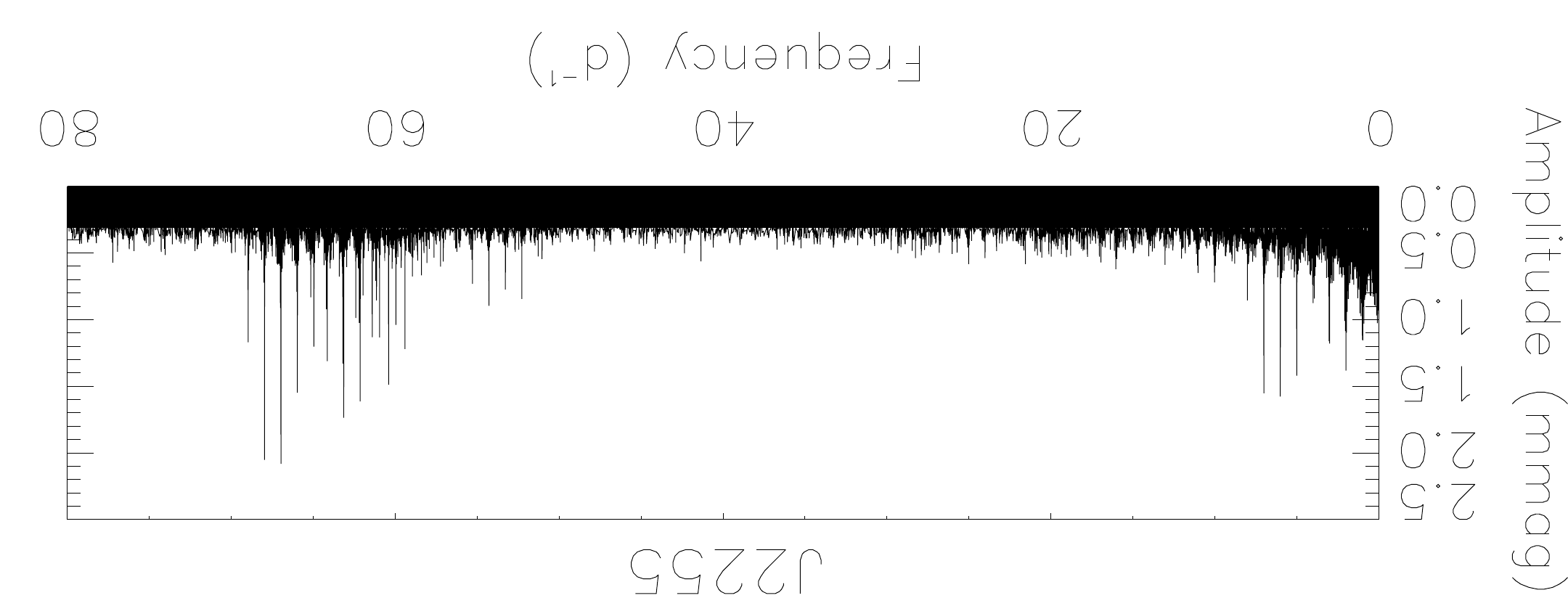}\hfill
      \includegraphics[angle=180,width=80mm, trim= 23mm 21mm 37mm 110mm,clip]{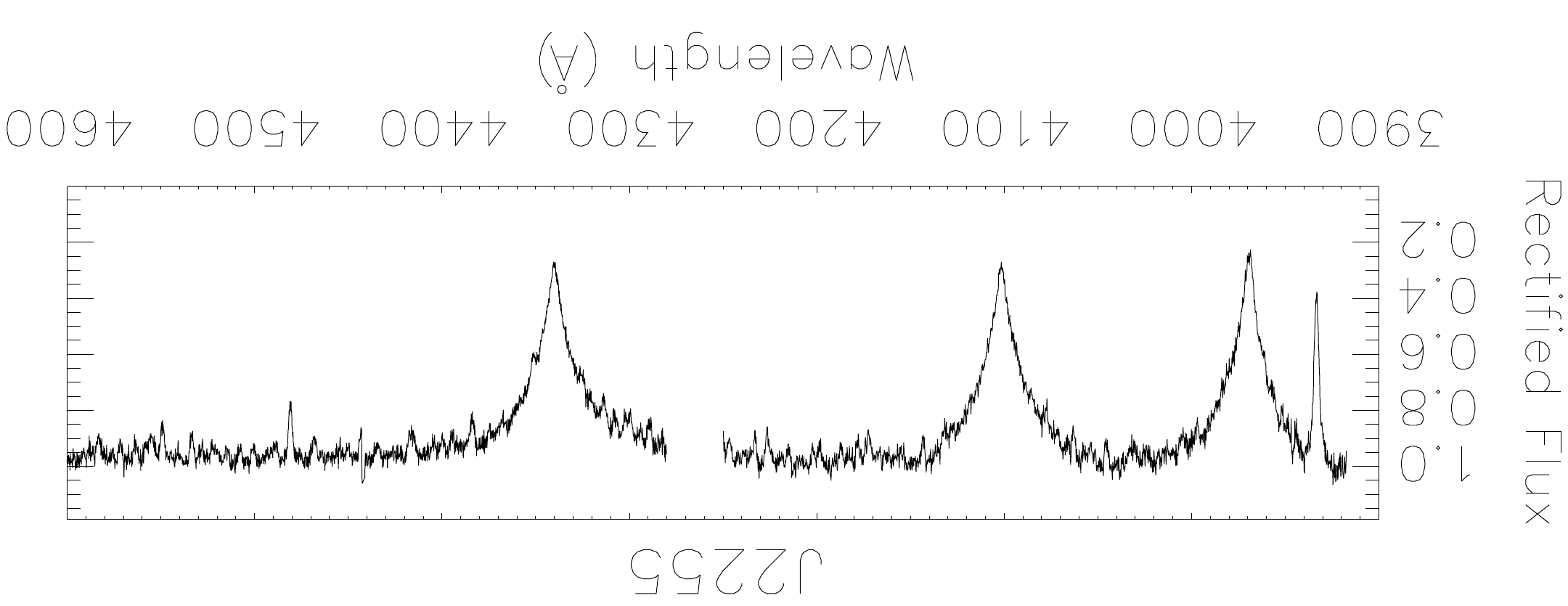}\hfill

      \caption{Continued.}
  \end{minipage}
\end{figure*}
\newpage

\setcounter{figure}{0}
\begin{figure*}
  \centering
  \begin{minipage}{170mm}
    \centering

      \includegraphics[angle=180,width=80mm, trim= 23mm 21mm 37mm 110mm,clip]{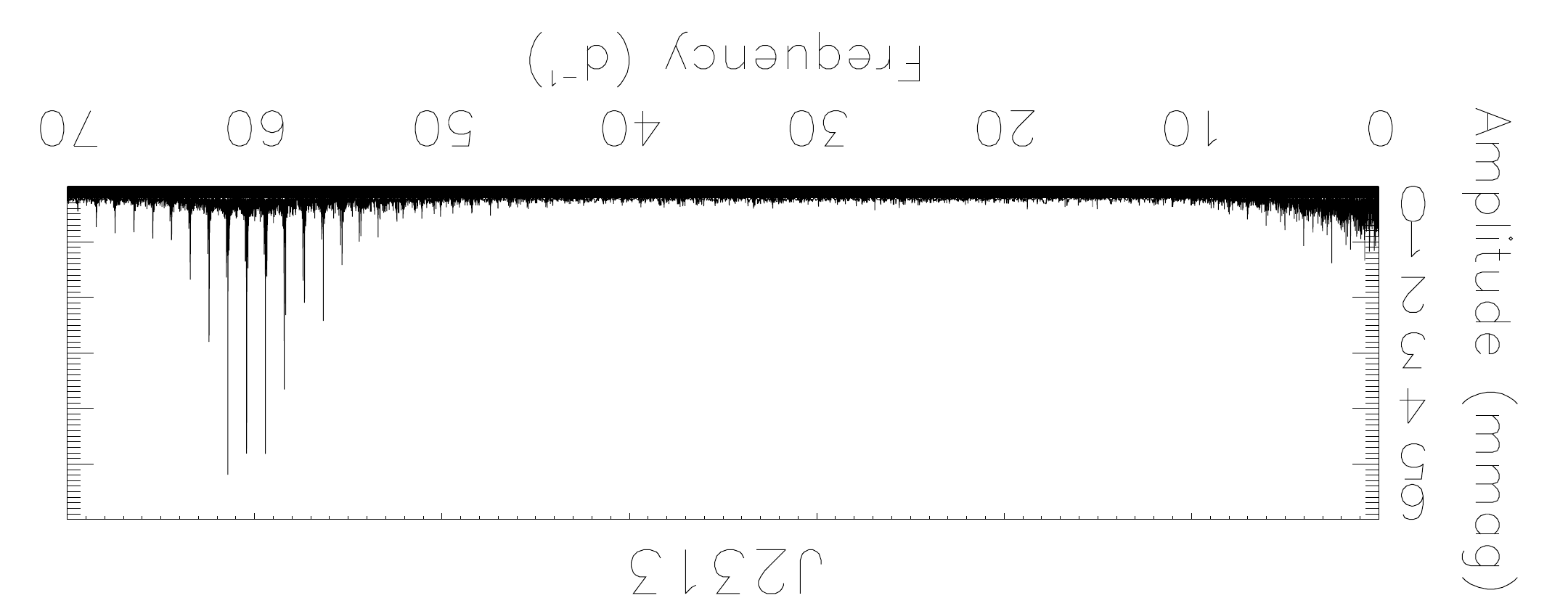}\hfill
      \includegraphics[angle=180,width=80mm, trim= 23mm 21mm 37mm 110mm,clip]{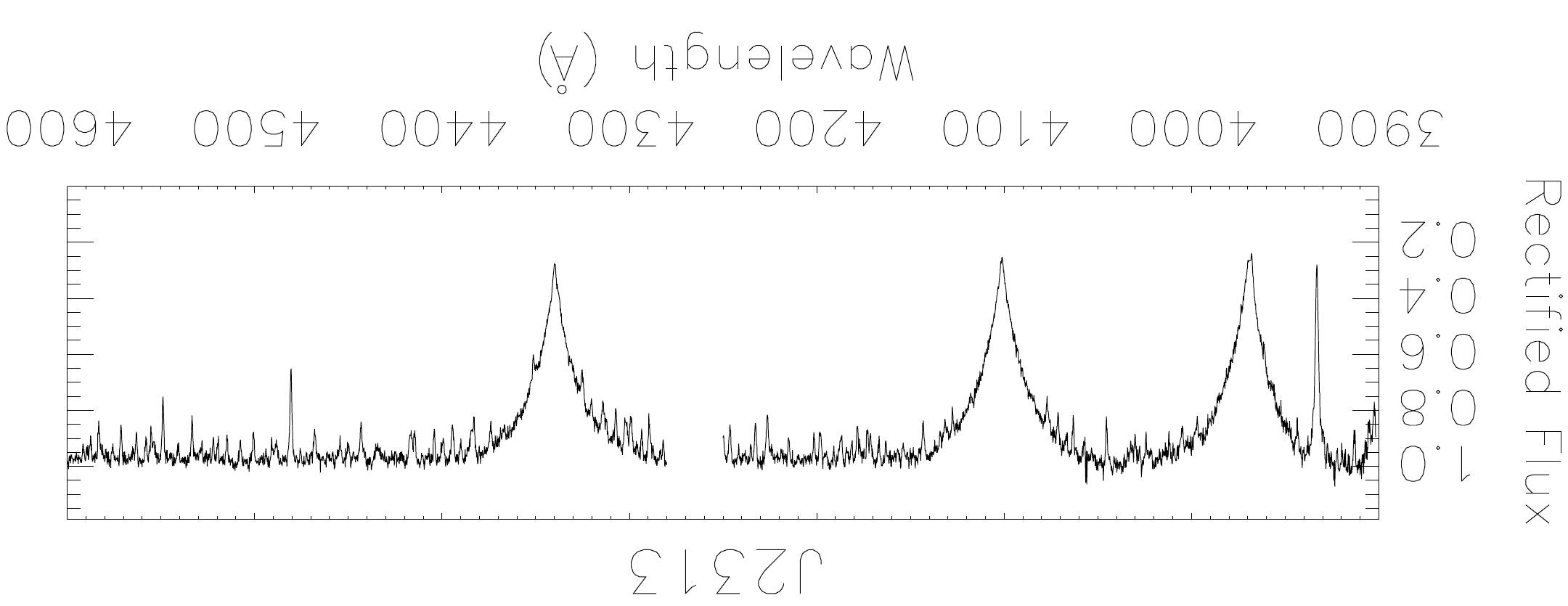}

      \includegraphics[angle=180,width=80mm, trim= 23mm 21mm 37mm 110mm,clip]{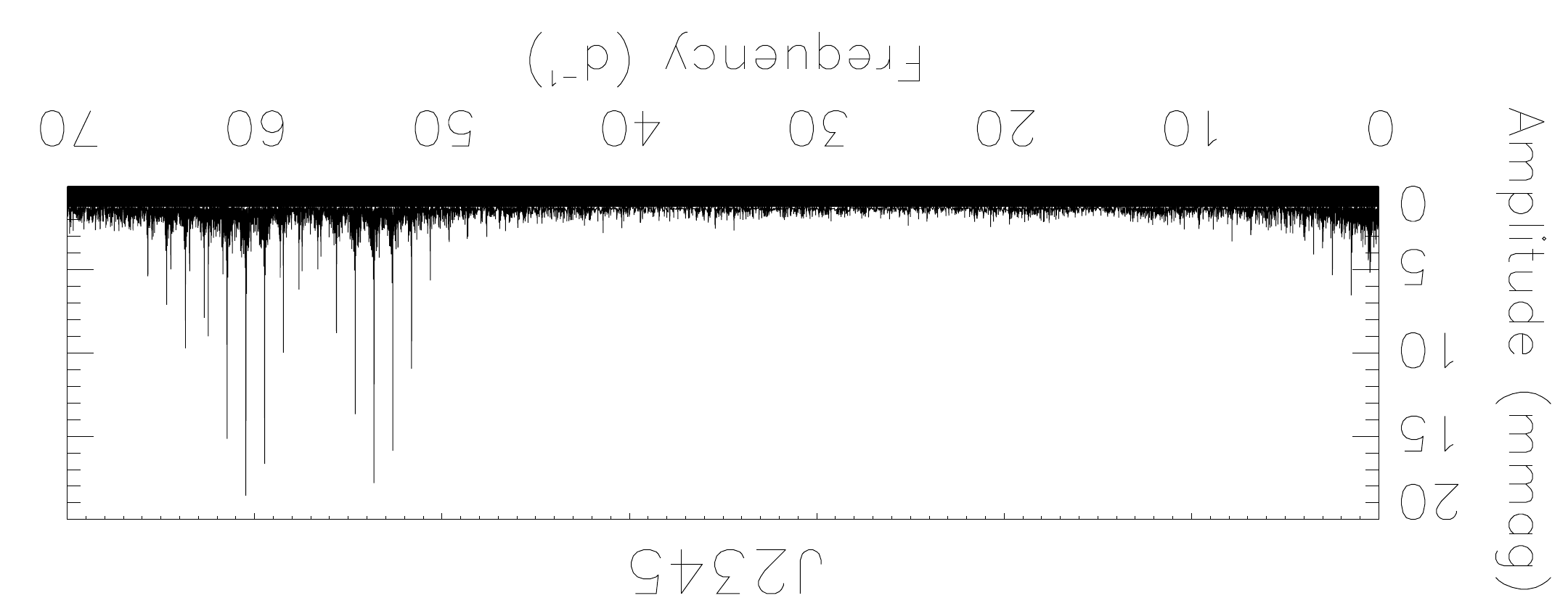}\hfill
      \includegraphics[angle=180,width=80mm, trim= 23mm 21mm 37mm 110mm,clip]{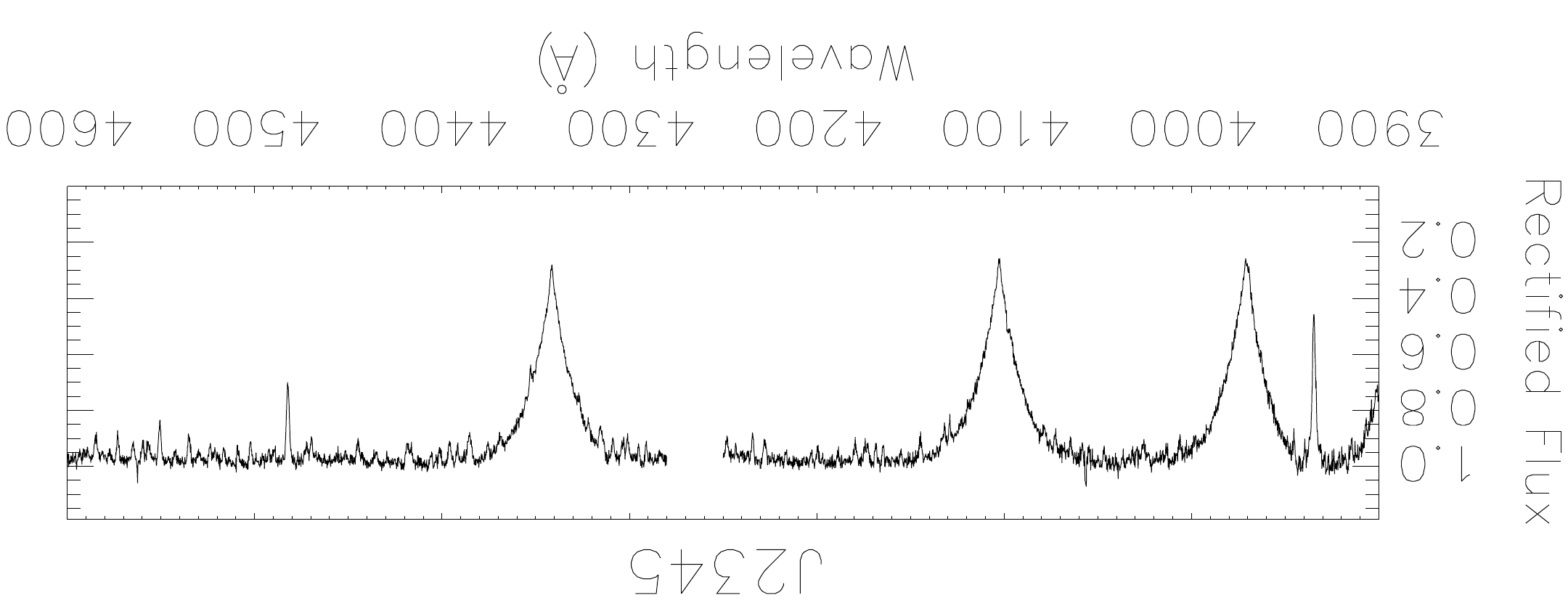}

      \caption{Continued.}
  \end{minipage}
\end{figure*}

\label{lastpage}

\end{document}